%% file: ms.tex
\documentclass[preprint,flushrt]{aastex}
\usepackage{graphicx,natbib,morefloats}
\citeindextrue
\newcommand{\WG}{{\bf W}\ensuremath{_G}}
\newcommand{\PKS}{PKS\ 2155-304}
  
\newcommand{\Hnaught}{\ensuremath{H_0}}
\newcommand{\z}{\ensuremath{z}}
\newcommand{\zi}{\ensuremath{z_{\rm i}}}
\newcommand{\bb}{\ensuremath{b}}
\newcommand{\ang}{~\mbox{\AA}}
\newcommand{\percmthree}{\ {\rm cm}\ensuremath{^{-3}}}
\newcommand{\persecond}{\ {\rm s}\ensuremath{^{-1}}~}
\newcommand{\persecondno}{\ {\rm s}\ensuremath{^{-1}}}
\newcommand{\percmtwo}{\ {\rm cm}\ensuremath{^{-2}}~}
\newcommand{\percmtwono}{\ {\rm cm}\ensuremath{^{-2}}}

\newcommand{\perhzno}{\ {\rm Hz}\ensuremath{^{-1}}}

\newcommand{\persrno}{~{\rm sr}\ensuremath{^{-1}}}
\newcommand{\nomang}{{\rm m}\mbox{\AA}}
\newcommand{\noang}{\mbox{\AA}}
\newcommand{\mang}{~{\rm m}\mbox{\AA}}
\newcommand{\Wcutoff}{133\mang}
\newcommand{\Ang}{\ang\ }
\newcommand{\Mang}{\mang\ }
\newcommand{\Nno}{\ensuremath{\mathcal{N}}}

\newcommand{\Wno}{\ensuremath{\mathcal{W}}}
\newcommand{\W}{\Wno~}
\newcommand{\Ws}{\Wno s}
\newcommand{\Wi}{\ensuremath{\Wno_i}}
\newcommand{\nW}{\ensuremath{n(\Wno)}}
\newcommand{\nWi}{\ensuremath{n(\Wi)}}
\newcommand{\subH}{\ensuremath{_{\rm HI}}}
\newcommand{\Wstar}{\ensuremath{\Wno_\star}\ }
\newcommand{\Wstarno}{\ensuremath{\Wno_\star}}

\newcommand{\etno}{et~al.}
\newcommand{\et}{\etno\ }
\newcommand{\eti}{\etno}
\newcommand{\etl}{\et}

\newcommand{\taueff}{\ensuremath{\tau_{\rm eff}}}
\newcommand{\dtaudz}{d\taueff/\dz}
\newcommand{\hsfi}{\ensuremath{h^{-1}_{70}}}

\newcommand{\zsolar}{\ensuremath{Z_{\odot}}}
\newcommand{\hone}{\ion{H}{1}\ }

\newcommand{\Jno}{\ensuremath{J_0}}
\newcommand{\Jm}{\ensuremath{J_{-23}}}
\newcommand{\Jo}{\Jno~}
\newcommand{\Jnuz}{\ensuremath{J_{\nu}(\z)}}
\newcommand{\Jnuv}{\ensuremath{J_{\nu}}}
\newcommand{\sig}{\ensuremath{\sigma}\ }
\newcommand{\signo}{\ensuremath{\sigma}}
\newcommand{\onesig}{1\signo\ }

\newcommand{\threesig}{3\signo\ }
\newcommand{\foursig}{4\signo\ }
\newcommand{\about}{\ensuremath{\sim}}
\newcommand{\nokmsno}{{\rm km~s}\ensuremath{^{-1}}}
\newcommand{\kmsno}{~\nokmsno}
\newcommand{\nommsno}{Mm\persecondno}
\newcommand{\mmsno}{~\nommsno}
\newcommand{\mms}{\mmsno\ }

\newcommand{\kms}{\kmsno\ }

\newcommand{\lya}{Ly\ensuremath{\alpha} }
\newcommand{\lyano}{Ly\ensuremath{\alpha}}
\newcommand{\lyb}{Ly\ensuremath{\beta} }

\newcommand{\Nhno}{\ensuremath{N_{\rm HI}}}
\newcommand{\logNh}{\ensuremath{\log{\left[\Nhno\right]}}}
\newcommand{\Nh}{\Nhno\ }
\newcommand{\lowrange}{\ensuremath{12.3 \leq  \logNh  \leq  14.0}}
\newcommand{\highrange}{\ensuremath{14.0 \leq  \logNh  \leq 16.0}}
\newcommand{\upperrange}{\ensuremath{13.5 \leq  \logNh  \leq 16.0}}
\newcommand{\fullrange}{\ensuremath{12.3 \leq  \logNh  \leq 16.0}}

\newcommand{\nofh}{\ensuremath{n_{\rm H}}}
\newcommand{\fhI}{\ensuremath{f_{\rm HI}}}
\newcommand{\Nmin}{\ensuremath{N_{\rm min}}}
\newcommand{\nNh}{\ensuremath{n(\Nhno)}}
\newcommand{\iNh}{\ensuremath{I(\Nhno)}}
\newcommand{\nNhno}{\ensuremath{n(\Nhno)}}
\newcommand{\zgamma}{\ensuremath{(1+\z)^\gamma}}
\newcommand{\zgammai}{\ensuremath{(1+\zi)^\gamma}}
\newcommand{\nz}{\ensuremath{n(\z)}}
\newcommand{\nzi}{\ensuremath{n(\zi)}}
\newcommand{\nh}{\Nh}
\newcommand{\myb}{25\kmsno}

\newcommand{\cgs}{ {\rm ergs} {\rm cm}\ensuremath{^{-2}}\persecondno\perhzno\persrno}
\newcommand{\bobs}{\bb}

\newcommand{\bmsd}{\ensuremath{b_{\rm obs}}}
\newcommand{\definite}{\ensuremath{SL \ge 4\signo}}
\newcommand{\expanded}{\ensuremath{SL \ge 3\signo}}
\newcommand{\tent}{\ensuremath{3\signo \le SL < 4\signo}}
\newcommand{\real}{\ensuremath{SL \ge 4\signo}}

\newcommand{\cDz}{\ensuremath{c\Delta}\z}
\newcommand{\Dz}{\ensuremath{\Delta}\z}
\newcommand{\delz}{\ensuremath{\delta}\z}
\newcommand{\DW}{\ensuremath{\Delta}\Wno}
\newcommand{\DNh}{\ensuremath{\Delta}\Nhno}
\newcommand{\Dv}{\ensuremath{\Delta v}}
\newcommand{\recom}{\ensuremath{\alpha^{(A)}_H}}
\newcommand{\photo}{\ensuremath{\Gamma_{\rm HI}}}
\newcommand{\dz}{{\rm d}\z}
\newcommand{\pz}{\partial\z}
\newcommand{\Nobs}{\ensuremath{N_{\rm obs}}}
\newcommand{\Nran}{\ensuremath{N_{\rm ran}}}
\newcommand{\Nsym}{C}
\newcommand{\Nofz}{\ensuremath{\Nsym_\z}}
\newcommand{\NofNh}{\ensuremath{\Nsym_{\rm HI}}~}
\newcommand{\NofNhno}{\ensuremath{\Nsym_{\rm HI}}}
\newcommand{\NofW}{\ensuremath{\Nsym_\mathcal{W}}}
\newcommand{\betaNh}{\ensuremath{\NofNh \Nhno^{-\beta}}}
\newcommand{\gt}{\ensuremath{>}}
\newcommand{\lt}{\ensuremath{<}}
\newcommand{\Wlya}{\Wno$_{\lyano}$}
\newcommand{\meanz}{\ensuremath{\langle \z \rangle}}

\newcommand{\zrange}{\ensuremath{0.002 < \z < 0.069}}

\newcommand{\zem}{\ensuremath{z_{\rm em}}}
\newcommand{\prox}{c\zem -- 1,200\kms}
\newcommand{\proxno}{c\zem -- 1,200\kmsno}
\newcommand{\lowz}{low-\z\ }
\newcommand{\highz}{high-\z\ }
\newcommand{\lowzya}{low-\z\ \lya}
\newcommand{\dn}{{\rm d}\Nno}
\newcommand{\dNh}{{\rm d}\Nh}

\newcommand{\pNh}{\partial\Nh}

\newcommand{\dW}{{\rm d}\Wno}
\newcommand{\pNhno}{\partial\Nhno}
\newcommand{\pW}{\partial\Wno}
\newcommand{\dndz}{\dn/\dz\ }
\newcommand{\dndzover}{\ensuremath{{\dn(\Nhno) \over \dz}}}
\newcommand{\dndzno}{\dn/\dz}
\newcommand{\dndzzero}{(\dndzno)$_{\z=0}$}
\newcommand{\zonelog}{\ensuremath{\log\left[1+\z\right]}}
\newcommand{\dndzlog}{\ensuremath{\log\left[\dndzno\right]}}

\newcommand{\dndW}{\dn/d\Wno}

\newcommand{\dndnh}{\dn/\dNh\ }

\newcommand{\dnhdWover}{\ensuremath{{\dNh \over \dW}}}

\newcommand{\Mcl}{\ensuremath{M_{\rm cl}}}
\newcommand{\phio}{\ensuremath{\phi_0}}
\newcommand{\cprime}{A}
\newcommand{\dtwondnhdzover}{\ensuremath{{\partial^2 \Nno \over \pz~\pNh}}}
\newcommand{\dtwondnhdz}{\ensuremath{\partial^2\Nno/\pz~\pNh}}
\newcommand{\dtwondnhdzno}{\ensuremath{\partial^2\Nno/\pz~\pNhno}}
\newcommand{\dtwondzdnhno}{\dtwondnhdzno}

\newcommand{\dtwondzdnh}{\dtwondnhdz}
\newcommand{\dtwondWdz}{\ensuremath{{\partial^2\Nno/\pz~\pW}}}
\newcommand{\dtwondWdzover}{\ensuremath{{\partial^2 \Nno \over \pz~\pW}}}

\newcommand{\cz}{\ensuremath{cz}}

\newcommand{\lognh}{\logNh}
\newcommand{\kimrange}{\ensuremath{13.1\lt\lognh\lt14.0}}

\newcommand{\Exp}[2]{\ensuremath{#1\times10^{#2}}}
\newcommand{\bvalue}{\bb-value\ }
\newcommand{\bvalues}{\bb-values\ }
\newcommand{\bvalueno}{\bb-value}
\newcommand{\bvaluesno}{\bb-values}
\newcommand{\lam}{\ensuremath{\lambda}}

\newcommand{\Dlam}{\ensuremath{\Delta\lam}}
\newcommand{\Plam}{\ensuremath{P(\lam)}}
\newcommand{\Zsi}{\ensuremath{Z_{\rm Si}}}
\newcommand{\Zsolar}{\ensuremath{Z_{\odot}}}

\newcommand{\CIII}[1]{\ion{C}{3}~\lam#1}
\newcommand{\CIV}[1]{\ion{C}{4}~\lam#1}
\newcommand{\FeII}[1]{\ion{Fe}{2}~\lam#1}
\newcommand{\SII}[1]{\ion{S}{2}~\lam#1}
\newcommand{\SiII}[1]{\ion{Si}{2}~\lam#1}
\newcommand{\Sithree}{\ion{Si}{3}~\lam1206.5}

\newcommand{\SiIII}[1]{\ion{Si}{3}~\lam#1}
\newcommand{\SFsixty}{\SiII{1260.4}\ +\ \FeII{1260.5}}
\newcommand{\NVdoublet}{\ion{N}{5}~\ensuremath{\lambda\lambda}1238, 1242}

 \newcommand{\omegab}{\ensuremath{\Omega_{\rm b}}}
\newcommand{\omegacl}{\ensuremath{\Omega_{\lyano}}}

\newcommand{\impact}{\ensuremath{p}}
\newcommand{\Nabs}{81}
\newcommand{\Npos}{30}
\newcommand{\Ntot}{111}

\begin{document}
\title{The Local \lya Forest. II. Distribution of \\
  \ion{H}{1} Absorbers, Doppler Widths, and Baryon Content
\footnote{Based on observations with the NASA/ESA Hubble Space Telescope, obtained at the Space
Telescope Science Institute, which is operated by the Association of Universities for Research in
Astronomy, Inc. under NASA contract No. NAS5-26555.}}
\author{Steven V. Penton, J. Michael Shull\footnote{Also at JILA, University of Colorado
and National Institute of Standards and Technology.}, and John T. Stocke}
\affil{Center for Astrophysics and Space Astronomy, Department of Astrophysical and Planetary Sciences,
University of Colorado, Boulder CO, 80309}
\email{spenton@casa.colorado.edu, mshull@casa.colorado.edu, stocke@casa.colorado.edu}
\shorttitle{Statistics of the \lowya Forest}
\shortauthors{Penton, Shull, \& Stocke}
\begin{abstract}In Paper~I of this series we described observations
of 15 extragalactic targets taken with the Hubble Space Telescope+GHRS+G160M grating 
for studies of the \lowzya forest.  We reported the detection of
\Ntot\ Ly$\alpha$ absorbers at significance level (SL) $\geq 3 \sigma$, \Nabs\ with \real,
in the redshift range \zrange, over a total pathlength
$c \Delta z = 116,000$\kms ($\Delta z = 0.387$).
In this second paper, we evaluate the physical
properties of these \lya absorbers and compare them to their \highz
counterparts. The \real\ distribution of Doppler parameters is similar to that
at high redshift, with $\langle $\bobs $\rangle =
38.0 \pm 15.7$\kmsno.  The true Doppler parameter may be somewhat lower, owing to component blends and
non-thermal velocities.   The distribution of equivalent widths exhibits
a significant break at $\Wno \leq 133$\mang, with an increasing number of
weak absorbers (10\Mang$< \Wno < 100$\mang). Adopting a curve of growth with
$\bb = 25 \pm 5$\kms and applying a sensitivity correction as
a function of equivalent width and wavelength,
we derive the distribution in column density,
\Nhno$^{-1.72 \pm 0.06}$ for $12.5 \leq \lognh \le 14.0$.
We find no redshift evolution within
the current sample at $z < 0.07$, but we do see a significant decline in
\dndz compared to values at $z > 1.6$.
Similiar to the high equivalent width ($\Wno \gt 240\mang$) absorbers, the number density of 
low-\W absorbers at \z=0 is well above the extrapolation of \dndz from $\z \gt 2$, but
we observe no difference in the mean evolution of \dndz between absorbers of  high ($\Wno \gt 240\mang$) and 
low ($\Wno \le 100\mang$) equivalent width absorbers. While previous work has suggested slower 
evolution in number density of lower-\W absorbers, our new data do not support this conclusion. A consistent evolutionary
pattern is that the slowing in the evolution of the low column
density clouds occurs at lower redshift than for the higher column density clouds.
A $4-5\sigma$ signal in the two-point correlation function
of \lya absorbers for velocity separations $\Delta v \leq 150$\kms is
consistent with results at high-$z$, but with somewhat greater amplitude.
Applying a photoionization correction, we find that the \lowzya
forest may contain $\sim20$\% of the total
number of baryons, with closure parameter
$\Omega_{Ly\alpha} = (0.008 \pm 0.001) h_{70}^{-1}$,
for a standard absorber size and  ionizing radiation field.
Some of these clouds appear to be primordial matter, owing to the
lack of detected metals (\ion{Si}{3}) in a composite spectrum, although current limits on composite metallicity are not strong.
\end{abstract}
\section{Introduction}\label{sec:obs}
Since the discovery of the high-redshift \lya forest over
25 years ago, these abundant absorption features in the spectra
of QSOs have been used as evolutionary probes of the intergalactic
medium (IGM), galactic halos, large-scale structure, and chemical 
evolution.  Absorption in the \lya forest of \hone (and \ion{He}{2}) is 
also an important tool for studying the high-redshift universe
 \citep{ME90,Shapiro94,Fardal98}.  A comparison of the \hone and \ion{He}{2}
absorption lines provides constraints on the photoionizing background
radiation, on the history of structure formation, and on internal
conditions in the \lya clouds.  In the past few years,
these discrete \lya lines have been interpreted in the context of
N-body hydrodynamical models \citep{Cen94,Hernquist96,Zhang97,Dave99}
 as arising from baryon density
fluctuations associated with gravitational instability during structure formation.  
The effects of hydrodynamic shocks,
Hubble expansion, photoelectric heating by AGN, and galactic
outflows and metal enrichment from early star formation must all
be considered in understanding the IGM.

At high redshift, the rapid evolution in the distribution of 
\lya absorption lines per unit redshift, 
$d{\cal N}/dz \propto (1+z)^{\gamma}$ ($\gamma \approx 2.5$ 
for $z \geq 1.6$), was consistent with a
picture of these features as highly ionized ``clouds'' whose numbers
and sizes were controlled by the evolution of the IGM pressure, the
metagalactic ionizing radiation field, and galaxy formation.
However, the rapid evolution of the \lya forest stopped
at $z < 1.6$.   One of the delightful spectroscopic surprises from the
{\it Hubble Space Telescope} (HST) was the discovery of \lya 
absorption lines toward the quasar 3C~273 at $z_{\rm em} = 0.158$ by 
both the Faint Object Spectrograph \citep[FOS, ][]{Bahcall91} and 
the Goddard High Resolution Spectrograph \citep[GHRS, ][]{Morris91,Morris93}.  
The number of these absorbers was found to be far in excess of their expected number
based upon an extrapolation from high-\z\ (see e.g. \citet{Weymann98} and section 5 therein).
This evolutionary shift is probably a result of the collapse and 
assembly of baryonic structures in the IGM \citep{Dave99} 
together with the decline in the intensity of the 
ionizing radiation field \citep{Haardt96,Shull99b}. 
Detailed results of the \lya forest evolution in the redshift 
interval $0 < z < 1.5$ are described in the FOS Key Project papers:  
the three catalog papers  \citep{Bahcall93,Bahcall96,Jannuzi98} 
and the evolutionary analysis study \citep{Weymann98}.

However, the HST/FOS studies were primarily probes of strong \lya
lines with equivalent widths greater than 0.24 \AA.  
A great deal more information about the low-$z$ IGM can be gained
from studies of the more plentiful weak absorbers.   Realizing the 
importance of spectral resolution in detecting weak \lya
absorbers, the Colorado group has engaged in a long-term program with
the HST/GHRS, using the G160M grating at 19 km~s$^{-1}$ resolution,
to study the very low-redshift ($z < 0.07$) \lya forest.
Earlier results from our study have appeared in various research papers 
\citep{Stocke95,Shull96,pks} and reviews \citep{Shull97,Shull99a}.

The current series of papers discusses
our full GHRS dataset.  In Paper~I \citep{PaperI}
we described our HST/GHRS observations and catalog of \lya
absorbers.  In Paper II (this article)
we describe the physical results from our program, including
information on the physical parameters and nature of the 
low-redshift \lya forest. A discussion of the relationship between
the \lya absorbers discovered in our GHRS program and the nearby
galaxy distribution as mapped using available galaxy redshift survey
data will be presented in Paper~III \citep{PaperIII}.  We believe that the low-redshift 
\lya forest of absorption lines, combined with information
about the distribution of nearest galaxies, can provide
a probe of large-scale baryonic structures in the IGM, some of which may be remnants of physical conditions
set up during galaxy formation.
 In Tables~1 and~2 we present the basic data for the definite ($\geq 4\sigma$) and
 possible ($3-4\sigma$) \lya absorbers from Paper~I. 
In general, scientific results will be determined for only the $\geq 4\sigma$ absorbers in Table~1, with data
using the expanded ($\geq 3\sigma$) sample (Tables~1 and 2) shown only as corroborating.
 Only the absorbers
 judged to be intervening (``intergalactic'') are included ; see
 Paper I for detailed criteria and analysis. We exclude all Galactic metal-line
 absorbers and absorbers ``intrinsic'' to the AGN, with  $(cz_{AGN} - cz_{abs}) < 1200$\kmsno; see Paper~I. 
The information in  Tables~1 and~2  by column is: (1) name of target AGN; (2) absorber
 wavelength and error in \ang; (3) absorber recession velocity and error quoted
 non-relativistically as \cz~in\kmsno; (4) observed \bvalue (\bmsd) and
 error in \kmsno; (5) resolution-corrected \bvalue and error in\kmsno; (6) rest-frame equivalent
 width and error in\mang; (7-10) estimated column densities in cm$^{-2}$ assuming
 \bvalues of 20, 25, 30\kms and the value from column (5). 
 Detailed descriptions of  the determination of values in columns (1-6) can be found in Paper~I.
As described in detail in Paper~I, the uncertainties for the \W values in column 6 are the uncertainties
in the Gaussian fit to each feature and not the significance level (i.e., typically $\Wno/\DW \ge $ significance level (SL)).
Further discussion of the correction of the \bmsd-values (columns 4 and 5) can be found in \S~\ref{sec:Ob}.
 Descriptions and justification for values in columns (7-10) can be found in \S~\ref{sec:REW}.

In this paper, we analyze physical quantities  derivable from the measured properties of the intergalactic
\lowzya lines of Paper~I.  In \S~\ref{sec:Ob}, we discuss the results and limitations
of the \bvalue determinations for our \lowzya detections. In \S~\ref{sec:REW}, we discuss the basic properties of
our measured rest-frame equivalent width (\Wno) distributions  and compare them to higher-\z\ distributions. 
In \S~\ref{sec:Onh}, we  estimate \hone column densities (\Nhno) for our \lya absorbers and discuss their
distribution, \dndzzero, relative to similarly derived values at higher redshift. 
In \S~\ref{sec:Z}, we discuss the \z\ distribution of the \lowzya
forest, as well as the cumulative  Lyman continuum opacity of 
these absorbers and the \z\ evolution of the number density of lines,  \dndzno. 
In  \S~\ref{sec:TPCF}, we analyze the cloud-cloud two-point correlation function (TPCF) for \lowzya clouds, 
and in \S~\ref{sec:metals}, we explore the limits on metallicity of the \lowzya forest. In \S~\ref{sec:omega}, we
estimate its contribution, \omegacl, to the closure parameter  of the Universe in  baryons, \omegab, inferred from D/H.
Section~\ref{sec:conc} summarizes the important conclusions of this investigation.
\input{table1.tex}
\input{table2.tex}
\section{Observed \bvalue Distribution}\label{sec:Ob}
Doppler widths (\bvaluesno) are estimated from the velocity widths (\WG = \bmsd/$\sqrt{2}$) of our
Gaussian component fits. As such, they are not true
measurements of the actual \bvaluesno, as when fitting Voigt profiles, but rather velocity
dispersions assuming that the absorption lines are not heavily saturated. This is a particularly good 
assumption for the large number of low-\W lines (i.e., $\Wno \lt 75$\mang), but it becomes increasingly suspect for the
higher \W lines.

The GHRS/G160M produced spectral resolution elements (REs) with Full Widths at Half Maximum (FWHM) of \about19\kmsno. 
The line spread function (LSF) of the HST+GHRS/G160M is approximately Gaussian with $\sigma_G \about 8.06$\kms for
both pre- and post-COSTAR data \citep{Gilliland92, Gilliland93}. As discussed in Paper~I, to improve the
robustness of our Gaussian component fitting we smooth our data with the LSF.
The  measured \bvalue (\bmsd) is actually the convolution of the instrumental profile, our pre-fit smoothing, and
 the observed \bvalue (\bobs) of the absorber. The \bvalues add in quadrature, 
$\bobs^2=\bmsd^2-2*\sigma_G^2=2(\WG^2-\sigma_G^2)$, where \WG\ is the
Gaussian width of the fitted absorption component.
Therefore, we are hampered in detecting absorptions with \bvalues near or below $\bb = \sqrt{2}~\sigma_G = 11.4\kmsno$.
Tables~1 and~2 present \bvaluesno,
rest frame equivalent widths (\Wno), and estimated \hone\ column densities for our
\real\ and \tent\ \lya samples, respectively. The \hone\ column densities are estimated  assuming \bb~= 20, 25, 30\kmsno, 
and  the individual corrected \bvalue for each absorber. 

Motion of the target in the GHRS large science aperture (LSA) during
our lengthy exposures can broaden the line spread function (LSF) or  modify the wavelength scale of our spectra, causing us to
overestimate the \bvaluesno.  Our subexposures are generally of insufficient
signal-to-noise to perform accurate cross-correlations to minimize the wavelength scale degradation. Weymann \etl (1995) performed
such cross-correlations on one GHRS G160M spectrum of \objectname[]{3C~273} in our sample. 
For the 1220\Ang and 1222\Ang features in 3C~273,
they measured \bvalues of $40.7 \pm 3.0$\kms and
$34.3 \pm 3.3$\kmsno, respectively.  As indicated in the first two entries of Table~1, we measure much larger
\bvalues of 69$\pm$5 and 72$\pm$4\kmsno. We believe these differences arise from target motion within the
LSA, causing larger \bvaluesno, although the line center and
\W measurements are unaffected. We use the Weymann \etl (1995) \bvalues for these two features to compute \nh in
column~10 of Tables 1 and 2, instead of
our values, although these are pre-COSTAR values which will not be used statistically herein.
 Other spectra, such as those of \objectname[]{Markarian 335}, \objectname[]{Markarian 501}, 
and \objectname[]{\PKS}\ (pre-COSTAR only), also seem to
be affected by this degradation (``jitter"). The fact that all of the suspected exposures were taken before the 1993 HST servicing
mission  leads us to speculate that spacecraft jitter, due to wobble introduced by thermal gradients
across the solar panels during passage across the terminator, may be responsible for the spectral smearing we observe.
In addition to installing COSTAR, 
this servicing mission corrected the jitter problem caused by spacecraft wobble \citep{Bely93,Brown93}.  
Motions in the target aperture can cause spectral
motion on the detector that can broaden spectral features and increase measured \bvalues (an offset of 1 arcsec in the LSA
corresponds to \about70\kms on the spectrum). 
All wavelengths in Tables~1 and 2 are LSR values, and all velocities
are non-relativistic values relative to the Galactic LSR, as explained in Paper~I.

 Figure~1 shows our  \bvalue distribution. Grey boxes indicate definite (\real)
\lya \bvaluesno, while the black boxes indicate possible (\tent) detections. Together, these two samples form our
``expanded'' sample (\expanded).  The \bvalue distribution in the top panel of Figure~1 shows
a  secondary peak at $47.5 < \bb < 72.5$\kms not present in the other observed \bvalue distributions.  This peak is
produced primarily (16 of 22 absorbers, or 73\%) by absorption features in pre-COSTAR data that appear to suffer from systematic
broadening.  In addition, we interpret a second population (5 of 22 absorbers, or 23\%)
as possible blended absorption features that we were not able to resolve fully. 
One post-COSTAR absorption feature appears unblended and thus truly broad.
 These blended features will be discussed further
in \S~\ref{sec:TPCF} when we analyze the linear two-point correlation of our absorbers. The second panel of
Figure~1 shows only those \bvalues obtained after the 1993 HST servicing mission.
  
In Table~3, we report statistics for \bobs, with and without 
the pre-COSTAR absorbers included in the sample.
Table~3 contains the following information by column: (1) the reference
papers for the \bvalue sample; (2) the redshift range over which \bvalues
were determined; (3) the mean redshift of the absorber sample; (4) the
observed wavelength range in\Ang from which the absorber data were
extracted; (5) the spectral resolution of the observations in\kmsno; 
(6) the median \bvalue of the sample in\kmsno; (7) the
mean \bvalue of the sample in\kmsno; and (8) the standard deviation of
the mean in\kmsno. Where the referenced work reports two samples,
the \bvalue results in columns (6-8) are reported for both samples (one
in parentheses), with the two samples described in footnotes to Table~3.
Our post-COSTAR  definite (\definite) distribution of \bobs\ has a mean of
$38.0\pm15.7$\kms and a median of 34.8\kmsno, while our pre-COSTAR  definite distribution has a mean of
$58.6\pm15.9$\kms and a median of 60.6\kmsno. 
Table~3 also compares our results with other studies at higher redshift and is ordered by decreasing mean redshift, \meanz. 
For studies up to \meanz = 3.7, Table~3 reports the redshift range, \meanz, wavelength~($\lambda$) range, resolution element (RE),
median \bb, mean \bb, and standard deviation ($\sigma_b$) of the mean \bb~for the various \lya absorption samples.
The third through seventh panels  of Figure~1 show the higher-\z\ (1.6 \lt \z \lt 4.0) \bb\ distributions from \citet{Lu96}, 
\citet{Carswell84}, \citet{Carswell91}, \citet{Kulkarni96}, and \citet{Khare97}. Statistics for these distributions are tabulated in
Table~3. \citet{Lu96}'s observations of the \lyano-forest at $3.4<\z<4.0$, shown in the third panel of Figure~3, 
were taken with the Keck HIRES spectrograph, with a resolution element of 6.6\kmsno. 
As expected, the \citet{Lu96}  distribution extends to lower \bvalues than ours.
This \highz distribution contains
412 absorbers with a peak at \bb=23\kms and a median value of \bb=27.5\kmsno.   
Like the \highz distribution of \citet{Lu96}, the \lowz distributions show an increasing number of
absorbers at decreasing \bvaluesno, until one approaches the resolution limit. \citet{Lu96} report a detectable turnover in the
\bvalue distribution below 20\kms with almost no absorbers at \bb~$\le 15$\kmsno.  
Because these Keck data have three times higher resolution than HST/GHRS, features that are blended in our 
data would be resolved by  Keck. 
\input{table3.tex}

A more accurate comparison to our data can be made with
\bvalue distributions taken at comparable resolution \citep[e.g., ][]{Carswell84}.
As indicated in Table~3, the \bvalues of  \citet{Carswell84,Kulkarni96,Khare97} were obtained with spectral resolutions
similar to the GHRS/G160M ($\sim$19\kmsno). For direct comparison to our \bvaluesno, the second-to-last panel in
Figure~1 displays the combined \bvalue distributions of these three studies and is labeled ``C84KK''. 
%A KS-test between the \bvalues of the second panel (COSTAR only) and the C84KK 
%\bvalues indicate that they are drawn from the same parent
% population at the 86\% confidence level, while a K-S comparison 
%to the \citet{Lu96} sample gives a confidence level of 91\%,
%comparable to or higher than the K-S
%test results comparing any of the higher redshift samples to each other. 
The final panel of Figure~1 
displays the combined \bvalue distributions of other studies 
\citep{Lu96,Carswell84,Carswell91,Kulkarni96,Khare97} and is labeled ``All''.
Comparing our \bvalue distributions to the samples listed here, we conclude that
at the 2\sig level, the high and low-\z\ \bvalue distributions appear to differ, with the \lowz distribution containing somewhat
broader lines.
Comparisons via KS tests to determine if our \real\ post-COSTAR distribution is drawn from the
same parent sample as the other and combined samples range from highly unlikely (C84KK, 1\% and All, 0.3\%) to 
probable (Carswell 1991, 30\%). However, due to the variety of methods used in the observations, data reduction, and profile fitting,
 we cannot be certain that the observed modest \bvalue evolution is real at this time.
A consistent dataset with higher spectral resolution and high S/N is required; HST/STIS echelle data would be ideal.

It has been proposed \citep{Kim97} that there is evolution of \bvalues with redshift. 
Although the three \citet{Kim97} entries in Table~3
do show a tendency to higher \bvalues (broader absorption features) at lower \z, 
this is not confirmed by the other studies listed in the table. Our
results, at significantly lower \z\ than any other study in Table~3, do not support significant evolution.
Neither does the \bvalue distribution found at $1.2 < \z < 2.2$ by \citet{Savaglio99} in a single sightline show
any signs of the evolution suggested by \citet{Kim97}.
Instead, a small 
decrease in \bvalues from \meanz = 3 to \meanz = 0 is found by ``effective equation of state'' models of the IGM  fitted to \lya line-width data 
\citep{Ricotti00,Bryan00}.

The measurement of the \bvalue distribution is different from the measurement of other distributions such as \W and \Nhno, because
 it is difficult to correct for incompleteness (what we cannot detect). 
For example, consider a portion of a spectrum  with a \foursig detection limit of 100\mang.
When calculating the \W distribution, we cannot detect features in this region with $\Wno \lt 100$\mang. 
Therefore, we eliminate this pathlength for $\Wno \lt 100$\Mang when
 determining the true rest-frame equivalent width
(number-density) distribution, \Nno(\Wno)/\Dz(\Wno). Here, \Dz(\Wno) is the available redshift pathlength for detection of
absorption features as a function of \Wno. 
But, in our hypothetical spectral region, we cannot distinguish between a non-detection of a narrow absorption feature of
\bb=20\kms and a broad one of \bb=80\kmsno, if both produce features of \W$<$ 100\mang. In addition,
it is possible to misidentify very broad, low-\W features as continuum undulations.

The accurate measurement of \bvalues is important in determining the actual \hone column densities (\Nhno) of the saturated
\lya absorbers, since each \bvalue produces a different curve of growth for the upper range of column densities
in Tables~1 and 2. 
The only reliable method of deriving \bvalues for such weak lines is from higher Lyman lines such as \lyb \citep{Hurwitz98,Shull00}.
For example, \citet{Hurwitz98} found that  the \lyb absorption strength observed by  ORFEUS II was in strong disagreement with  the predicted values
based on two \lya Voigt profiles in the GHRS spectrum of 3C~273 sightline associated with the intracluster gas 
of the Virgo supercluster \citep{Morris91,Weymann95}. 
\citet{Hurwitz98} observed \lyb at 1029.11 and 1031.14\ang, corresponding to the first two \lya features of
Table~1, with \lyb equivalent widths of 145 $\pm$ 37\Mang and 241 $\pm$ 32\mang,
respectively.  Weymann \etl (1995) 
measured the \lya \bvalues of these features as  $40.7 \pm 3.0$\kms and $34.3 \pm 3.3$\kmsno, respectively. 
The \lyb observations of \citet{Hurwitz98} imply \bvalues as low as 12\kmsno.
Similar trends towards lower \bvalues are found in initial \lyb studies with the Far Ultraviolet Spectroscopic Explorer (FUSE)
satellite \citep{Shull00}.  This disagreement in estimating \bvalues can be understood 
if  the \lya absorption profiles include non-thermal broadening from cosmological expansion and infall, 
and multiple unresolved components. 
\citet{Hu95} reach the same conclusion, based upon \z\about3 Keck HIRES QSO spectra. The analysis of \citet{Hu95}
implies that the average \z\about3 \lya absorber could be well represented by \about3 components, each having \bb\about15\kmsno.

Therefore, we are faced with a dilemma: should we use our approximate
\bvalues inferred from the line widths, a constant value based upon the better known \highz distribution, or an adjusted constant value considering the
\lyb results ?  In \S~\ref{sec:Onh}  we  compare the \Nh results from these three alternatives
and discuss the merits of each.
\begin{figure}[htp] \epsscale{0.67}
\plotone{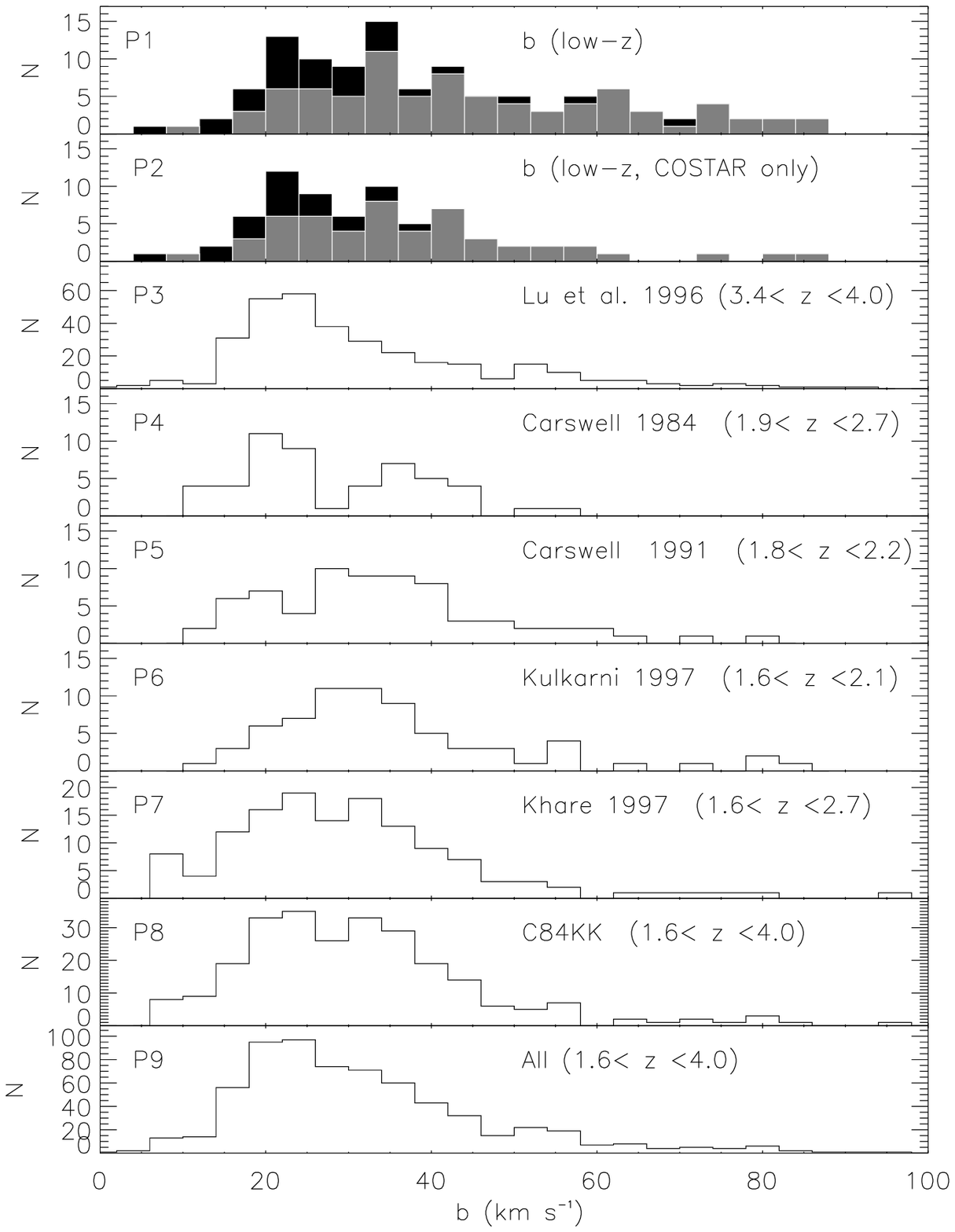}
	\caption{\label{blow_vs_high} 
Top panel (P1) shows the \lowz distribution of line widths, \bobs.
 Two features in the 3C~273 spectrum have been adjusted in accordance
with Weymann \etl (1995). P2 shows only \bvalues obtained with HST+COSTAR. 
In P1 and P2, grey and black boxes indicate our \real\ and \tent\ samples,
respectively. P3-P7 show higher-\z\  distributions from various sources
(Table~3). Panel P8,``C84KK'', combines all the
higher-\z\  distributions with resolutions comparable to ours (\about19\kmsno). 
P9, ``All'', combines panels P3-P7.}
\end{figure}
\clearpage
\section{Observed Rest-Frame Equivalent Width Distribution}\label{sec:REW}
In Figure~2 we display the rest-frame equivalent width (\Wno) distribution for  all of our
detected \lya features. The solid grey boxes in Figure~2 represent our definite (\real) sample, while the
black boxes represent our possible (\tent) sample.
 As expected, we detect an increasing number of absorbers at decreasing \Wno, down to our detection limit. As discussed
in the previous section, our spectra are of varying sensitivity and wavelength coverage. This observed \W
distribution is not the true \W distribution, because we have not yet accounted for incompleteness. To
determine the true \W distribution, we must normalize the \W density  by the available pathlength \Dz(\Wno). 

The pathlength, \Dz, is actually a function of \W  and \z, since our spectra have varying
\foursig detection limits across the waveband and each spectrum covers a different waveband (\z\ range). Without this
important sensitivity correction, \Dz(\Wno,\z), any interpretation of the \W distribution is premature. Previous
studies, such as the HST/FOS Key project \citep{Bahcall93}, avoided this problem by considering the \W distribution
only above a universal minimum \W detectable in all portions of all spectra in the sample. This forces
\Dz(\Wno,\z) to be a constant pathlength, so that the \W  distribution is a true representation of the detected absorbers.
However, this procedure eliminates information about the lowest, most numerous, \W absorbers. Because our GHRS program probes the lowest
equivalent widths of any \lowz \lya program to date, it is important to use all available data for the weakest absorbers.
In this section, we will explore the \W distribution without regard to \z\ and neglect any evolution of \W with \z\ (we find no evolution
in numbers of \lya absorbers with redshift over our small \Dz~range). We will examine the \z\ distribution
(\zrange) and the \W evolution with \z\ in later sections.

In Paper~I, we discussed
the available pathlength for each sightline in detail.  In Figure~3, we display the available pathlength
(\Dz) in terms of \W for each spectrum. The solid line in each panel indicates the full
observational pathlength, uncorrected for the regions of spectra not available for \lya detection due to
Galactic, HVC, intrinsic, and non-\lya intervening absorption lines as well as our ``proximity effect'' limit.
This ``proximity limit'' ($cz_{AGN} - cz_{abs} \gt 1,200$\kms) eliminates potential absorption systems
``intrinsic''  to the AGN (see Paper~I for a detailed description and justification). The
dot-dashed line indicates the ``effective'' or available pathlength after we remove the spectral regions unavailable for
\lya detection. 
In Figure~3, \cDz(\Wno)\ is given in units of Megameters per second (\nommsno\  or 10$^3$\kmsno).
Note that three targets (\objectname[]{I~ZW~1}, \objectname[]{Q1230+0115}, and \objectname[]{Markarian 501}) 
have significantly poorer sensitivity than the rest
of our sample and should  be re-observed with HST, using the Space Telescope Imaging Spectrograph (STIS) 
or the Cosmic Origins Spectrograph (COS).
\begin{figure}[htbp]
	\plotone{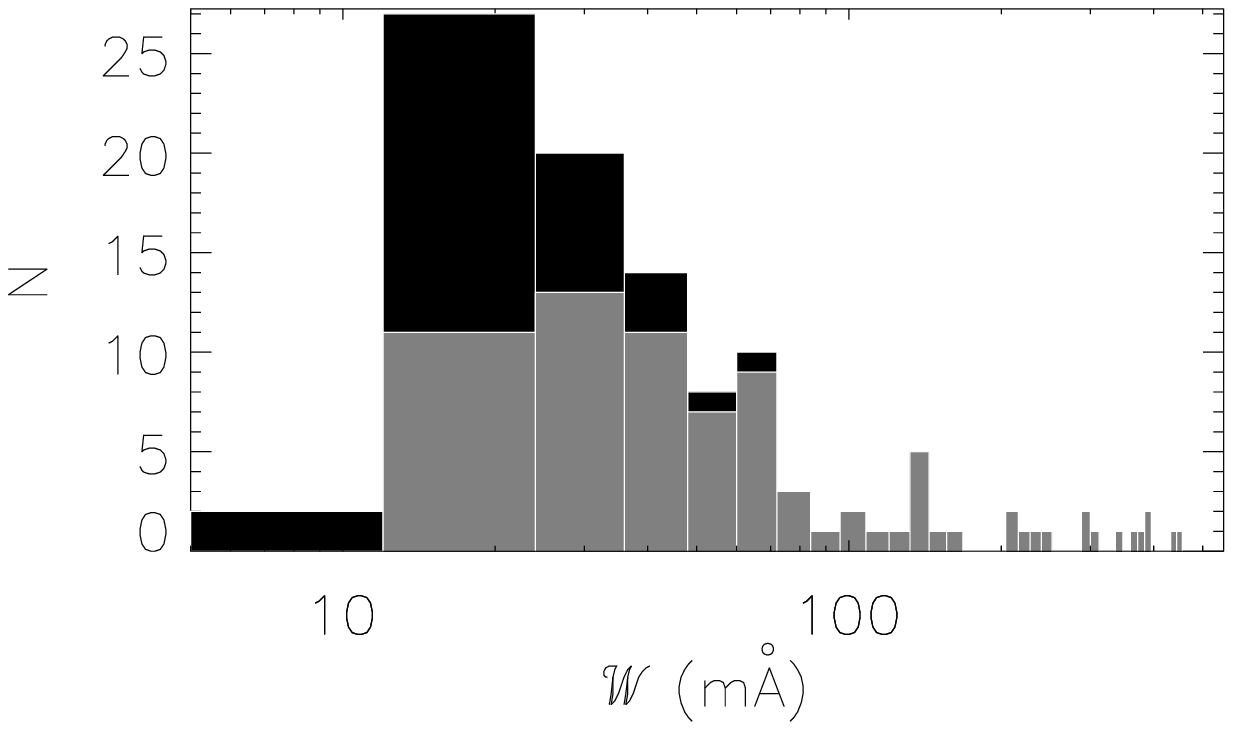}
	\caption{\label{ew_dist} Observed distribution in rest-frame equivalent width (\Wno)
 of our detected \lya absorbers. The solid  grey boxes represent definite (\real)
detections, while the black boxes indicate possible (\tent) absorbers. The \W bins are 12\Mang in width. This
distribution has not been corrected for the non-uniform wavelength and sensitivity coverage of our
observations. No intrinsic \lya absorptions are included in this distribution.}
\end{figure}
In Figure~4, we display the cumulative distribution of \Dz(\Wno) for
all spectra, for \real\ detections. In this cumulative plot, the solid line indicates
the uncorrected pathlength. Also indicated is the pathlength available for \lya detection after
correcting for the spectral obscuration due to Galactic and HVC features, intrinsic and
non-\lya intervening absorbers (Intrinsic+Non-\lyano), and our ``proximity limit''. Indicated by the lowest (dotted) line is
the cumulative available pathlength, for \real\ detections, after the indicated corrections have been made. The left
axis of Figure~4 indicates the available pathlength (\cDz) in units of \mmsno, while the right axis indicates
\Dz. Our maximum available pathlength (\cDz) is 116\mms (or \Dz=0.387) for all features with $\Wno \ge 150$\mang. 
Also indicated in Figure~4 is  our \Nh pathlength availability for a constant \bb=25\kmsno. This pathlength compares
to $\Dz\approx8$ at $\z \lt 0.3$ for the Key Project for strong absorbers ($\Wno \gt 240$\mang).
\begin{figure}[htbp]
\epsscale{0.7}	\plotone{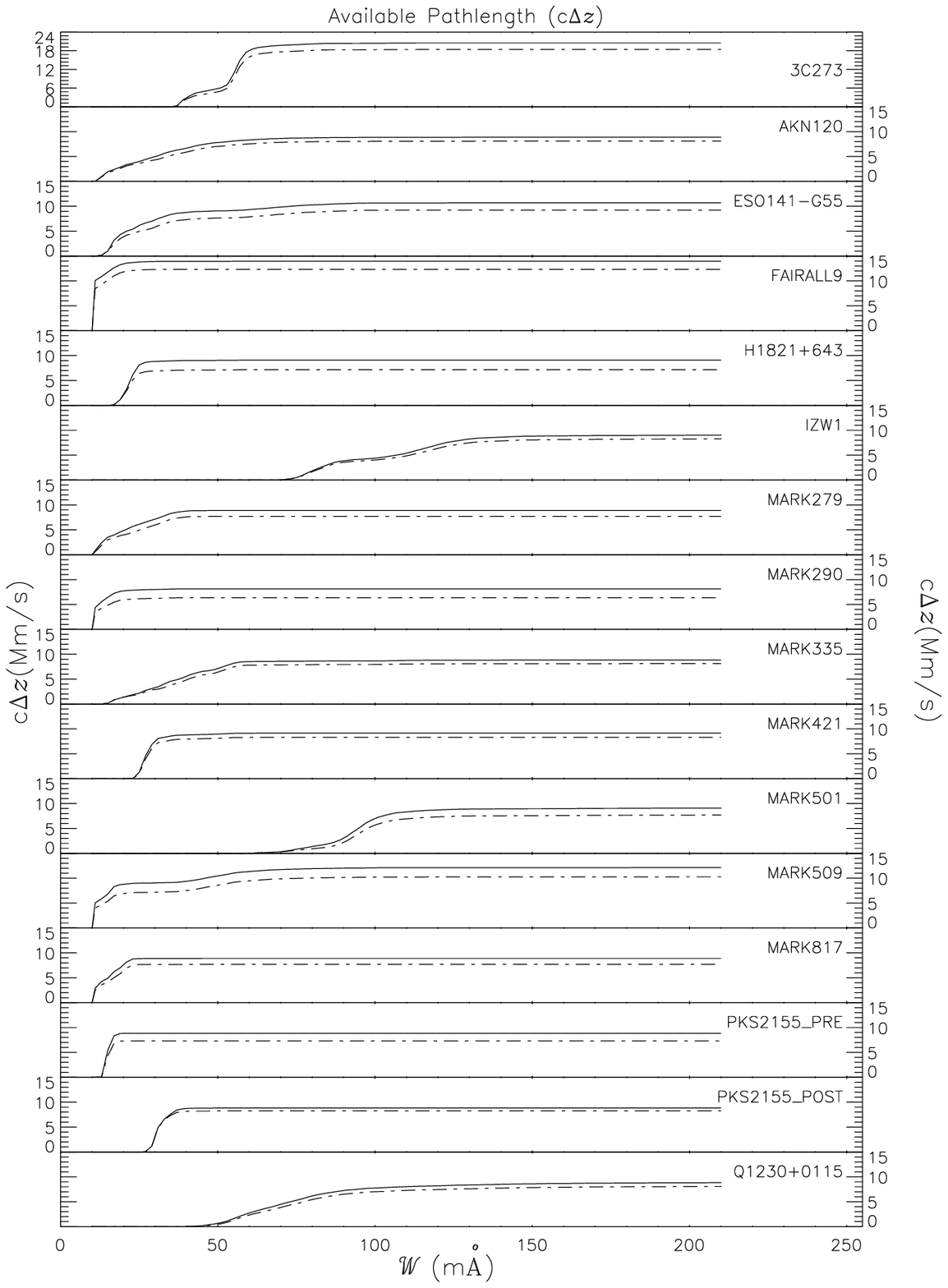}
	\caption{\label{master_corr} Each panel shows
the available pathlength, \cDz(\Wno) in\mmsno, for a sightline as a function of \foursig rest-frame equivalent
width (\Wno). The solid line is the full \cDz, uncorrected
for known Galactic, HVCs, intrinsic, or intervening features. Regions of  FWHM$\times$2 of each \gt\threesig non-\lya absorber are 
removed from the available \cDz~for detecting
\lya absorbers. 
The dashed line is the corrected \cDz, including the  removal of 10 pixels (\about0.2\ang) at each spectral edge and the
application of our \prox ``proximity limit''.}
\end{figure}
\begin{figure}[htbp]
\plotone{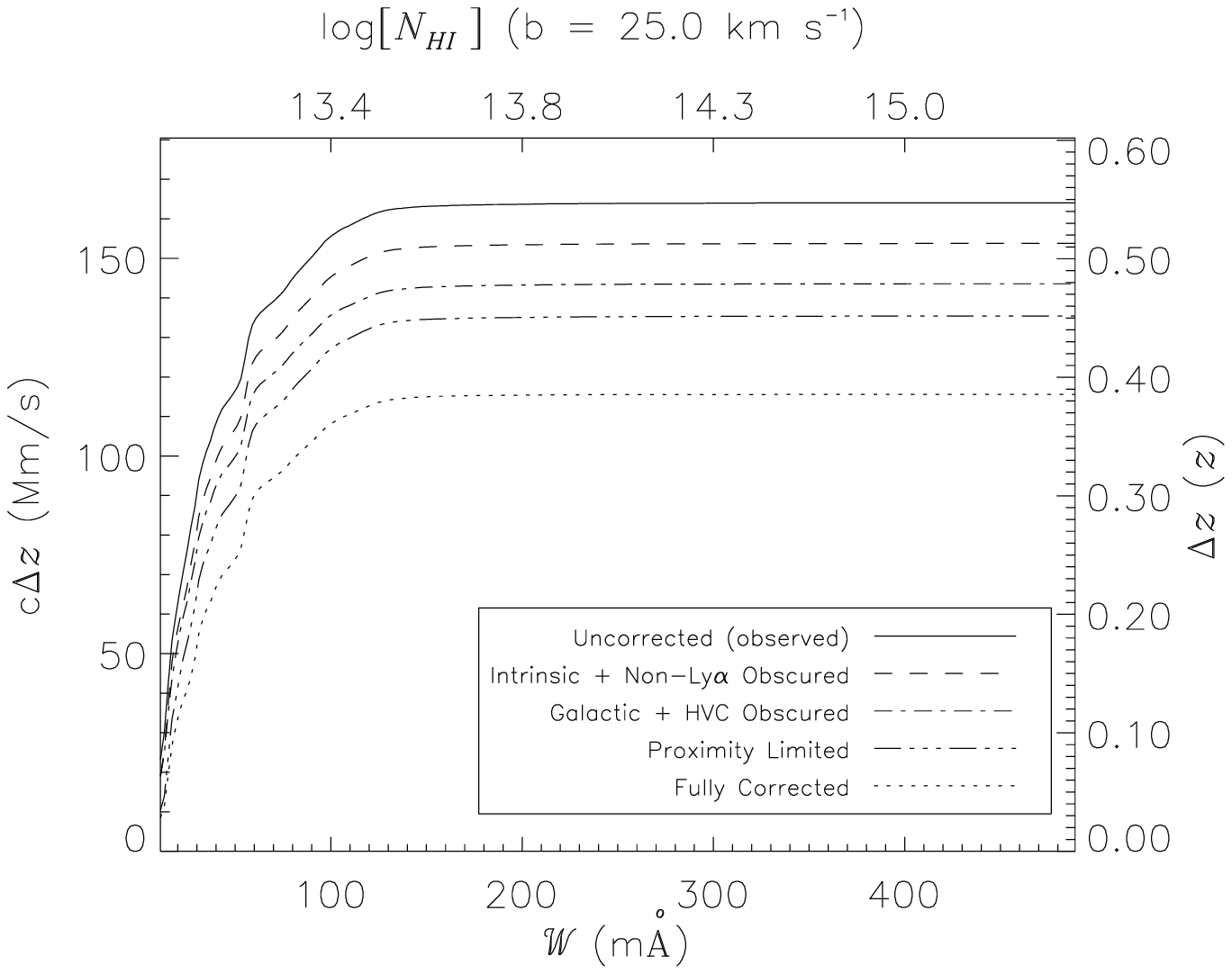}
	\caption{\label{ewsens_path}Cumulative available pathlength (\Dz) as a function of
\Wno, for \real\ detections. The left axis gives non-relativistic \cDz(\Wno) in units of \mmsno, while the right axis gives \Dz. 
The solid (upper) line indicates the observed, uncorrected, \Dz(\Wno). 
Dashed, dot-dashed, and dot-dot-dashed lines indicate \cDz(\Wno) after correcting for spectral obscuration due to non-\lya and
intrinsic absorption systems (Intrinsic+Non-\lyano), Galactic and HVC absorption systems (Galactic+HVC), and our \prox
``proximity'' limit, respectively and individually. The dotted line indicates the fully corrected ``effective'' \Dz(\Wno), after
the indicated spectral regions unavailable for intervening \lya detection have been removed. 
Top axis indicates the \logNh\ corresponding to \W for $\bb=25$\kmsno; for example, $\lognh = 12.26$ for $\Wno = 10$\Mang (left-hand border of bottom axis).}
\end{figure}
\subsection{The Low-\z\ Equivalent Width Distribution}\label{sec:ew_spectrum}
In Figure~5, we apply the effective pathlength correction of Figure~4 to
the  \Nno(\Wno) distribution of Figure~2. This procedure gives  the true detected \W number
density, 
\begin{equation}
\nWi=\Nno(\Wi)/\Dz(\Wi)\DW \approx \dtwondWdz~\vert_{\Wi},
\end{equation} corrected for the pathlength,
\Dz(\Wi), available to detect features at each \Wi, without regard to \z. The approximation that
\nW\ is equal to \dndW\ is limited by our sample size (\Nno) and by our bin size (\DW = 12\mang). 
However, this approximation is more than sufficient for our purposes.
\begin{figure}[htbp]
	\plotone{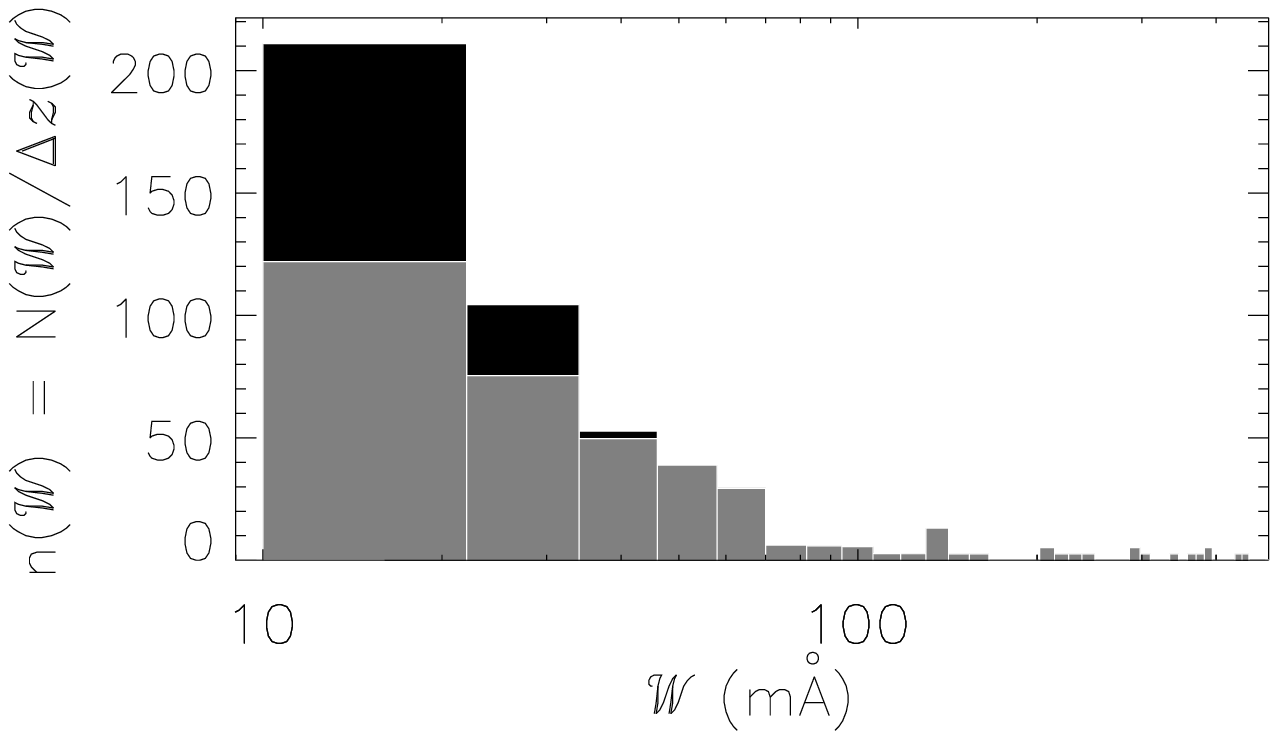}
	\caption{\label{ew_corrected}
Rest-frame equivalent width (\Wno) number distribution, \nW, corrected for available pathlength (\Dz).
The corrected number density,
\nW=\Nno(\Wno)/\Dz(\Wno), is the observed \W distribution corrected for the  non-uniform sensitivity function of our observations.
The solid grey boxes represent definite (\real) detections, while the black boxes indicate possible (\tent) absorbers. The \W bins are
12\Mang in width.}
\end{figure}
It has been shown \citep{Sargent80,Young82,Murdoch86,Weymann98}
that  \dtwondWdz, the pathlength-corrected number density of lines with respect to \Wno, is well modeled by $\nW=\left(\NofW
/\Wstar\right) \exp\left(-\Wno/\Wstar\right)$, 
where \Wstar\ is the characteristic equivalent width and \NofW\ is the characteristic line density
per unit redshift.  In Figure~6, \nW\ is plotted versus \W in natural logarithmic form, for both our
 definite (\real) and our expanded (\expanded) \lya samples. Our \W bins are 12\Mang in width, and our upper \W cutoff is \Wcutoff.
Also indicated by dashed lines in Figure~6 is our least-squares fit to $\ln\left[\nW\right] = \ln\left[\NofW/\Wstar\right] - \Wno/\Wstar$ for
$10\mang \lt \Wno \lt \Wcutoff$. 
For our definite sample, \Wstar$=27\pm 5$\Mang and \NofW$=520 \pm 179$\mang.  
For our expanded sample over the same \W interval, \Wstar$=23\pm 3$\Mang and \NofW$=733 \pm 205$\mang.

	For comparison, at higher redshift, \citet{Sargent80} obtained 
$\Wstarno = 362 \pm 21$\mang, $\NofW = 154 \pm 11$\mang, 
while \citet{Young82} obtained $\Wstarno = 232 \pm 34$\mang, $\NofW = 135 \pm 20$\Mang in the redshift 
range $1.7 \lt \z\ \lt 3.3$ for \lya absorbers with $\Wno \ge 160$\mang. 
At first comparison, our results seem inconsistent with these results. However, as has been noted
  \citep{Murdoch86,Carswell84,Atwood85}, the parameters \Wstar and \NofW\ 
are highly dependent on the spectral resolution of the observations, mostly because at higher resolution
blended lines break up into components. One sample that approximates our resolution and sensitivity is that of
\citet{Carswell84}. Based upon 20\kms resolution optical data, \citet{Carswell84} obtained 
$\Wstarno = 70 \pm 20$\Mang for their unblended, $\Wno \gt \Wcutoff$, \lya sample.
\begin{figure}[htbp]
	\plotone{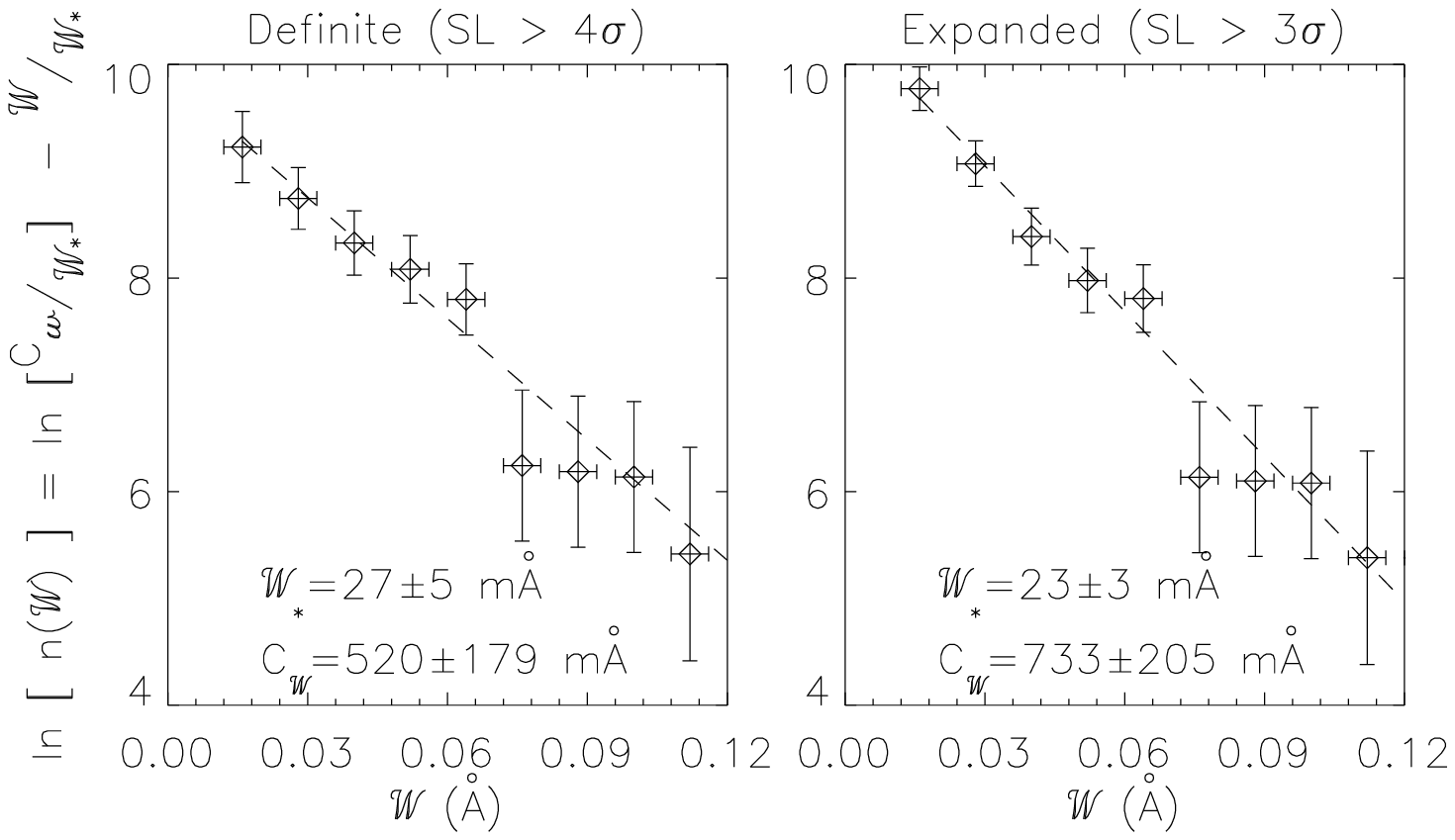}
	\caption{\label{dndW} 
The differential distribution, \nW = \dtwondWdz, plotted in
natural logarithm form.  The \nW~error bars are based upon
$\sqrt{N}$ statistics. The left panel indicates the distribution for our definite (\real) \lya absorbers only. 
The right panel includes the distribution for our expanded (\expanded) \lya sample. 
The \W bins are 12\Mang and only \lya absorbers with $\Wno \lt \Wcutoff$\ are included.}
\end{figure}

As indicated in \citet[][their Figure 3]{Murdoch86}, there is evidence for a break in the slope of
\nW\ below 200\mang.  As shown in Figure~7,  we confirm this break, even though our statistics
in this region (\W\gt200\mang) are poor. For \bb=25\kmsno, this corresponds to a break at \lognh\about14.0.
The reality of this break will be discussed further in \S~\ref{sec:Onh}, which focuses on the \Nh distribution of the \lowzya forest.

At first consideration, the presence of this break is intriguing. However, one must understand the nature of this discontinuity 
in relation to the \lya curve of growth to gauge its significance. As pointed out by \citet{Jenkins91} and \citet{Press93}, 
this upturn in the number of low \W clouds can be explained by the transition from saturated
 to unsaturated lines. The form of \nW =$\left(\NofW / \Wstarno\right) \exp{\left(-\Wno/ \Wstarno \right)}$ is designed to model absorbers in the logarithmic 
or flat part of the curve of growth.
As discussed in \S~\ref{sec:Onh}, it has been shown that at higher column densities, \lya absorbers appear to follow the distribution
$\nNh =\dtwondnhdzno = \NofNhno \Nhno^{-\beta}$.
On the linear portion of the curve of growth, where  $\Wno = \cprime \Nhno$,
\begin{equation}{
	\nW=\dtwondWdzover=\dtwondnhdzover \dnhdWover = {\NofNhno \Nhno^{-\beta} \over \cprime} ~~ {\rm~,~or}}\end{equation}
\begin{equation}{ \label{Cbeta} \nW ={\NofNhno \left(\Wno/\cprime\right)^{-\beta}  \over \cprime} =\cprime^{\beta-1}~\NofNhno~
\Wno^{-\beta}.}\end{equation}
As derived in  equation~(\ref{EWs}) for \lyano, A = \Exp{54.43}{-13}\Mang for \Nh in $10^{13}$ \percmtwo and \W in\mang.

In Figure~7 we add our absorbers with $\Wno \gt \Wcutoff$\ to the \definite\ $\ln{[\nW]}$ distribution of Figure~6. We
increase the upper \W bin size from 12\Mang to 42\Mang to compensate for our poor statistics, so that there is at least one absorber in each bin.
Above \W = \Wcutoff, we obtain  $\Wstarno = 259 \pm 180$\Mang and $\NofW = 138 \pm 129$\mang. Below \Wcutoff, we fit by equation~(\ref{Cbeta}), using the
least-squares method, and find $\beta = 1.89 \pm 0.39$ for our \definite\ sample ($\beta = 2.20 \pm 0.37$ for our \expanded\ sample).
The  value of \NofNh is not well constrained  due to the exponential nature of this fitting
method, but the best fit gives $\NofNh \sim \Exp{4}{14}$.
\begin{figure}[htbp]
\epsscale{0.7} 	\plotone{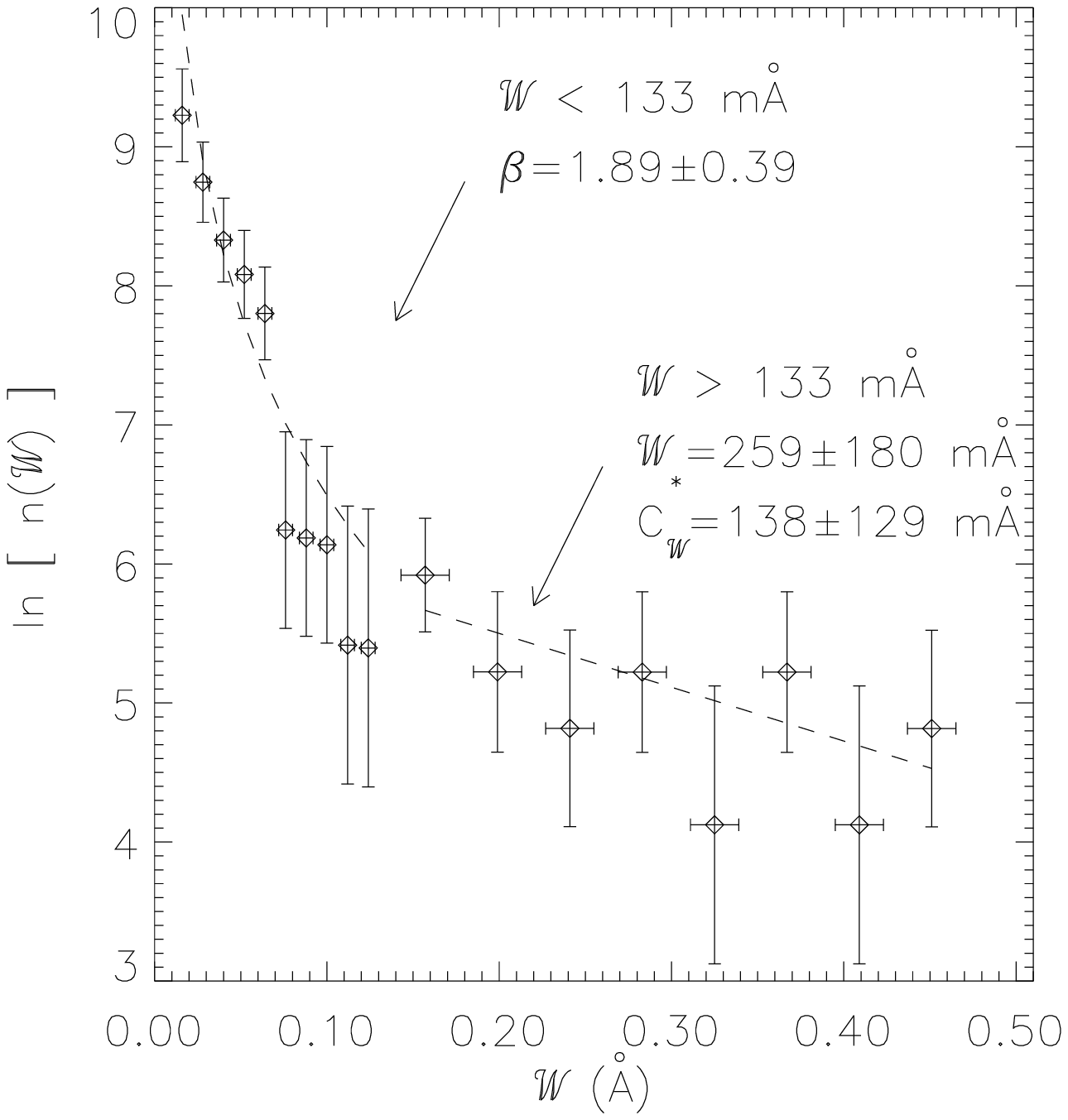}
	\caption{\label{dndW_big}  The differential distribution, \nW = \dtwondWdz, of \lya
absorbers with respect to \W is plotted in natural logarithm form, for our \definite\  \lya sample. 
The \nW~error bars are based upon
$\sqrt{N}$ statistics.  The \W range has been expanded to include all detected \lya absorbers. 
Below \W = \Wcutoff, the \W bins are 12\Mang in width; above
\Wcutoff\ the bins are 42\Mang wide. Horizontal (1$\sigma$) error bars are set to 1/3 of the \W bin widths.}
\end{figure}
\subsection{Integrated \dndzno\ Results}
By integrating \dtwondWdz\ from \W to $\infty$, we can determine the number density of lines per
unit redshift, \dndz, stronger than \Wno. As stated previously, we have assumed no evolution with \z\ over our small
range for \lya detection (\zrange). Because of our very \lowz range, these values for \dndz are a good approximation of \dndzzero. 
Figure~8 shows \dndz, defined as 
\begin{equation} \dndz (\W >\Wno_i) = \int_{\Wno_i}^\infty  \dtwondWdzover~\dW .\end{equation}
The vertical axis of Figure~8 gives \dndzzero\  in terms of both \W (lower axis) and \lognh\ (upper axis, assuming that all absorbers are single
components with \bvalues of 25\kmsno). 

For comparison, Table~4 gives \dndzzero\  from other HST
absorption-line studies, along with our \lowz results for their observational \W limits (\Wno$_{min}$). The detection limits of the various
HST spectrographs and gratings are indicated by the dashed vertical lines in Figure~8. The unlabeled vertical line on the far left of Figure~8 in
the GHRS/G140L+G160M merged sample of \citet{Tripp98}. In all cases, our
values for \dndzzero\ at various limiting \W are slightly greater than those derived from higher limiting \W studies at \lowz, but are still within the \onesig error ranges. 
The value of \dndz at the present epoch (\z=0) is an important constraint for  hydrodynamical cosmological models attempting to reproduce the
observed baryon distribution  of  the Universe \citep{Dave99}. The \dndzzero\ values of our GHRS survey have the lowest measured \W values 
to date. In \S~\ref{sec:dndzz}, these values will be used to place important constraints upon the \z-evolution of \dndzno.
\begin{figure}[htbp]
	\plotone{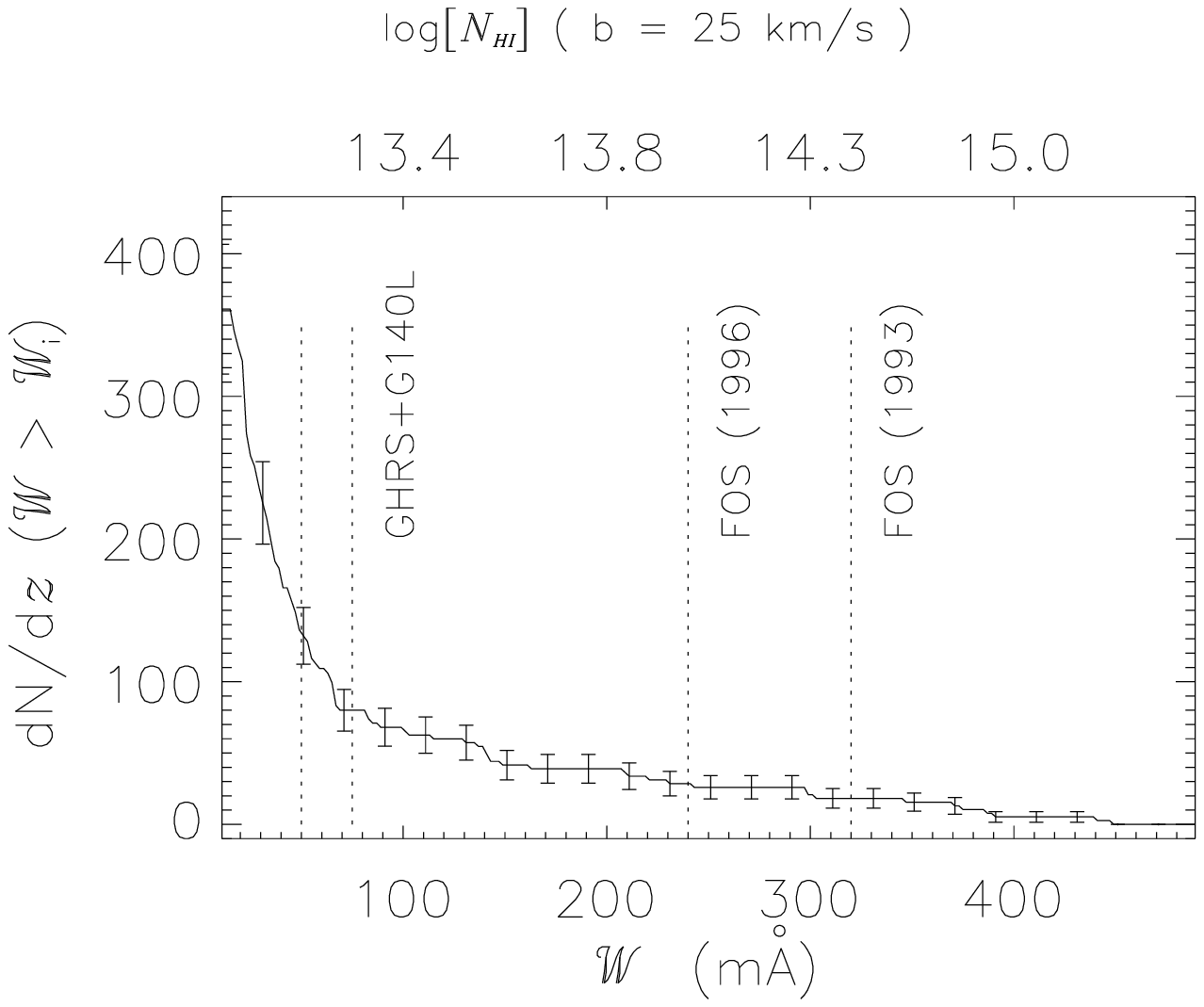}
	\caption{\label{int_dndz}
Integrated \dndz above \Wno: $\dndz (\W > \Wno_i) = \int_{\Wno_i}^\infty  (\dtwondWdz) \dW$, for our \definite\ sample.
 For clarity, we plot every third error bar. 
The \W bins are 12\mang, connected by the solid line. Also shown by dotted vertical lines are the minimum
\W limits for the various HST \lya surveys presented in Table~4. The unlabeled vertical dashed line on the far left
corresponds to the \protect{\citet{Tripp98}} G140L+G160M combined sample (see Table~4). 
The horizontal axes begin at \W = 10\Mang or \lognh = 12.26.}
\end{figure}
\input{table4.tex}
\subsection{Rest-Frame Equivalent Width versus \bb\ Distribution}
 In Figure~9, we plot the observed \W distribution versus the resolution-corrected  \bobs\ values.
 In this figure, the \Nabs\ definite absorbers are plotted as small dots while the \Npos\ possible (\tent) \lya detections are plotted as diamonds. 
In neither sample do we detect any correlation of \W with respect to \bobs\ (in the definite sample the correlation coefficient is zero). 
Since very broad weak features are indistinguishable
from small continuum fluctuations, we may not be sensitive to features in the lower
right corners (large \bvalues and low \Wno) of Figure~9.
A completely homogeneous distribution would be created in Figure~9 by adding \about 15 absorbers at $\Wno \le 30$\Mang and 
$\bb = 60-100$\kmsno. At the spectral resolution of these observations these absorbers have maximum deflections of
$\lt 15\%$ of the base continuum level and thus would be extremely difficult to distinguish from a slight undulation in the continuum.
Therefore, given the absence of reliable \bvalues for each absorber and no obvious \bb-\W correlation, we will assume a single
\bvalue when calculating \Nhno.  Figure~9 shows that there is no obvious bias in making this single
\bvalue choice.
\begin{figure}[htbp]
	\plotone{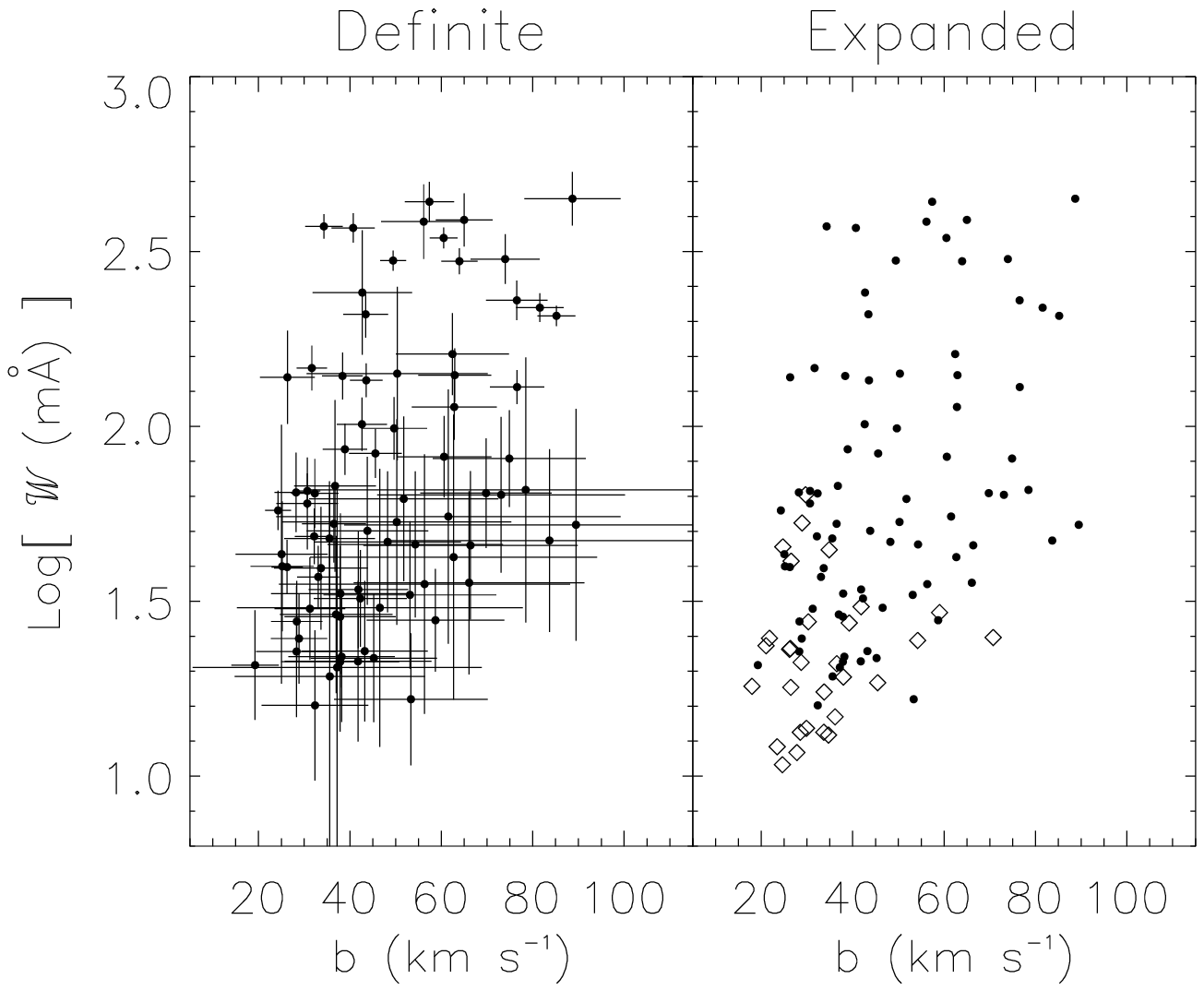}
	\caption{\label{ew_b_dist}
Distribution of rest-frame equivalent width (\Wno) versus \bvalue of our detected \lya absorbers. The left panel displays the distribution of
the definite (\real) sample with error bars.  The right panel shows the \real\ sample as dots, combined with the \tent\ sample displayed as
diamonds, forming our expanded sample. The absence of absorbers in the lower right corners is related to an insensitivity to very broad weak
features, which could be interpreted as continuum fluctuations.}
\end{figure}
\clearpage
\section{Observed H~I Column Densities}
\label{sec:Onh}
In Table~1, we estimated the neutral hydrogen column density (\Nhno) of each detected \real\ (definite) \lya absorber from its equivalent
width, assuming a
single component with a \bvalue of 20, 25,  30\kmsno, and  the measured  \bobs\ values. Table~2 presents \Nh results for our
 \tent\ (possible) \lya sample.
As described in Paper~I and in \S~\ref{sec:Ob}, the observed \bvalues have been corrected for the HST+GHRS/G160M instrumental profile and
our pre-fit smoothing.
Columns 4 and 5 of Tables~1 and 2 indicate the measured \bvalues (\bmsd) and the corrected \bvalues (\bobs), respectively.
Figure~10 compares the results on \logNh\ of our sample for these various \bvaluesno.
\begin{figure}[htbp]
\plotone{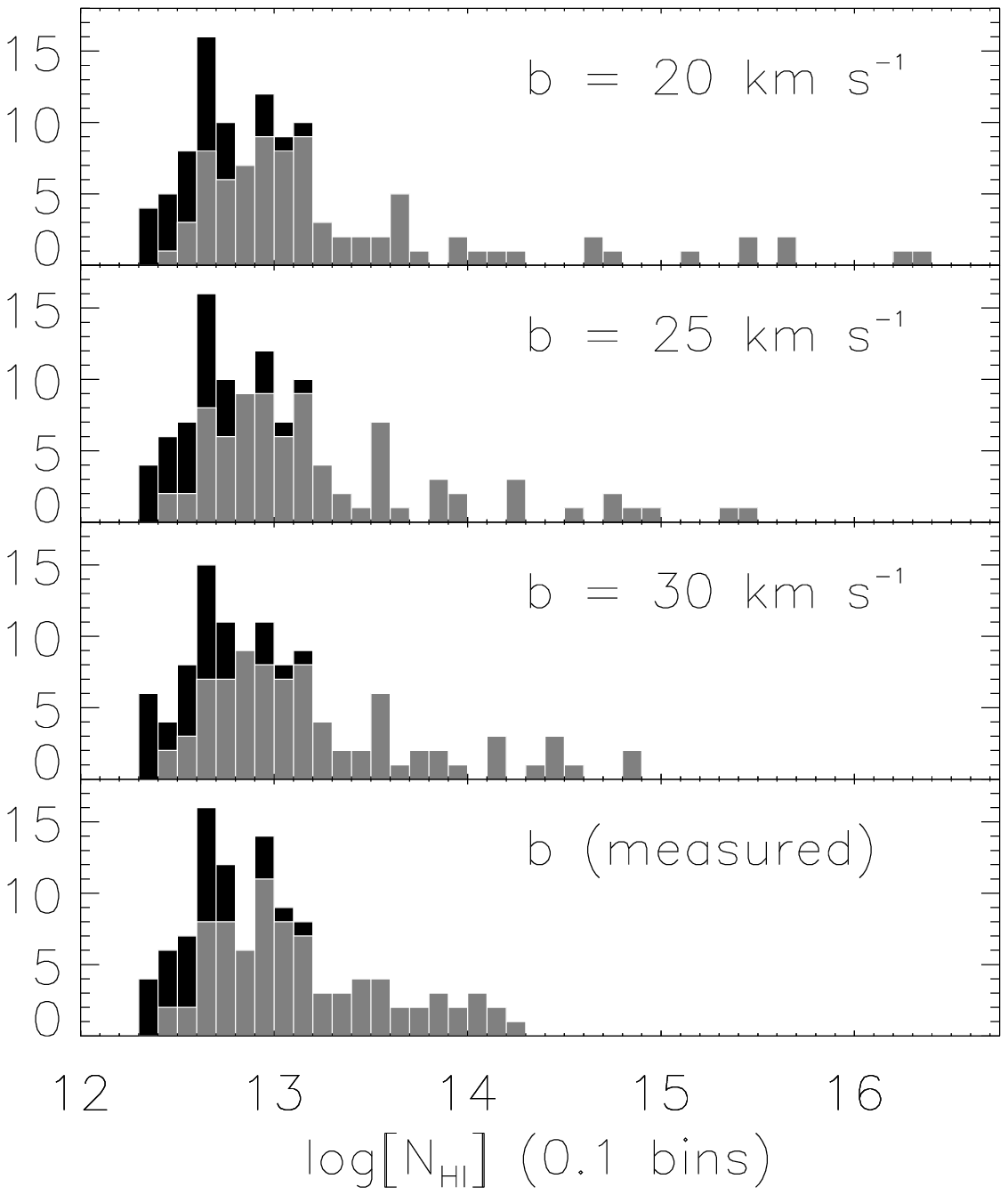}
	\caption{\label{nh_dist} \hone column densities, \nh, of our observed
\lowzya absorbers for various \bvaluesno. The top three panels show the distribution based upon all
features having a single \bvalue of 20, 25, or 30\kms as indicated. Bottom panel shows the
distribution based upon \bvalues measured from Gaussian components and corrected by the HST+GHRS/G160M resolution element.
 The grey absorbers are our definite (\real) sample, while the black absorbers are our possible (\tent) sample.}
\end{figure}
Below \logNh\about13, all the \Nh distributions of Figure~10 are similar, because the \lya absorbers are on the
linear part of the curve of growth where \Nh does not depend on the \bvalueno. 
Above \logNh = 14, the \lya absorbers are partially saturated for \bb=25$\pm$ 5\kmsno. For Gaussian line profiles, the optical depth
at the line center is given by:
\begin{equation}{
	\tau_0 = \left(0.303\right) \left[ \Nh \over 10^{13}\percmtwo \right] \left[ 25\kmsno \over \bb \right] \ \ {\rm .}}\end{equation}
\begin{figure}[htbp]
	\plotone{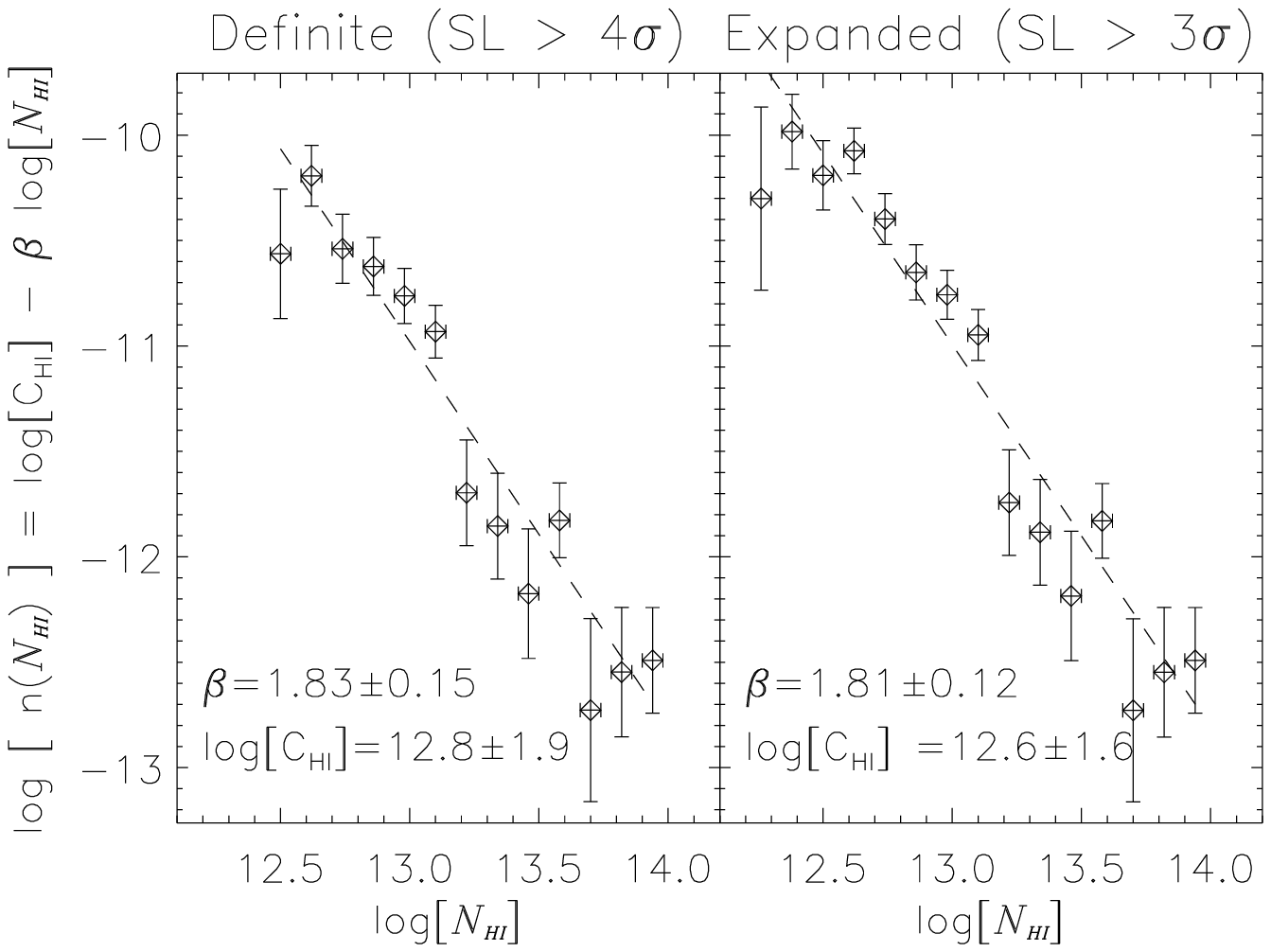}
	\caption{\label{d2ndzdnh} The low-\z\ \lya number distribution per unit redshift
and column density, \nNh=\dtwondzdnhno, for a constant \bvalue of \myb and $\lognh < 14.0$. The \nNh\ error bars are
based upon $\sqrt{N}$ statistics. The left panel indicates the distribution for our definite (\real) \lya absorbers only. The right panel includes the
distribution for our expanded (\expanded) \lya sample.}
\end{figure}
On the logarithmic or flat-part of the curve of growth, \Nh depends on the \bvaluesno. 
Above \logNh\about18.5, the \lya absorbers develop damping wings, and
 \Nh  is once again insensitive to the \bvalueno.  None of our detected \lya absorbers has \logNh \gt 17.
While we believe this to be a typical range of true values for our detections, clearly there is some uncertainty in the
individual values of \nh at $\lognh \ge 13.5$. 

The \Nh number density per unit redshift and column density is often modeled by a power-law distribution:
\begin{equation}{
 \dtwondnhdzover \sim {N\left(\Nhno\right) \over \Dz\left(\Nhno\right)\DNh} \equiv \nNhno=\betaNh~~.}\end{equation}
In Figure~11, we display $\log\left[\nNhno\right]$ for both our definite (\definite) and 
expanded (\expanded) \lya samples over the range
\lowrange. Also indicated in Figure~11 are the least-squares fits to $\log\left[\nNhno\right] = \log\left[\NofNhno\right]-\beta$
\logNh. 
Using a \bvalue of \myb\ for all features, we obtain
$\beta =1.83 \pm 0.15$ and $\log\left[\NofNhno\right] = 12.8 \pm 1.9$ for our definite sample and 
$\beta =1.81 \pm 0.12$ and $\log\left[\NofNhno\right] = 12.6 \pm 1.6$ for our expanded sample 
over the range \lowrange. 
 There is no evidence for  a turnover of \nNh\ below $\Nh= 10^{12.5}$\percmtwo in either our definite or expanded \lya samples.
 The determination of $\beta$ and \NofNh is insensitive to \bvalue below $\lognh  \leq  14$,
since all absorbers are then on or near the linear portion of the \lya curve of growth, 
which is independent of \bvalueno.

Above \lognh\about14, the \lya absorbers become partially saturated, and the choice of \bvalue becomes important in determining
\Nhno. As shown in Figure~12,  we detect a break in the $N^{-\beta}_{\rm HI}$\ power-law above \lognh\about14, which is
possibly related to saturation or poor line statistics. 
However, \citet{Kulkarni96} detect a similar break in the region of \nh\about10$^{14.5}$\percmtwo
for \lya absorbers at $1.7<\z<2.1$.  
Figure~12 compares the power-laws for the two column density regimes, \lowrange\ and \highrange, for both
our definite (\real) and expanded samples (\expanded), again for a constant \bb=25\kmsno. 
Since none of the strong absorbers (\lognh $\gt$ 14.0) are tentative detections (\tent), the results for the definite and
expanded samples are identical with $\beta = 1.04 \pm 0.39$ and $\log\left[\NofNhno\right] = 1.5 \pm 5.7$. 
The large  uncertainties are a reflection of the poor number statistics. 
\input{table5.tex}
\begin{figure}[htbp]
	\plotone{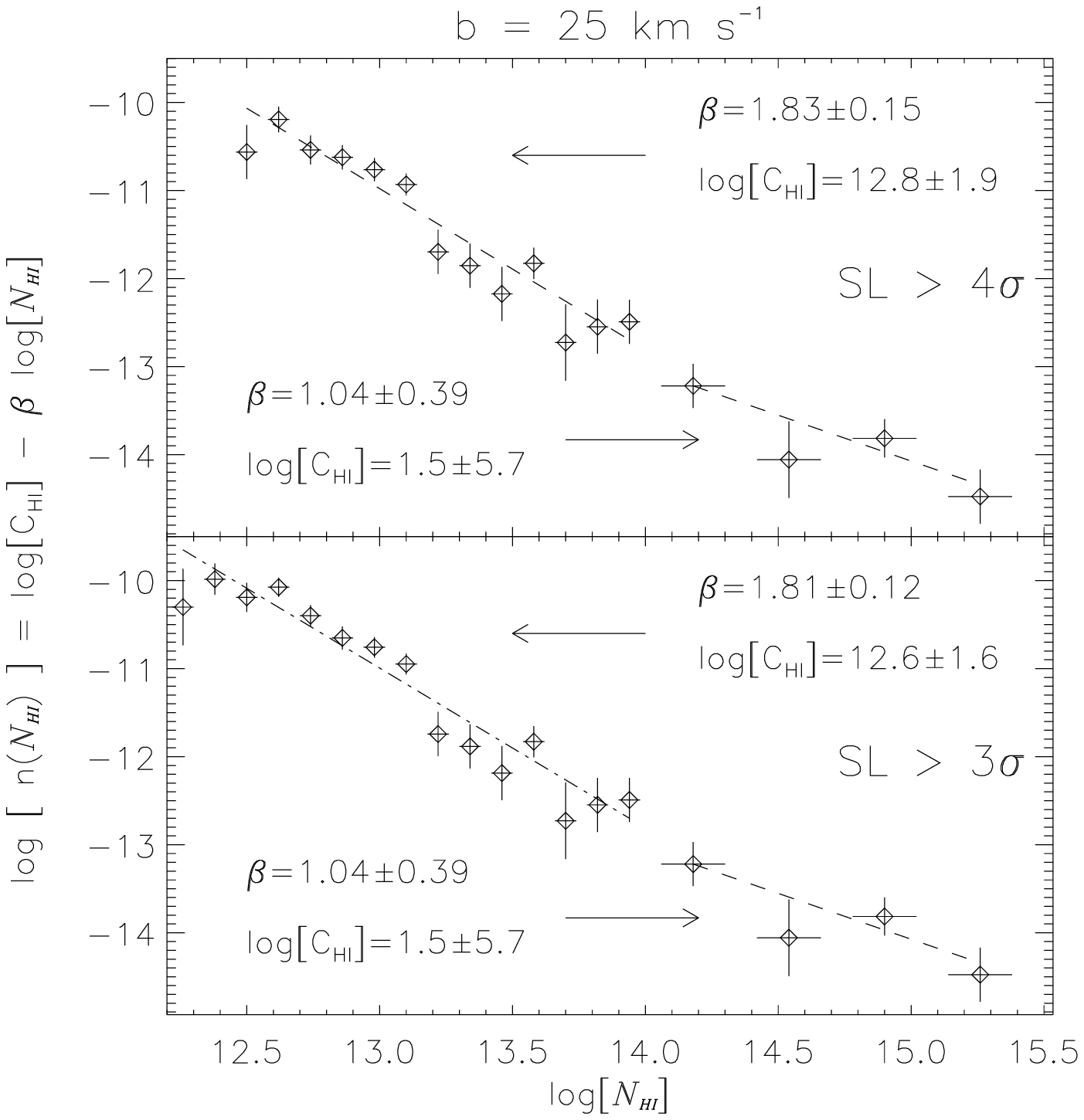}
	\caption{\label{d2ndzdnh_full} The differential \lowzya number distribution
per unit redshift and column density, \nNh=\dtwondzdnh, over the range \fullrange\  for a constant \bvalue of \myb. The \nNh\ error bars are
based upon $\sqrt{N}$ statistics. The upper panel indicates the distribution for definite (\real) \lya absorbers only. The lower panel includes
the distribution for expanded (\expanded) \lya sample.  Indicated on each panel are the best-fit power-laws for each \logNh\ regime,
\lowrange\ and \highrange. In the lower \Nh regime, the \logNh\ bins are 0.12, while in the
upper regime the bins are 0.40.}
\end{figure}
To obtain better determinations of $\beta$ and \NofNh, we evaluate the integrated \nNh,
\begin{equation} \label{iNh} \iNh =\int_{\Nh}^{\infty} \betaNh \dNh = {\NofNh \over 1-\beta }~\Nhno^{1-\beta}. \end{equation}
Table~5 compares our results for both the differential, \nNh, and integrated, \iNh, determinations of $\beta$ and \NofNh
for \bvalues of 20, 25, 30\kms and \bobs. This table is divided into three \lognh\ ranges over which least-squares determinations of
$\beta$ and \NofNh are performed: \lowrange, \highrange, and the
full range \fullrange. Results are given for both our definite and expanded \lya samples. 
By column Table 5 includes: (1) sub-sample name; (2) \bvalue assumed in\kmsno;
 (3-6) best-fit values for $\beta$ and $\log\left[\NofNhno\right]$ in cm$^{-2}$ 
for both differential and integral counts; and (7) the number of absorption lines in each subsample. 
Note that the results over
the range \highrange\ are the same for the definite and expanded samples since these data sets are identical over this
\lognh\ range. When using our measured \bvaluesno, all features in this column density range occur in the same \lognh\ bin.
Hence we cannot report values for this column density range for these \bvaluesno.
Using the integrated \iNh\ rather than the differential distribution \nNh\ to determine $\beta$ and
\NofNh is more robust since it uses the cumulative
\Nh distribution, instead of the individual \Nh bin values in the parameter determination. Using the integrated distributions for \bb=\myb, we
obtain $\beta = 1.72 \pm 0.06$ and $\log\left[\NofNhno\right] = 11.3 \pm 0.7$ for our definite sample and 
$\beta = 1.81 \pm 0.05$ and $\log\left[\NofNhno\right] = 12.4 \pm 0.7$ for our 
expanded sample over the range \lowrange.

As indicated in Table~5, the values for $\beta$ that we obtain over the 
column density range \lowrange\ are consistently \gt 1.7.  
This value is in agreement with our results of \S~\ref{sec:REW}, where we fitted \nW $\propto$ \Wno$^{-\beta}$ and 
obtained $\beta = 1.89 \pm 0.39$.
We derive our best values from the integrated \definite\ sample, using a constant value \bb=25\kmsno:
	\begin{equation}{ \beta = 1.72 \pm 0.06 ~,~  \log\left[\NofNhno\right] =11.3 \pm 0.7 ~{\rm for}~12.5 \leq  \logNh  \leq  14.0 ,}\end{equation}
	\begin{equation}{ \beta = 1.43 \pm 0.35 ~,~  \log\left[\NofNhno\right] =7.4 \pm 5.2 ~{\rm for}~\highrange\ {\rm ,}}\end{equation}
	\begin{equation}{ \beta = 1.66 \pm 0.06 ~,~  \log\left[\NofNhno\right] =10.5 \pm 0.8 ~{\rm for}~12.5 \leq  \logNh  \leq  16.0 .}\end{equation}
These results are in contrast with the recent higher-\z\ results of \citet{Kim97}, \citet{Lu96}, and \citet{Hu95}, who find that
 $\beta$\about1.4  adequately describes \nNh\ for 12.3\lt\lognh\lt14.3 over the redshift range 2.17\lt\z\lt4.00.
In conjunction with other studies between \z=4 and \z=2, these authors claim that the \nNh\ break appears to strengthen and move to lower column 
densities with decreasing redshift.
 Furthermore, \citet{Kim97} report
a break in \nNh\ above $\lognh \ge 14.3$ that steepens ($\beta=1.7-1.8$) and 
moves to lower column density with decreasing redshift (2.17\lt\z\lt4.0). 
In contrast, at low-\z\ we see a flattening in \nNh\ above \lognh=14.0. 
It is possible that the apparent flattening in our data is an artifact of our selection of a constant
\bvalue for all \lya absorbers. If the stronger \lya absorbers had larger \bvaluesno, 
then the inferred column densities of these features could be much smaller.
This would leave us with fewer features above \lognh \gt 14, and therefore  no evidence for a break in \nNh\  at $\lognh = 14.0$. 
In \S~\ref{sec:dndnh_z}, we will examine
the possibility that the break in \nNh\  is a consequence of the \z\ evolution of \dtwondzdnh.
\clearpage
\section{Observed Redshift Distribution}
\label{sec:Z}
In this section, we examine the redshift distribution of the \lowzya forest. In particular, we want to examine the evidence
for structure at specific redshifts, or for any evolution of \dndz with \z. 
We begin with Figure~13, a presentation of the observed number distribution per redshift bin without any sensitivity correction, \Nno(\z),
 for both our definite and expanded \lya samples. 
The left axis corresponds to the displayed histograms of \Nno(\z). Our redshift range, \zrange, is divided into
 12  redshift bins, $\delta$\z = 0.0056. Individual \real\ \lya absorbers 
are plotted as pluses, whose \Wno(\nomang) is shown on  the right vertical axis.
Individual \tent\  \lya absorbers are plotted as `{\bf x}'s. 
Like the \Nno(\Wno) distribution, the \Nno(\z) distribution is not the true
redshift distribution, because we have not corrected for the varying wavelength and sensitivity coverage of our
observations. 

We show the correction for our varying wavelength and sensitivity coverage, \Dz(\z), in Figure~14.
In this figure, the panels show various stages of 
correction for the cumulative distribution of \Dz(\z)\ for \real\ detections in all sightlines. The upper panel of
Figure~14 presents the full pathlength availability of our sample. 
The left axis gives \cDz(\z) in \mmsno, while the right axis is in redshift units.
Subsequent panels present the available pathlength after removing portions of our 
wavelength coverage due to obscuration by extragalactic non-\lya
features and intrinsic features (Intrinsic+Non-\lyano), by Galactic+HVC features, and by our \prox proximity limit. 
The bottom panel presents the pathlength available
after removing all spectral regions not suitable for detecting intervening \lya absorbers. The available pathlength
in Figure~14 approximately represents \Dz(\z) for features with \Wno\gt 150\mang. As presented in
Figure~4,  our pathlength is approximately constant at \cDz(\Wno)=0.387  for \Wno\gt 150\mang.
 \begin{figure}[htbp]
\epsscale{0.8}
	\plotone{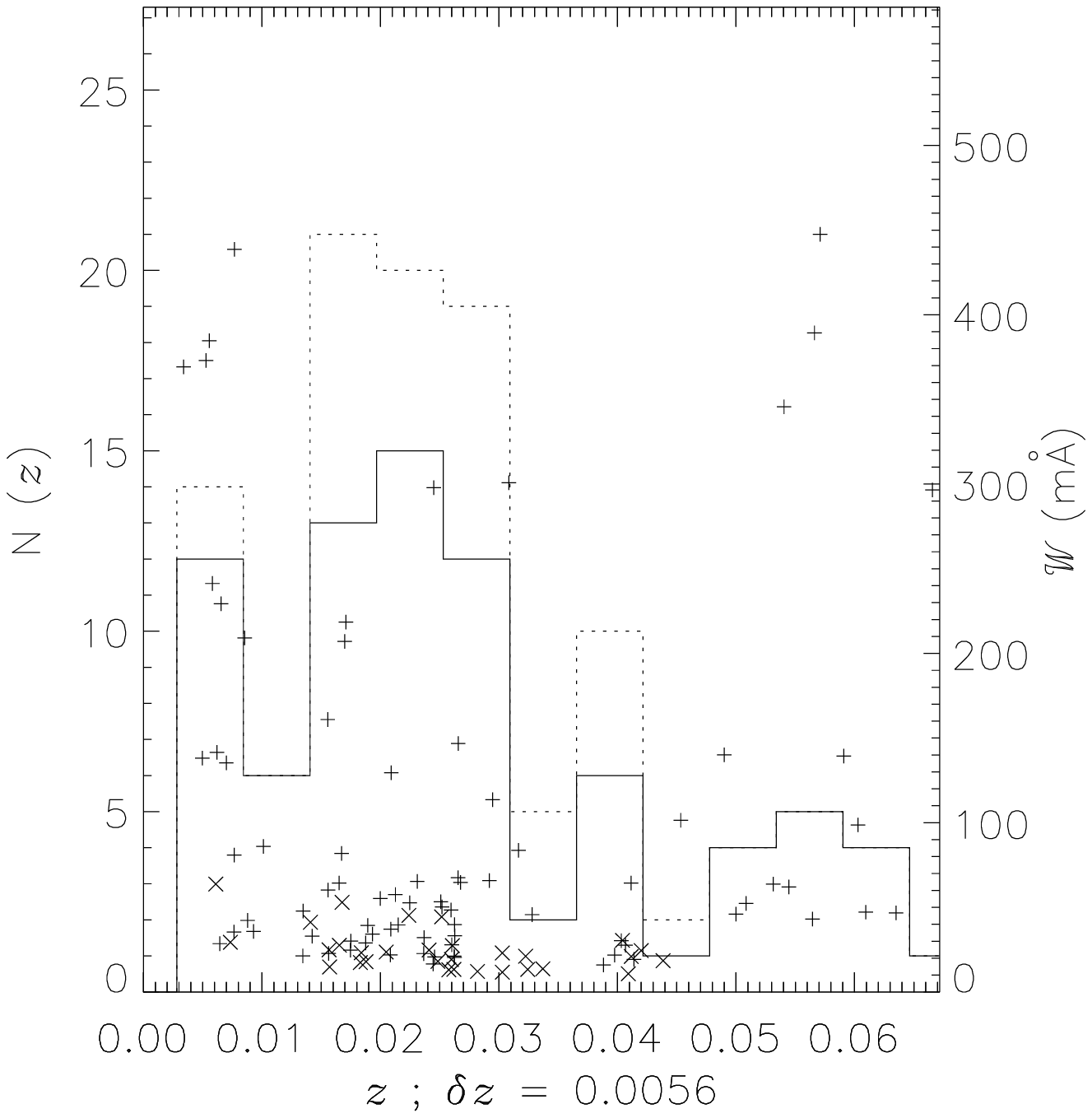}
\caption{\label{ew_redshift} Histogram of \Nno(\z) and distribution of rest-frame equivalent
width (\Wno) versus redshift. The histograms correspond to the left axis and indicate \Nno(\z), 
the number of absorbers per redshift bin
(\delz). The solid histogram indicates the distribution of the 
definite (\real) \lya sample, while the dotted distribution is for the expanded (\expanded)
\lya sample. The pluses (`+') and {\bf x}'s  correspond to the right 
axis and indicate \W of the individual \lya absorbers in the
 definite and possible (\tent) samples, respectively.}
\end{figure}
\begin{figure}[htbp]
\epsscale{0.8} \plotone{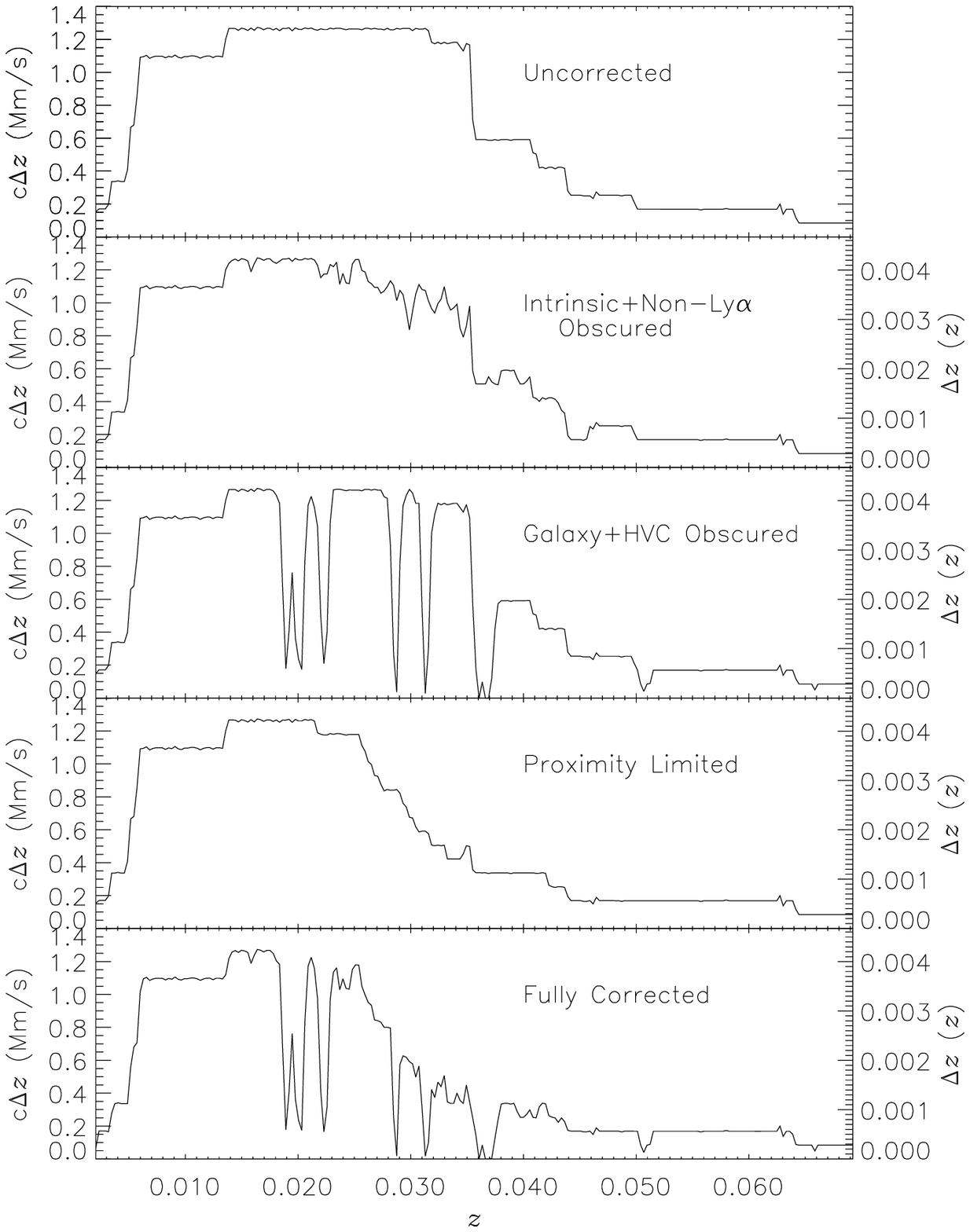}
	\caption{\label{zsens_path}Cumulative available pathlength (\Dz) for all sightlines as a
function of \z, for features with $\Wno \gt 150$\mang. 
Left axis gives \cDz(\z) in units of \mmsno, while the right axis gives \Dz(\z) in terms of
\z. The upper panel presents the full, uncorrected \Dz(\z). 
Subsequent panels present \Dz(\z) after removing portions of our
wavelength coverage due to obscuration by extragalactic non-\lya features, Galactic + HVC features, and the
(\proxno) proximity limit. The bottom panel presents our \cDz(\z) for detecting
intervening \lya absorbers after applying these corrections.}
\end{figure}
To properly characterize the \lowzya absorber distribution in redshift, one must accurately account
for varying available pathlength as a function of both  \Nh and \z\ of the observations. 
Figure \ref{full_cpath} displays 
the combined two-dimensional  sensitivity function, \cDz(\Nhno,\z), for  GHRS/160M observations as a function of \z\ and \Nhno, 
after accounting for the aforementioned corrections and obscurations. 
\begin{figure}[htbp]
\epsscale{0.8} \plotone{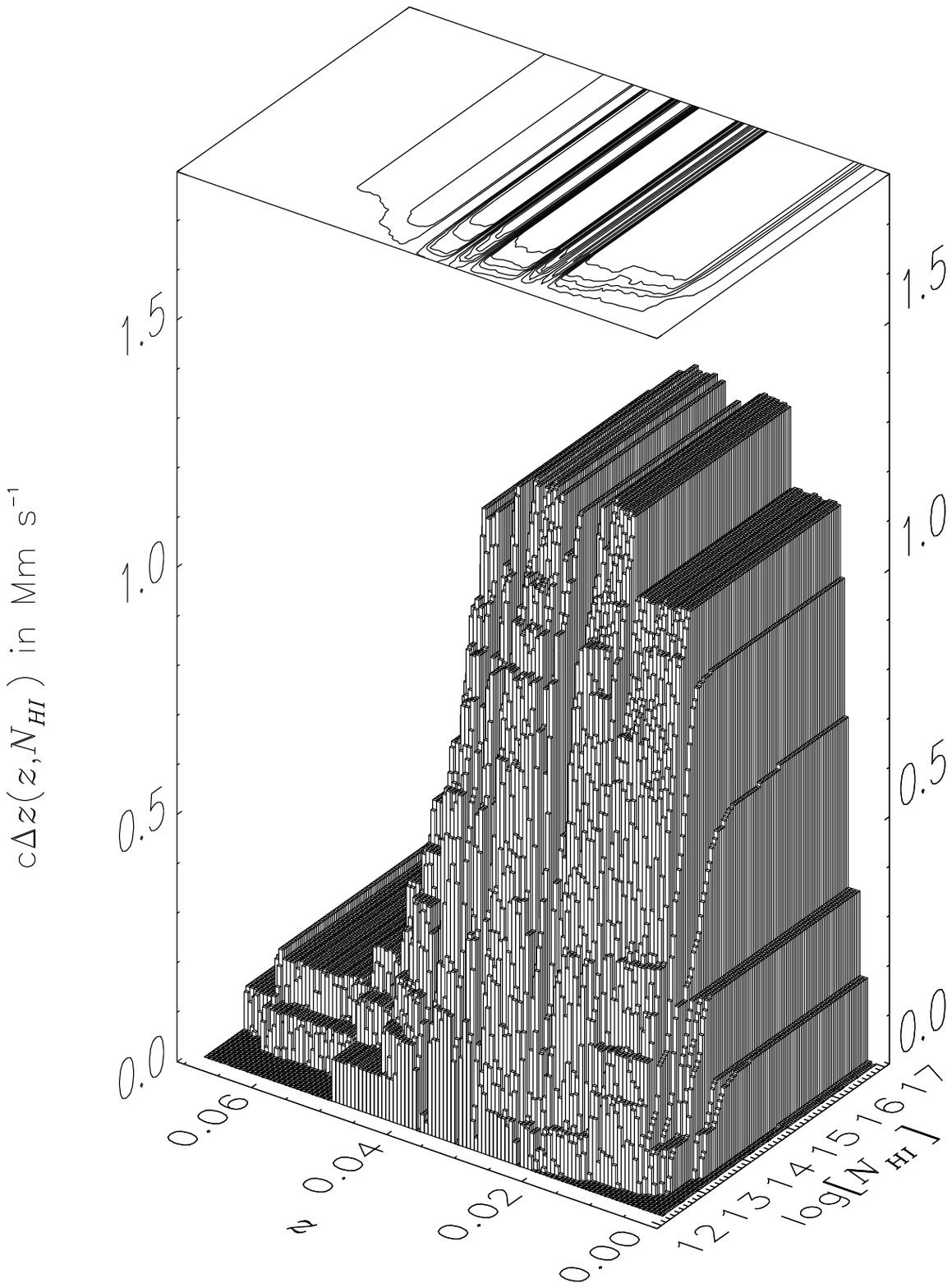}
	\caption{\label{full_cpath} Bivariate pathlength sensitivity correction, \cDz(\Nhno,\z). The
sensitivity function corrects \dtwondzdnh by accounting for varying regions of  \Nh sensitivity and \z\ coverage. 
The vertical axis indicates the available pathlength 
(in \mmsno) for a given column density, \Nhno, and redshift (wavelength). When
viewed from the column density axis, this figure reduces to the dotted line in Figure~4, and when viewed from the redshift axis,  
the bottom plot of Figure~14 is recovered.
}\end{figure}

Unfortunately, we do not have enough statistics (absorbers) to fully analyze the \Nh versus \z\ distribution over
the small \z-range of our sample. However, there is no obvious trend over our local pathlength.
We therefore assume no \z\ evolution \citep{Weymann98} of the column density distribution for our \lya sample 
($\beta$ is constant for all \z\ in our observed range).

\begin{figure}[htbp]
\epsscale{0.8}
\plotone{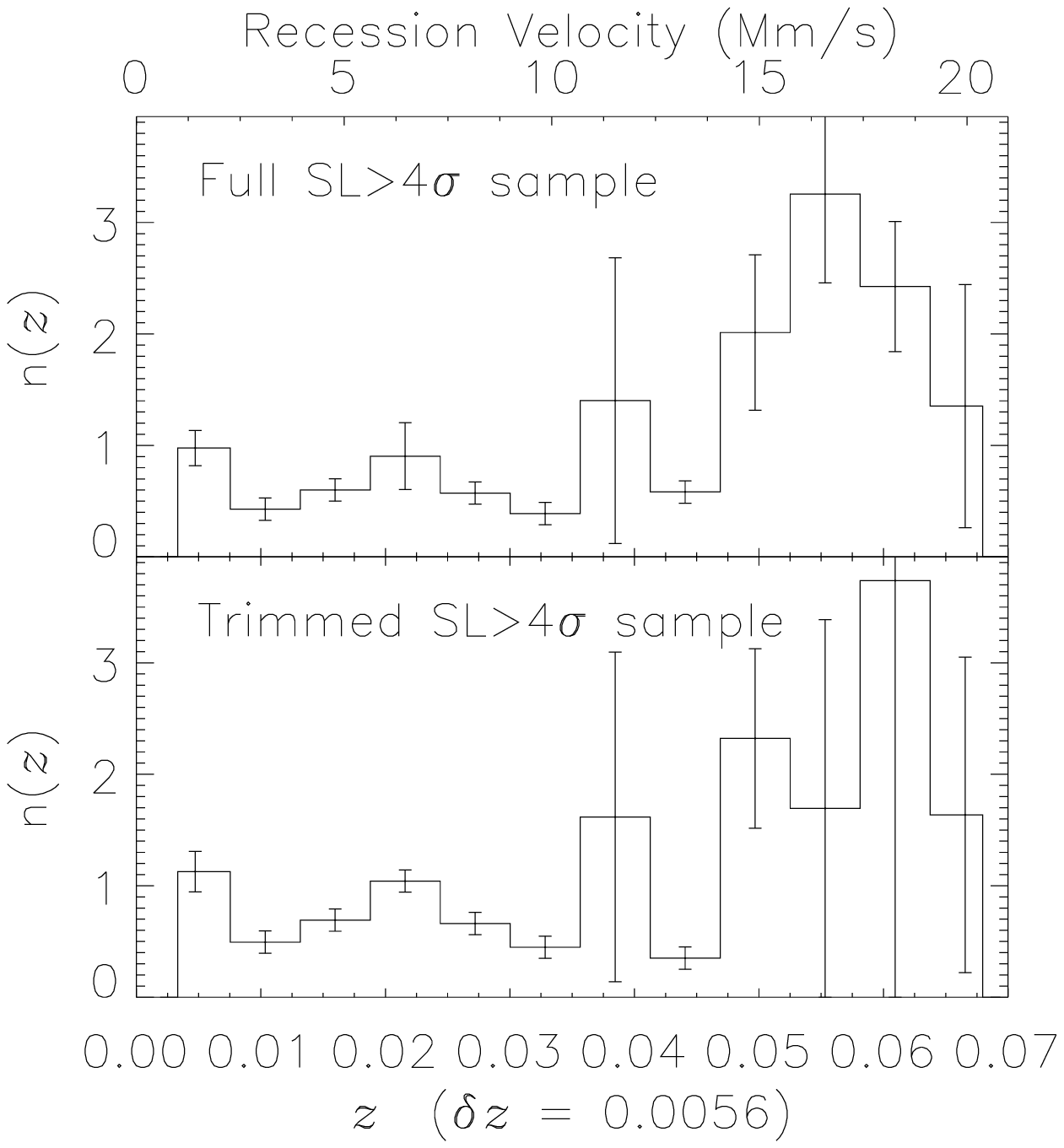}
	\caption{\label{dndz_z} 
The distribution, \nzi\ or normalized \dndz per unit redshift, is plotted versus \z\ for
the definite (\real) sample in the upper panel, and for the trimmed (\real) sample in the lower panel. Error
bars are based upon Poisson statistics.
The excess in \nzi\ above \z=0.05 in the upper panel is due to a complex of lines in the \PKS\ sightline at $\lambda > 1280\ang$.
We specifically selected this sightline to observe this complex of lines, therefore, we trimmed this portion of
the \PKS\ sightline to remove any bias introduced by selecting a sightline, which we previously knew contained an
strong complex of \lya absorbers. No obvious \z\ evolution of \dndz in observed over our small redshift range.
The bin size is \delz=0.0056. For comparison, a value of \nz =0.5 corresponds to \dndz\about90, integrated over all \Nh to which we
are sensitive.}
\end{figure}

It is customary to describe the pathlength-corrected number dependence of \lya absorbers as  a power-law in redshift, $\nz=\Nofz \zgamma$. Because our spectra vary in wavelength coverage and \Nh sensitivity, we
must correct \nz~for  incompleteness. In particular,  we know that there exists a distribution, 
\begin{equation}{\nNh= {\Nno(\Nhno) \over \Dz(\Nhno)\DNh} =\betaNh,}\end{equation} that describes the \Nh distribution, 
and we know the pathlength availability, \Dz(\Nhno,\z), as a function of both \z\ and \Nhno. 
%(Figure~\ref{full_cpath}).
Therefore, the \z\ distribution evaluated at each redshift bin, \nzi, can be corrected for incompleteness by, 
\begin{equation}
	\label{eqNz}
\int_{\Nmin}^\infty  \nNh \nzi d\Nh =\int_{\Nmin}^\infty 
{\partial ^2 \Nno(\Nhno,\zi) \over \partial z\, \partial \Nhno}\dNh,
\end{equation}
\begin{equation}
\approx  \int_{\Nmin}^{\infty} {\Nno(\Nhno,\zi) \over \Dz(\Nhno,\zi)\DNh(\Nhno)}  d\Nhno
\end{equation}
or, 
\begin{equation} \label{nzi}
 \nzi = { {\int_{\Nmin}^{\infty} {\Nno(\Nhno,\zi) \over \Dz(\Nhno,\zi)\DNh(\Nhno)} d\Nh} \over {\int_{\Nmin}^{\infty} \nNh d\Nh} }
\end{equation} 
where, 
\begin{equation} \nNh = \betaNh . \end{equation}
The separability of the bivariate distribution, $\partial ^2 N/\partial z\, \partial \Nhno$, 
assumed in equations~(\ref{eqNz}) and (\ref{nzi}) implies that
there is no \z\ evolution of the \Nh distribution of our
\lya absorbers within the small \Dz\ of our GHRS observations. Under this assumption, the integrals in equation~(\ref{nzi}) can be combined, 
and the observed redshift distribution of absorbers, \nz, can be
expressed as:
\begin{equation} \label{nzi_final}
	\nzi= \int_{\Nmin}^{\infty} {  \Nno(\Nhno,\zi) \over \betaNh \Dz(\Nhno,\zi)\DNh(\Nhno)}  d\Nh = \Nofz \zgammai .
\end{equation} 
%In terms of the familiar derivative forms, this can be expressed as 
%\begin{equation} \label{deriv_dndz}
%	\nzi = { \dtwondnhdzi \over \dndNhi} = \Nofz \zgammai .
%\end{equation}
Our goal is to search for the variations in \dndz with redshift, accounting for the sensitivity limits of our spectra.
Thus, $\nzi$ in Figure~\ref{dndz_z} shows $\dndzno(\z_i)$ normalized by the expected \dndz given our observed \dndnh distribution and sensitivity limits for a
uniform distribution of clouds over our redshift range. For all \z\ bins, \Nmin\ is set to $10^{12.3}$ \percmtwono, although most bins
have no pathlength at this column density.

The top panel of Figure~\ref{dndz_z} displays \nzi\ for the \real\ sample. Error bars in this figure are based upon Poisson statistics.
The increase in features at \z=0.055-0.065 is due to 
a single cluster of lines in the upper portion of the \PKS\ spectrum. 
We specifically selected this sightline to observe this complex of lines, therefore, we trimmed this portion of
the \PKS\ sightline to remove any bias introduced by selecting a sightline which we previously knew contained an
strong complex of \lya absorbers.  The trimmed distribution is shown in the lower panel of Figure~\ref{dndz_z}.
Above \z=0.05, we have very little pathlength after removing the upper portion of the \PKS\ spectrum from our sample.
The large uncertainty in \nzi\ at \z=0.04 is due our reduced pathlength in this bin due to the presence of Galactic \SII1259 and \SFsixty.
We see no compelling evidence for any \z\ evolution over our small range in redshift ($\gamma$ is consistent with 0; $\gamma = -0.01 \pm 0.02$).
\subsection{Redshift Evolution of \dndzno.} \label{sec:dndzz}
Figure~\ref{comp_dndz}~displays \dndz over the redshift interval $0\lt\z\lt3$, 
for several studies over two \W ranges: \Wno$\ge$240\Mang
(\lognh \gt 14 for \bb=\myb) and $60\mang\le\Wno\le240\mang$ (\kimrange\ for \bb=25\kmsno). 
The lower distribution in Figure~\ref{comp_dndz} is data normalized to $\Wno \ge 240$\Mang by \citet{Weymann98}, while the upper
distribution corresponds to absorbers in the range \kimrange. 
The data points indicated by squares (\zonelog \lt 0.4) were obtained as part 
of the HST/FOS Key Project \citep{Weymann98}, and while normalized to
$\Wno \ge 240$\mang, does contain a few absorbers below that limit, down to 60\mang. 
The stars and diamonds correspond to ground-based data (\zonelog \gt 0.4), taken
with an equivalent width limit of $\Wno \gt 360$\mang, reported by \citet{Lu91} and \citet{Bechtold94}. 
Since the distribution of \dndz
at $\Wno \gt 240$\Mang can be described by a single power-law in \z, the \highz data were scaled by 
 Weymann \etl (1998) to be consistent with the HST/FOS Key Project $\Wno \gt 240$\Mang 
 data for $\z \lt 1.5$. 
The two low-\z\ points indicated by filled circles are taken from this survey, 
one point for each of the two sensitivity ranges. 
The solid lines indicate the best fits to the \Wno \gt 240\Mang data above 
and below \z = 0.4, and have slopes of
$\gamma = 0.16\pm 0.16$ \citep[\zonelog \lt 0.4, ][]{Weymann98} and $\gamma=1.85\pm0.27$ 
\citep[\zonelog \gt 0.4, ][]{Bechtold94}. 
The solid triangle represents the mean of the \citet{Kim97} data (open triangles), 
for which $\gamma = 1.19\pm0.45$ (dash-dot line). While the difference between $\gamma=1.85$
for $\Wno \gt 240\Mang$ and $\gamma=1.19$ for lower-\W
is suggestive of a slower evolution at high-z for the lower \W
absorbers, the large error bars on the  \citet{Kim97} data mean that
this is far from definitive at this time. However, the \citet{Savaglio99} point, using the QSO in the Hubble
Deep Field-South, is suggestive of a somewhat slower evolution as well, but again with large error bars.

The interpretation of the high-\W data follows the numerical modeling of \citet{Dave99}.
The break in \dndz\ at \z\about0.4 can be explained by the expansion of the Universe, 
combined with a rapid decline in the metagalactic
ionizing flux at \z \lt 2. The expansion of the Universe causes \lya absorbers to decrease 
in density with decreasing redshift in the absence of gravitational confinement, 
resulting in a rapid decrease in recombinations and thus in the observed column density 
of \hone for any
initial baryon overdensity. Also, since there exists an inverse relationship between 
the number of clouds and \Nhno, as the Universe expands there are
fewer clouds at any given \Wno. 
At \z\lt2, the rapid decrease \citep{Shull99b} in the ionizing background intensity, \Jnuz, allows \lya clouds 
to become less ionized ($n_{\rm HI} \propto n_H^2 / \Jno$), 
which results in an increase in the amount of hydrogen detectable in \lyano. 
In other words, at \z\lt2, the decline in the recombination rate
caused by the expansion of the Universe is countered by a decreasing 
photoionization rate (\photo) due to the declining UV background, resulting in the
dramatic break in \dndz seen in Figure~\ref{comp_dndz} at $\log[1+\z] \about 0.4$. 
%This decrease is not sufficient to overcome the effects of the expansion of the
%Universe. However, it is sufficient to flatten the \z\ evolution of 
%\dndz below \zonelog\lt0.4 (\z \lt 2), as shown in Figure~\ref{comp_dndz}.

\subsection{An Important Technical Issue for Redshift Evolution of low $\Wno$
Absorbers}
\begin{figure}[htbp]
\epsscale{0.7}
	\plotone{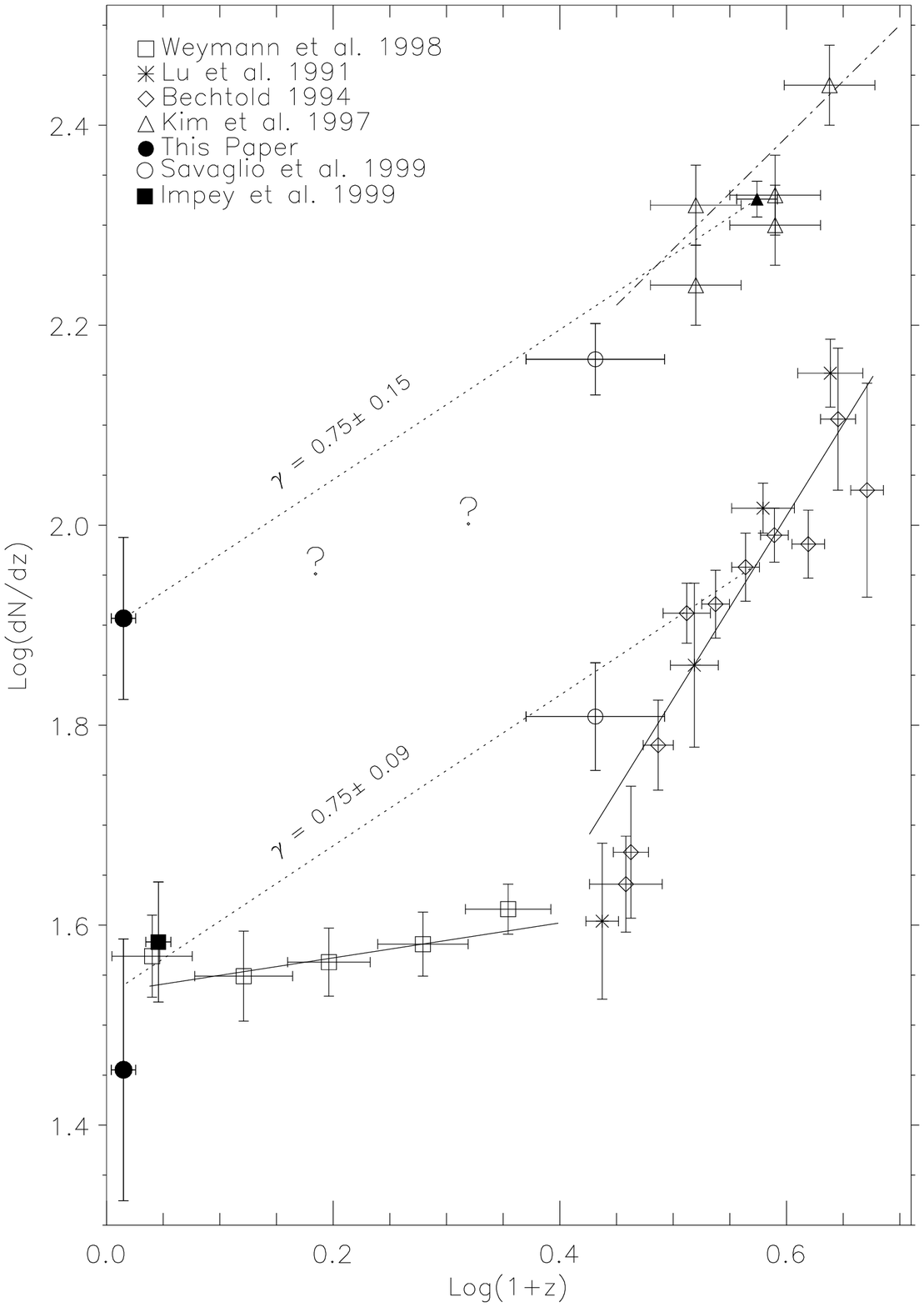}
	\caption{\label{comp_dndz} 
Comparison of \dndzlog\ versus \zonelog\ for two \Nh ranges. 
The lower distribution corresponds to $\Wno \ge 240$\Mang ($\Nh \ge 10^{14}$\percmtwo for \bb=25\kmsno). 
The upper distribution corresponds to absorbers in the range \kimrange. The
\z\about0 points (solid circles) are taken from our survey (Figure~8), 
for each of the two \Nh ranges.
Solid lines are taken from \protect{Weymann \etl (1998)} and have slopes 
of $\gamma = 0.16$ (\zonelog \lt 0.4) and 1.85 (\zonelog \gt 0.4). 
The filled triangle is the mean of the \protect{\citet{Kim97}} data (open triangles), for which $\gamma = 1.19$. 
Connecting this point (upper dotted line) to our datapoint with \kimrange\ at \z$\sim$0 gives 
$\gamma=0.75\pm0.15$. For the $\Wno \ge 240$\Mang sample of Weymann \etl (1998), 
the nearly identical result of $\gamma=0.75\pm0.09$ is obtained, 
indicating no difference in the overall \z-evolution of \dndz for these distributions.
The complete evolutionary picture for low-\Nh absorbers is
not available due to the lack of data for $0.1 \lt \zonelog \lt 0.4$, indicated by the question marks.}
\end{figure}

It has been suggested by cosmological hydrodynamical simulations \citep{Dave99} 
that one would expect lower-\W \lya absorbers to evolve more slowly (smaller $\gamma$). 
Evidence for this effect is seen in Figure~\ref{comp_dndz} for \zonelog \gt 0.4 in 
that the lower-\W distribution has a shallower slope \citep[$\gamma$= 1.19; ][]{Kim97} 
than the higher-\W distribution \citep[$\gamma$=1.85; ][]{Bechtold94}.
While this difference
is suggestive of a slower evolution at high-z for the lower $\Wno$
absorbers, the large error bars on the \citet{Kim97} and \citet{Savaglio99} data mean that
this result is not yet definitive.
A slower evolution for lower $\Wno$ absorbers has
also been reported in HST/FOS data at lower redshift by \citet{Weymann98}. 
Indeed, the best-fit value for the lowest-\W absorbers in the Key Project data has a very slight
negative evolution (i.e., fewer absorbers at higher-\z).
To test this hypothesis for the low-\W absorbers with \kimrange, we connect the mean point of 
\citet[][upper dashed line]{Kim97} to our \z=0 data point (large filled circle in Figure~8) 
giving a slope $\gamma = 0.75 \pm 0.15$. 
A similiar analysis for the $\Wno \ge 240\Mang$ sample (connecting the mean point of \citet{Bechtold94} to the extrapolation
of the \citet{Weymann98} best-fit line to \z=0) produces a nearly identical result, 
$\gamma = 0.75 \pm 0.09$, indicating no distinguishable difference in the  overall \z-evolution of \dndz for 
these two distributions. In some ways, the much larger redshift difference between
these data points yields a much more secure result than either the
\citet{Kim97} or Weymann \etl (1998) analyses alone and also one that apparently differs
from these earlier papers. But because these other works are measuring only portions
of the redshift evolution, while our two-point measurement is for the full \z~range, these
results can differ and still not be contradictory (i.e., the ``breakpoint'' from faster to slower evolution could 
occur at lower redshifts for the low-\W absorbers).

Nevertheless,  we suspect that the source of this apparent discrepancy may arise from the manner in which these earlier
analyses have fitted the observed line densities. Specifically, \citet{Dave99} have extended 
the $\nW=(\NofW /\Wstarno) \exp\left(-\Wno/\Wstar\right)$ relationship to inappropriately low
\W values.  As discussed in \S~\ref{sec:ew_spectrum}
and as shown in Figure 7, below $\Wno$s of about
133\mang, where the absorption lines are on the linear 
portion of the curve of growth, the \nW\ 
relationship is better expressed as $\nW \propto \Wno^{-\beta}$.
Thus, the \citet{Dave99} fitting technique results in an overestimate of \Wstar and 
an underestimate of $\gamma$ by assuming an exponential distribution of absorber numbers rather
than the steep power-law, which is observed.  
While this overestimate almost certainly occurs in the \citet{Dave99} analysis, the HST/FOS Key Project
line list does not contain enough absorbers at $\Wno \le 150\Mang$ to cause a significant overestimate
by assuming an exponential distribution in \Wno.  
The fact that \citet{Weymann98} find similiar \Wstar values for all their analyzed subsamples is 
suggestive that this is, in fact, the case. On the other hand, the \citet{Dave99} synthetic line list almost certainly
has many absorbers that are on the linear portion of the curve-of-growth.
This would explain why the \citet{Dave99} analysis of their simulations
 significantly overpredicts (factor of 2.3) the number of 
$\Wno \ge 50$\Mang \lya absorbers found by \citet{Tripp98}, and would
presumably overpredict the number density of low \W absorbers that we have found as well. 
 
The amount of this underestimate of $\gamma$ increases with decreasing redshift in the 
\citet{Dave99} analysis (their Figure 8), casting doubt on their
prediction that weaker \W features should evolve faster in \dndzno. Because
this overestimate of the line density and attendant underestimate of
$\gamma$ are most important at $\Wno \lt 200$\mang, it is the lowest \W point
in Figure~9 of \citet{Dave99}, and possibly in  Figure~7 of Weymann \etl (1998)
that are too low. At higher \Wno, where the inferred equivalent 
widths are not affected by this technical issue, both the \citet{Dave99} simulated
data and \citet{Weymann98} Key Project data still show marginal evidence for slower evolution.
Thus, if the proper analysis were
made of both the simulations and the local \lya data, they would still
agree. However, there would be little evidence that the higher and lower
\W absorbers evolve at different rates between \z=3 and 0. This is
exactly what our data show in Figure~\ref{comp_dndz}.  
When our data are combined with that of \citet{Kim97} and \citet{Weymann98},
there is no conclusive evidence for a difference in evolutionary rates between high- and low-\W absorbers. 

The lower \W results are complicated 
by the lack of \dndz data in the range \kimrange\ 
for $0.1 \lt \zonelog \lt 0.4$. 
This redshift range ($0.25 \le \z \le 1.5$) is indicated in Figure~\ref{comp_dndz} 
by the question marks.
%For example, using our value for \dndz ($\Wno \gt 240$\mang), we arrive at $\gamma=0.91\pm0.12$,
%suggesting the possibility of differing \z-evolutions for different column density regimes,
%but the number of absorbers in our high-\W sample is small and thus not
%nearly as compelling statistically as the Key Project data points at low-z. 
Two HST+STIS cycles 8 \& 9 projects (B.~Jannuzi,~PI) are scheduled to obtain spectra 
of sightlines in the redshift range, $0.9 \lt \z \lt 1.5$ to address this deficiency.
The improved sensitivity of HST+COS also should be able to provide data to clarify 
the evolution of weak \lya lines at redshifts $0 \lt \z \lt 1.5$.
Furthermore, the inclusion of our cycle 7 STIS observations (J. T. Stocke, PI) 
should double our number statistics at low redshift and help determine
the low-\W line density more precisely as well.
\subsection{Redshift Evolution of \dtwondzdnhno.} \label{sec:dndnh_z}
\begin{figure}[htbp]
\epsscale{0.8}
	\plotone{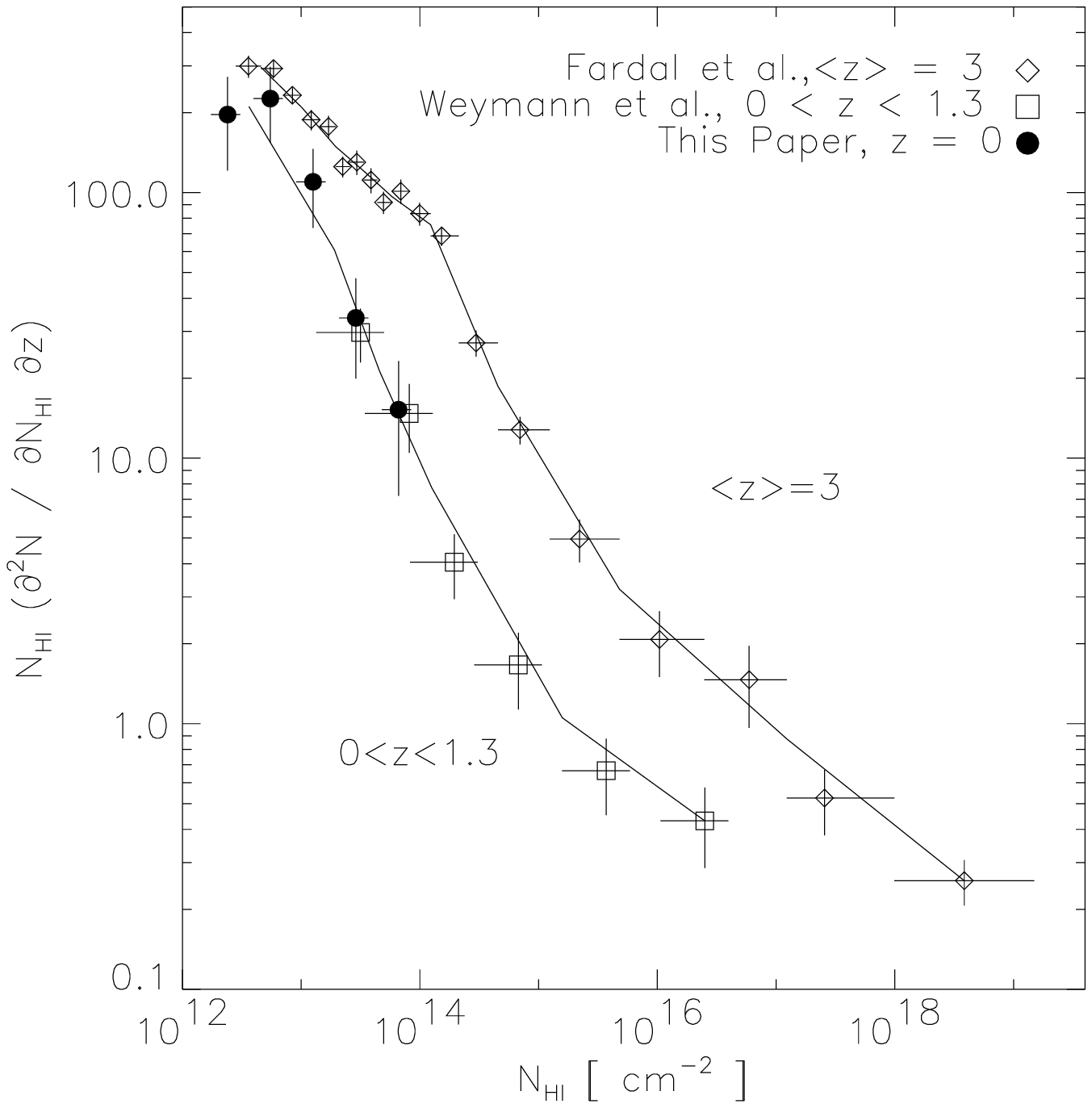}
	\caption{\label{dndnh_z} 
Comparison of the column-density distribution, \nh$\times \left(\dtwondnhdzno\right)$, 
for \z=0 (our data), for the HST/FOS Key Project at 0\lt\z\lt1.3 
\protect{\citep{Weymann98}} and for \meanz=3. 
For our data at \meanz=0, we used a constant value of \bb=30\kms  in determining the column densities to be consistent with the FOS data. 
We include only our  results below \lognh \lt 14, since our statistics become  poor at higher \Nhno.  
The \meanz =3 data are taken from Fardal, Giroux, \& Shull (1998) and references therein. 
The solid lines connect the displayed data with a two-point summation, and are indicative only.}
\end{figure}

In this section, we examine the redshift evolution of the bivariate number distribution 
with respect to redshift and column density, \dtwondnhdzno.  
If any redshift evolution is
detected, it could yield insight into the merging or dissipation  of \lowzya clouds. 
In Figure~\ref{dndnh_z}, we plot \dtwondzdnhno, multiplied by \Nh to expand the
structure near $\lognh \about 14$. We only include our  results below $\lognh \lt 14$, since 
above this column density value our statistics become  poor. However,
we combine our
$\lognh \lt 14$ results (\z\about0) with the HST+FOS Key project data of Weymann \etl (1998) in the redshift range $0 \lt \z \lt 1.3$.
Note that there is good agreement between the Weymann \etl (1998) lowest \Nh points and the highest \Nh points from our survey.
For comparison, Figure~\ref{dndnh_z} also includes data at higher redshift, $\langle\z\rangle = 3$, 
compiled by Fardal, Giroux, \& Shull (1998). 

The preliminary suggestion from Figure~\ref{dndnh_z} is that the 
distribution of \lya absorbers moves leftward and downward from \z=3 to the present.
The major factors governing the \z\  evolution of \dndz for the \lya forest are: 
(1) the expansion of the Universe; (2) the rapid decline
in \Jnuz\ at \z\lt2; and (3) the merger and dissipation of \lya absorbers. 
The expansion of the Universe will tend to disperse clouds, causing \dtwondnhdz to move
leftward in Figure~\ref{dndnh_z}.  In addition, the expansion is expected to reduce \dtwondnhdz at any given \Nhno,
since fewer clouds are available at higher column densities, which have been reduced in column density down to \Nhno. 
These are the same effects that drive \dndz to lower values above \z\gt2 (Figure~\ref{comp_dndz}). 
At \z\lt2, the rapid decline in the ionizing background, \Jnuz, will cause the ionized fraction in the \lya
absorbers to drop. This will increase the measured \Nhno, countering the effect of expansion and causing \dndnh to move 
to the right in Figure~\ref{dndnh_z}.
These same two factors cause the break in \dndzno\ (Figure~\ref{comp_dndz}) for the higher column density absorbers. These two factors alone can explain the
offset in Figure~\ref{dndnh_z} between the lower and higher redshift \dtwondnhdz distributions. Additionally, we expect that some  \lya
absorbers are merging into higher column density systems, or collapsing due to gravity. This will cause additional changes
in the \Nh structure of \dtwondnhdzno. 
 This evolutionary track should be considered preliminary, as more comprehensive data sets, with better
statistics over wider column density  and redshift ranges, are needed to explore the evolution of \dtwondnhdz in detail.
%These data should become available after 2003, with the installation of the Cosmic Origins Spectrograph (COS) on HST.
\subsection{The Opacity of the Low-\z\ \lya Forest}
 For Poisson-distributed \lya clouds \citep{Paresce80},  the effective continuum opacity is given by
\begin{equation} \label{eq_tau}
   \taueff (\nu_{obs},z_{obs},\z) 
        = \int_{z_{obs}}^{z} d\z \int_{0}^{\infty}
      \frac {\partial^2 {\cal N}}{\partial \Nh \partial \z}
      \left[ 1 - \exp(-\tau(\nu)) \right] \; d\Nh \; ,
\end{equation}
where $\partial^2 {\cal N}/\partial \Nh \partial \z$ is the bivariate
distribution of \lya absorbers in column density and redshift,
and $\tau(\nu) = \Nh \sigma\subH(\nu)$ is the photoelectric (Lyman continuum)
optical
depth at frequency $\nu=\nu_{obs}/(1+\z)$ through an individual absorber with column
density \Nhno.  For purposes of assessing the local attenuation
length, it is useful (Fardal \& Shull 1993) to use
the differential form of equation~(\ref{eq_tau}), marking the rate of change of
optical depth with redshift,
\begin{equation}
   \frac {d\taueff}{dz} = \int_{0}^{\infty}
   \frac {\partial^2 {\cal N}}{\partial \Nh \partial \z}
      \left[ 1 - \exp(-\tau(\nu)) \right] \; d\Nh \;.
\end{equation}
The attenuation length, in redshift units, is then given by the
reciprocal of \dtaudz. For simplicity, we calculate \dtaudz~at the Lyman edge (912\ang). 
This is a reasonable approximation due to the strong dependence of 
$\sigma\subH(\lambda)$ on $\lambda$. 
As \dtaudz~approaches and surpasses 1, it significantly affects the radiative transfer of the
metagalactic ionizing background.

\begin{figure}[htbp]
\plotone{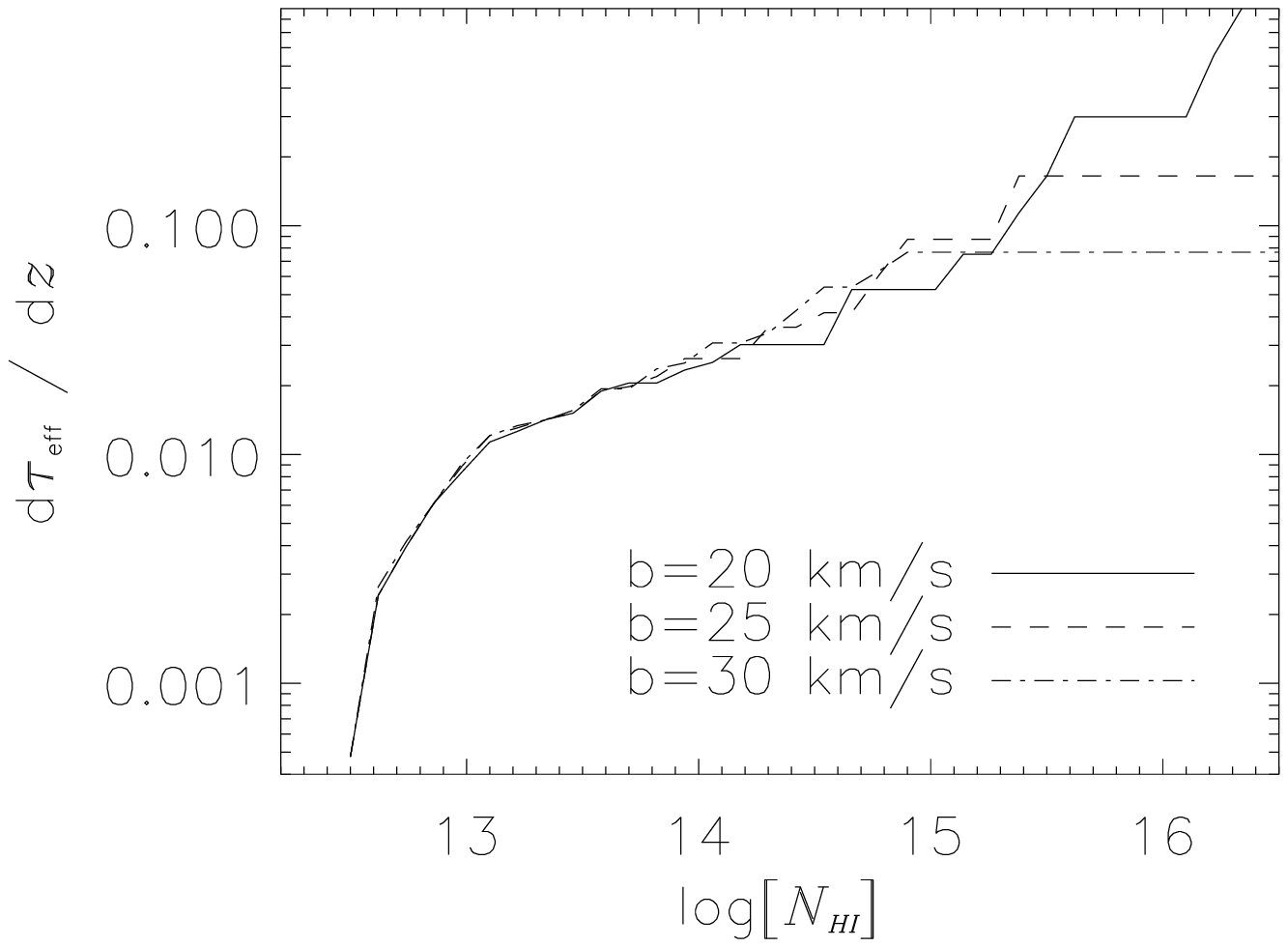}
\caption{\label{dtaueff_dz} Cumulative opacity, \dtaudz, at low redshift at  the Lyman edge
(912\ang) as a function of 
\Nhno,  calculated for three different \bvaluesno: 20, 25, and 30\kmsno. The crossover of the curves for 
\dtaudz~between 14.5\lt\logNh\lt15.5  arises from the \lya curve of growth and small number statistics (see text).}
\end{figure}

As shown in Figure~\ref{dtaueff_dz}, the cumulative opacity of \lowzya clouds is 
 \dtaudz\about0.01 for \logNh \ensuremath{\leq} 13, rising to \about0.1  for \logNh \ensuremath{\leq} 15.
  Figure~\ref{dtaueff_dz} gives \dtaudz~for three \bvaluesno, 20, 25 and 30\kms for all \lya absorbers. 
The crossover with respect to \bvalue  in the \dtaudz~curves  between 14.5 \ensuremath{\leq} \logNh \ensuremath{\leq} 15.5  
arises from the \lya curve of growth and  small number statistics. 
A lower assumed value of \bb~will systematically increase the inferred column density for lines  above  \lognh$\approx$14.
For our limited data set, this reduces the observed number of absorbers with \lognh =14.5--15.5
for \bb=20 and 25\kmsno.
Compared to \bb=30\kmsno, this causes a deficiency in \dtaudz\ for lower \bvalues in this limited
column density range. Once
 all absorbers are accounted for, the cumulative \dtaudz\ becomes larger for lower assumed \bvaluesno. 
At \logNh\gt15, \dtaudz~becomes uncertain due to both poor
number statistics in our sample and to the large dependence on \bvalueno. 
For a constant \bb=\myb\ for all \lowzya  absorbers, \dtaudz $\sim$ 0.2 for \logNh \ensuremath{\leq} 16. 
However, if \bb=20\kms is a more representative Doppler parameter,
\dtaudz\about0.4  for \lognh \ensuremath{\leq} 16. If, as inferred from higher redshift studies \citep{Hu95} and from ORFEUS
and FUSE \lyb data at \lowz \citep{Hurwitz98,Shull00}, some \lya clouds have
\bb\about15\kmsno, the cumulative \lya cloud opacity in the local Universe could 
approach unity for \lognh\about16. As Figure~\ref{dtaueff_dz} indicates,
\lya absorbers with 15\lt\lognh\lt17 probably dominate the continuum opacity of the \lowzya forest 
and could impact the level of the ionizing background
 \citep{Shull99b}. Characterizing the distribution of these absorbers accurately 
at \lowz will remain a challenge, even for HST+COS,
but would be very important in understanding the extragalactic ionizing background in the current epoch.
\clearpage
	\section{\lya Absorber Linear Two-Point Correlation Function}		\label{sec:TPCF}
The two-point correlation function (TPCF, $\xi$) can be estimated  from the pair 
counts of \lya absorption lines along each line of sight in our data according to :
\begin{equation}{\label{xi} \xi(\Dv) = {\Nobs(\Dv) \over \Nran(\Dv) } - 1~~.}\end{equation}
Here, \Nobs\ is the number of observed pairs and \Nran\ is the number of pairs that would be
expected randomly in the absence of clustering, in a given velocity difference bin, \Dv . 
We determine \Nran~ from Monte Carlo simulations based upon  
our determined number density, \dtwondnhdz, as well as the wavelength extent and
sensitivity limit of our observations. Like the pathlength normalization vector, 
we include only those portions of the spectra not obscured by Galactic lines,
non-\lya lines, and spectral regions blueward of  \prox of the target. 

At each position along the spectrum, the  probability of finding an absorber is calculated by:
\begin{equation}{ \Plam = \int_{\z} \int^{\infty}_{N_{min}(\lambda)} \dtwondnhdzover~\dNh \dz \approx \Dz(\lambda) \int^{\infty}_{N_{min}(\lambda)}
\betaNh~\dNh~,}\end{equation} 
where $N_{min}(\lambda)$ is based upon the sensitivity limit of the spectrum. 
The integral in \z\ can be replaced by the \z\ width of each pixel,
$\Dz(\lambda)$, since there appears to be no \z\ evolution between \zrange\ (i.e., $\gamma=0$).
The quantities $\beta$ and \NofNh were taken  from our expanded  sample over  \fullrange\  for \bb=\myb, and have
values of $1.71\pm0.06$ and $11.3\pm0.6$, respectively.
The probability, \Plam, is then compared to a uniformly distributed random number. 
If the probability exceeds the random number, an absorber is inserted into the Monte Carlo simulation at this position ($\lambda$). 
To correct for blending effects, once an absorber is inserted into the Monte Carlo simulation,
\Plam~is set to zero for the adjacent 12 pixels. 
This corresponds to 2.5 resolution elements or \about50\kmsno, since no pairs were observed at our resolution with
separations less than 50\kmsno. Undoubtedly, such closer pairs exist,  but at our resolution we are insensitive to them.
Because \Plam~depends exclusively on our observed distribution, any blended lines counted as a single absorption system will 
affect $\Nobs$ and $\Nran$ identically, leaving $\xi(\Dv)$ unchanged.
The mean \bvalue for our combined pre- and post-COSTAR \real\ sample is \about38\kmsno, corresponding to a Gaussian width
(\WG) of 27\kmsno. 
One would expect to begin resolving pairs separated by two Gaussian widths, which is in agreement with our
observed 50\kms cutoff. Our pre-COSTAR \bvalues have a higher median value of \about60\kmsno, or a Gaussian width \WG\about42\kmsno.
Therefore,  in our Monte Carlo simulations, we may slightly overestimate the number of random pairs $\le 70\kms$ in the
pre-COSTAR sample, or underestimate $\xi$ in our lowest velocity bin. 
For each sightline, we performed 1,000  simulations
($N_s$)  and combined them to form
\Nran. The error in \Nran(\Dv), denoted $\sigma_{ran}$, is taken to be
$\sqrt{\Nran(\Dv)}$.  For proper scaling of $\xi(\Dv)$, both \Nran(\Dv) and $\sigma_{ran}$ are normalized by $N_s$.

\begin{figure}[htbp]
\epsscale{0.8}
 \plotone{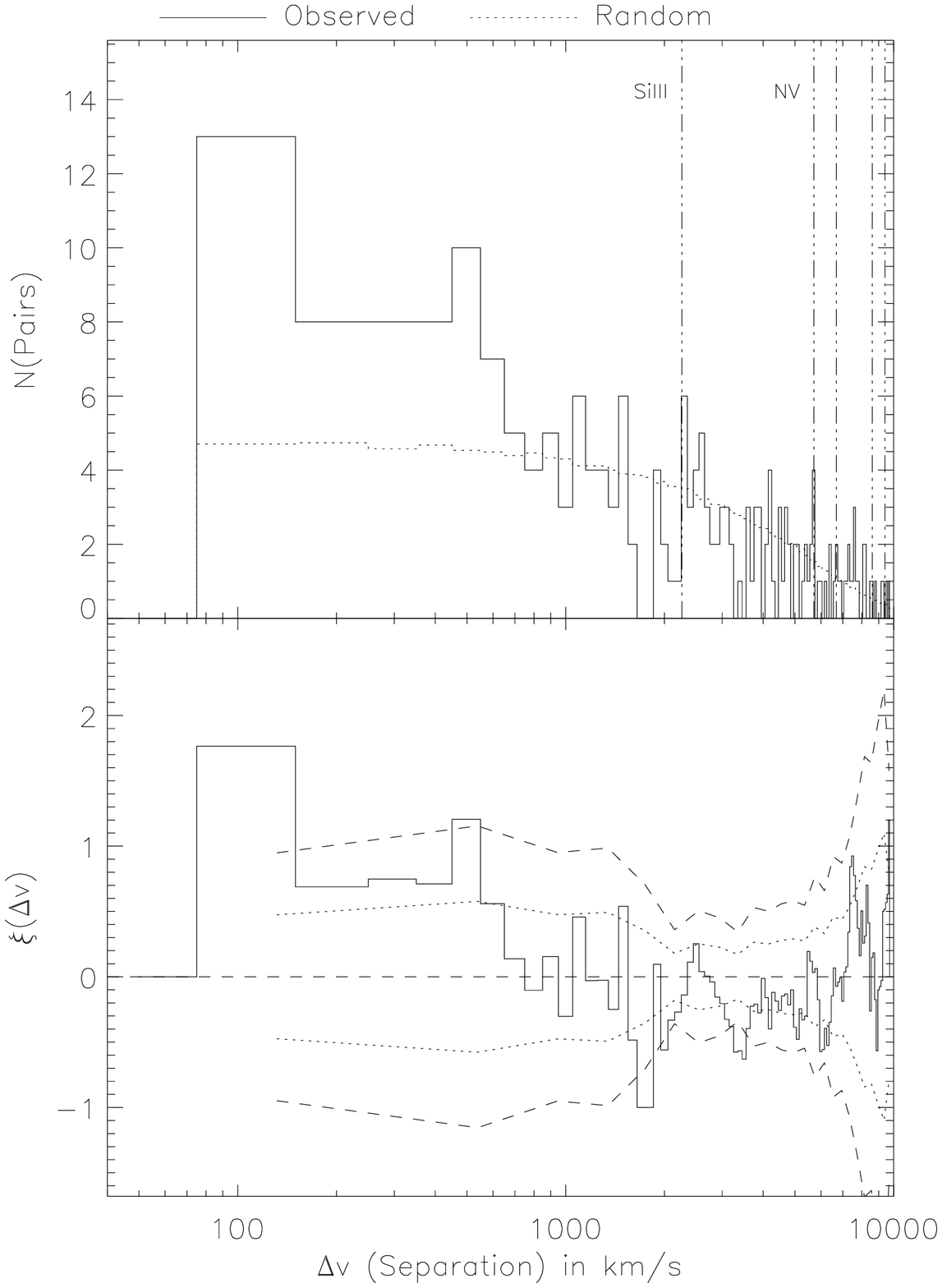}
 \caption{\label{twopoint_no} Two-point correlation function (TPCF, $\xi$) of the
\lya absorbers as a function of velocity pair separation, \Dv, in 100\kms bins. Upper panel (solid line)
gives the number of  observed \lya pairs,
\Nobs(\Dv), uncorrected for the varying wavelength coverage and spectral availability of our
observations. The vertical dot-dashed lines are the \Dv~separations between \lya
and nearby metal lines (\SiIII1206.5 and \NVdoublet\ are indicated, the unlabeled lines are S~II). The dotted line is the random
distribution, \Nran(\Dv), accounting for the varying available spectral regions. Bottom panel
gives the corrected TPCF, $\xi(\Dv)$. The 1 and 2\sig ranges of $\sigma_{\xi}$ are indicated by the short and long dashed lines, respectively, in the bottom panel. 
The $\lambda >$ 1280\Ang and $1226.76 < \lambda < 1227.16\Ang$ regions of the \PKS\ spectrum have been removed.}
\end{figure}

\begin{figure}[htbp]
\epsscale{0.85}
\plotone{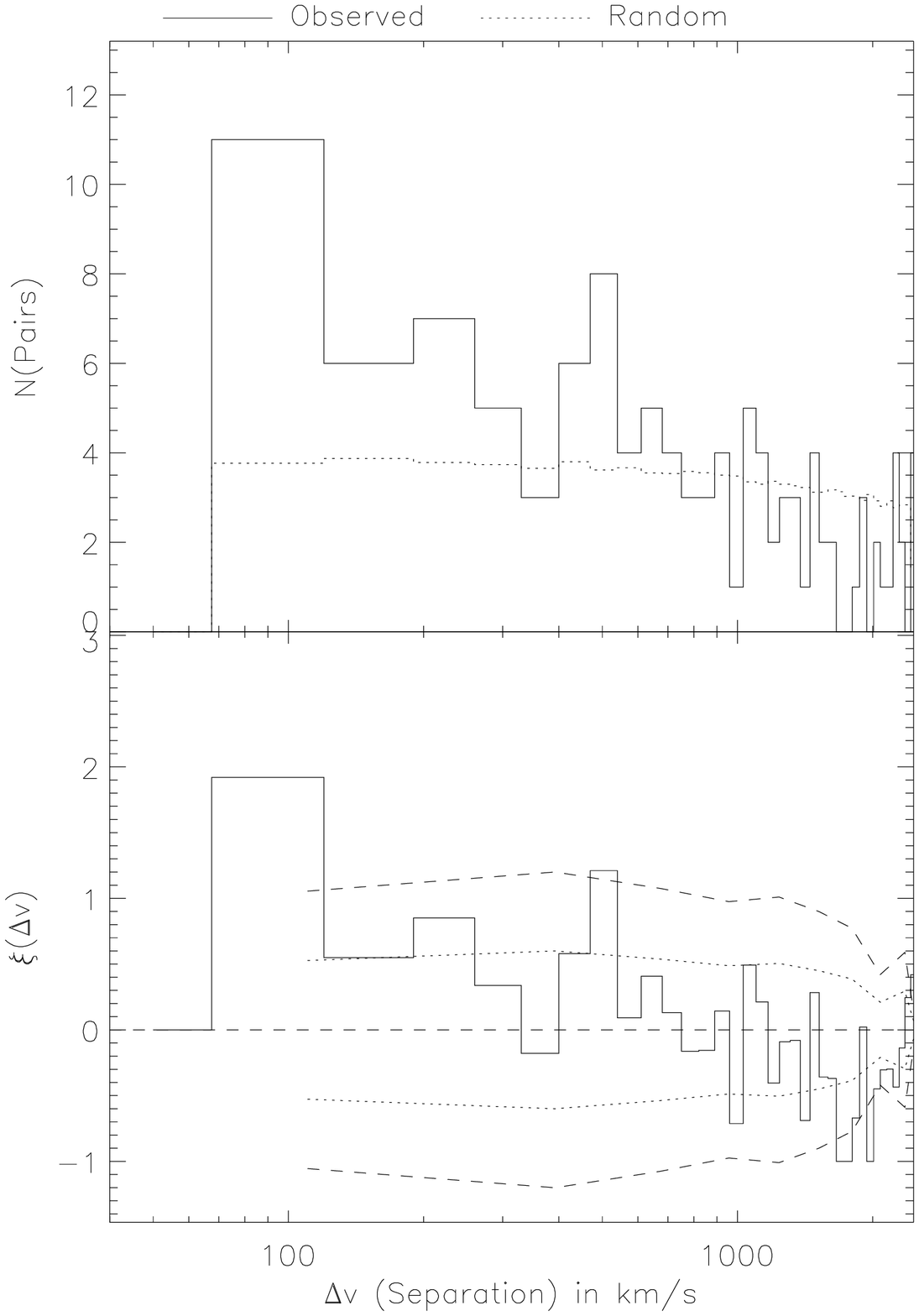}
\caption{\label{twopoint_no_70} Two-point correlation function as explained in Figure~\ref{twopoint_no}, with velocity 
bins of 70\kmsno; each bin is 1 \hsfi~Mpc.}
\end{figure}

Table~6 lists all absorber pairs with velocity separations of $50 \le \Dv \le 150$\kmsno, and
Figures~\ref{twopoint_no} and~\ref{twopoint_no_70} display the results of our TPCF analysis, $\xi(\Dv)$. 
Table~6 lists by column: (1) The central wavelength of the line pair; (2-3) the wavelength and 
rest-frame velocity separation of the pair; (4-5) the equivalent widths of the two absorbers; (6-7) the observed
\bvalues of the two absorbers; and (8) the target sightline. For this analysis we have excluded one \lya feature (and 
0.4\Ang of pathlength) at 1226.96\Ang in the \PKS~sightline because it could be \SiIII1206.5 
absorption related to the strong \lya line at 1236.43\ang. 
This line would have combined with another weak \lya line to produce a pair in Table~6.
We have visually inspected the other line pairs in Table~6 and find that all but two entries clearly are pairs of
distinct lines. The remaining two pairs (Fairall~9 and the $\lambda_{cen}$=1236.21\Ang pair in \PKS) could be broader
lines which have been subdivided by our profile fitting routines. Higher resolution spectra are required to be certain.
But, at most, the TPCF peak at $\Dv < 150\kms$ may be overestimated by 15\% (2/13). We have also removed the $\lambda > 1280\Ang$ 
portion of the \PKS~sightline owing to a strong cluster of lines \citep{pks}.  Based upon lower resolution data \citep{Bruh93}, 
the \PKS\ spectrum was obtained specifically to study this cluster of
absorbers around 1290\ang. Because of these special circumstances, we also exclude this portion of our sample from our TPCF analysis.

The distribution drawn in Figure~\ref{twopoint_no} with a solid line in the upper panel displays the number of 
observed \lya pairs with the indicated velocity separations (\Dv) uncorrected 
for the varying wavelength coverage  and spectral availability of our observations.
The velocity separations between any two absorbers along the same line of sight are calculated by:
\begin{equation}{ \Dv = {c \Dz \over 1+\meanz} \;,}\end{equation} where $\Dz =\z_2 - \z_1 = (\lambda_2 -\lambda_1$)/1215.67\Ang and 
\meanz = $ (\z_2 + \z_1)/2$.
One concern is that we might have misinterpreted weak metal lines as \lya absorbers, as mentioned above. The vertical dotted
lines in Figure~\ref{twopoint_no} indicate the rest-frame separations between \lya and expected metal lines 
(\Sithree, \NVdoublet, \SII1250, or \SII1253), 
indicating that this is not a concern.
%except perhaps for \Dv $\approx$ 2,300\kmsno.
The dotted line is the random distribution, \Nran(\Dv), which accounts for the varying sensitivity and wavelength coverage 
of our observations and  leads to $\xi(\Dv)$, which is displayed in the bottom panel. The velocity separation bins
in Figure~\ref{twopoint_no} are \cDz = 100\kmsno.

Figure~\ref{twopoint_no} shows an excess in $\xi$ at the lowest velocities, indicating clustering in the
local \lya forest, and broad deficits centered at 1850 and 3500\kmsno. Other features at larger separations are also
observed, but are less than 2\signo. Figure~\ref{twopoint_no_70} was constructed with bin size \Dv= 70\kms to show the effects of 
different binning of $\xi(\z)$. Calculating the significance of any peak or deficit in the TPCF is achieved 
by summing $\xi/\sigma_{\xi}$. We find a 3.6\sig excess in the velocity range 50--250\kms in the 100\kms bins, $\xi(50-150\kmsno)=1.9\pm0.5$,
and a 4.5\sig excess in the velocity range 50--260\kms with the 70\kms bins.
%\begin{figure}[htbp]
%\epsscale{0.8}
%\plotone{figure21.eps}
%\caption{\label{twopoint_no} Two-point  correlation function as explained in Figure~\ref{twopoint}. The upper portion
%($\lambda >$ 1280\ang) of the \PKS\ spectrum has been removed.  The velocity  bins are 100\kmsno.}
%\end{figure}

For comparison, Ulmer (1996) calculated the TPCF for \lya lines with $\Wno \gt 240$\Mang 
in the range $0 <\z < 1.3$ using data from the HST/FOS Key Project  \citep{Bahcall93,Bahcall96}. He concluded that the
$\Wno \gt 240$\Mang local \lya forest is clustered on the scale of $\Dv \lt 500$\kmsno, with amplitude 
$\xi$(250-500\kmsno) = $1.8^{+1.6}_{-1.2}$. 
The deficit in $\xi(\Dv)$ at \Dv\about1,700\kms is also present in the FOS data, but is unexplained by Ulmer (1996). 
Interestingly, when metal-line systems are included in Ulmer's analysis, this deficit mostly vanishes. 
At high redshifts, there have been numerous reports of marginal detections of excess power in the TPCF at $\Dv \le 150$\kms 
\citep{Chernomordik95,Hu95,Kulkarni96,Lu96},
 as well as reports of non-detections over the same scales \citep{Pettini90,Rauch92}.
 \citet{Impey99} recently constructed a TPCF for \z\about0 \lya lines that showed no excess power at $\Dv \ge 500$\kms using 
GHRS/G140L data.
Their TPCF was not sensitive to smaller velocity separations, owing to the lower resolution of the G140L.
 The 4\sig peaks in our \lowz TPCF figures at \Dv\about100\kms are at the same \Dv\ as
the 2-3\sig peaks at high redshift. Because this peak occurs at small separations, close to our resolution limit, 
it is the least likely feature to be caused by the limited wavelength coverage of our spectra. 
It is also unrelated to the velocity bin size, since  different bin sizes do not alter its significance appreciably. 
Since no other \lowz study has the velocity resolution required to see this peak
clearly, the line pairs in Table~6 would not be fully resolved by HST/GHRS G140L or FOS spectra like those used by
 \citet{Impey99} and Weymann \etl (1998). 
Therefore, we believe that the TPCF of the \lya forest at \lowz has the same general characteristics as at high-\z,
but with a slightly greater excess at low \Dv.

Due to the rather complex detection limit, and low pathlength, of our spectra, we are less confident of the 
reality of the possible excesses and deficits above \Dv=500\kmsno. 
These features do not appear to be present in either 
\highz \citep{Rauch92} or \lowz \citep{Impey99} studies. However, if these features are real, 
they may be related to the local widths of filaments (\Dv=500\kmsno) and galaxy voids
(\Dv=600 - 1,900\kmsno), since there is a possible excess of \lya clouds associated with filaments compared to voids 
\citep{Stocke95,Shull96}. 
%However, these features do not appear in galaxy-galaxy TPCFs along our sightlines (see Paper~III).

In HST cycle 7, our group has observed 13 sightlines with the STIS/G140M,
which has resolution comparable to GHRS/G160M.  These data
should approximately double the number of Ly$\alpha$ absorbers in our sample,
and thus will provide much improved information on the \lowz lya forest TPCF. 
We will report on the full-sample TPCF in a later paper \citep[][Paper~IV]{PaperIV}.
\input{table6.tex}
\clearpage
\section{Metallicity of the Low-$z$ Ly$\alpha$ Forest}
		\label{sec:metals}
One of the interesting measurements that may elucidate the \highz origin
of the low-redshift \lya absorbers is a search for heavy elements.
Weak \ion{C}{4} lines are detected at \highz
in 50--75\% of the \lya forest clouds with \nh \gt~$10^{14.5}$ cm$^{-2}$
\citep{Songaila96} and perhaps to even lower column densities \citep{Ellison00}. The \ion{C}{4}/\hone ratios suggest abundances of 
$10^{-2.5 \pm 0.5}$ times solar, and the \ion{Si}{4}/\ion{C}{4} ratios suggest that
these elements have been formed by early generations of massive stars 
and expelled by galactic winds or tidal stripping  \citep{Giroux97,Gnedin98}.
 Observing the metallicity of the low-$z$ absorbers would
help to understand the chemical evolution of the IGM, and would test
the prediction from numerical simulations \citep{Cen99b} that 
the IGM metallicity has increased to values of 10--20\% solar at the current epoch. 
 
Our first metallicity observations were performed on the cluster of strong  
\lya absorbers toward PKS~2155-304 \citep{pks}.  ORFEUS
observations along this sightline \citep{Appen95} suggested,
from Lyman continuum absorption, that $N_{\rm HI} = (2-5) \times 10^{16}$
cm$^{-2}$ at the redshift of the Ly$\alpha$ lines.  The saturated Ly$\alpha$
lines only provide a conservative range of $(3-10) \times 10^{14}$
cm$^{-2}$ from our HST/GHRS data.  Our recent measurements of higher Lyman series lines and the Lyman continuum edge with FUSE
show that $\nh \le 10^{16} \percmtwono$.
From the non-detection of \ion{Si}{3}~$\lambda1206.5$ and \ion{C}{4} $\lambda1548.2$
at their predicted positions, together with photoionization corrections,
we hope to improve the limits on (Si/H) and (C/H) using HST/STIS spectra obtained in cycle~8.

To explore the mean metallicity of the \lowzya forest, we constructed a composite spectrum for all \lya absorbers. 
Using a ``shift and stack''
method, we aligned all \definite~ \lya absorbers in the absorber rest frame and then co-added the spectra. 
Virtually all of these \lya absorbers lie well away from any predicted
metal-line position.  However, we have not co-added any portion of the
spectrum whose wavelength region (5\Ang around the metal-line position)
included a definite \lya\ absorber.  In one case, the \lya absorber
at 1226.964\Ang (2785 \kmsno) towards \PKS, has a slight chance
of being Si~III $\lambda1206.500$.  Although we believe this alternate 
identification is unlikely, based on photoionization and metallicity
considerations, we have not included it in our coaddition.   

Table~7 indicates the strong metal resonance lines in the wavelength 
region of our spectra and the upper limits on \Wno$_{metals}$ for each species. 
All reported \W values are \foursig detection limits in\mang, based upon the pixel-to-pixel variations in the
region of the expected metal line of the rest-frame composite absorber spectrum. Because the composite spectra include our pre-COSTAR
observations, the determined \bvalues of \lya are slight overestimates due to the spectral smearing of
these observations from target motion in the GHRS LSA (\S~\ref{sec:Ob}).
We divide our definite \lya sample into two sub-samples based upon the equivalent width (\Wlya) of the \lya absorber.
 Table~7  gives metallicity limits, in the low \W sample (\Wlya\lt\Wcutoff), in the  stronger absorbers (\Wlya \gt
\Wcutoff), and in the combined sample (column denoted ``all'').   This division is designed to
 separate saturated and unsaturated absorption lines based upon their \nW\ distribution (Figure~7). 
Searching the available rest-frame
spectral regions ($1138\Ang\lt \lambda \lt1298\ang$) for possible metal lines, 
we found no absorption  down to the 4\sig level in any \Wlya\ sub-sample.

The composite ultraviolet spectrum can be used to place an upper limit on the average metallicity
associated with our set of \lya absorbers.  
Unfortunately, the strong carbon lines, \CIII{977} and \CIV{1549}, lie outside our \lam-range. 
Most of the resonance lines of Table~7 have ionization thresholds similar to that of hydrogen,
 except for \Sithree\ which is 33.5 eV. The lines with $\sim$15 eV ionization thresholds
have ionization corrections (IC)  similar to that for hydrogen, which are quite high, resulting in  low predicted \Ws. 
 \ion{Si}{3} has a much lower IC, which equates to more \ion{Si}{3}  available for detection and a higher predicted
\Wno($\lambda1206.5$). Therefore, of the resonance lines in our waveband (see Table~7), our most promising limit on the
metallicity is that derived from \ion{Si}{3}, due to its higher ionization threshold. \input{table7.tex}

\begin{figure}[htbp]
\epsscale{0.75}
\plotone{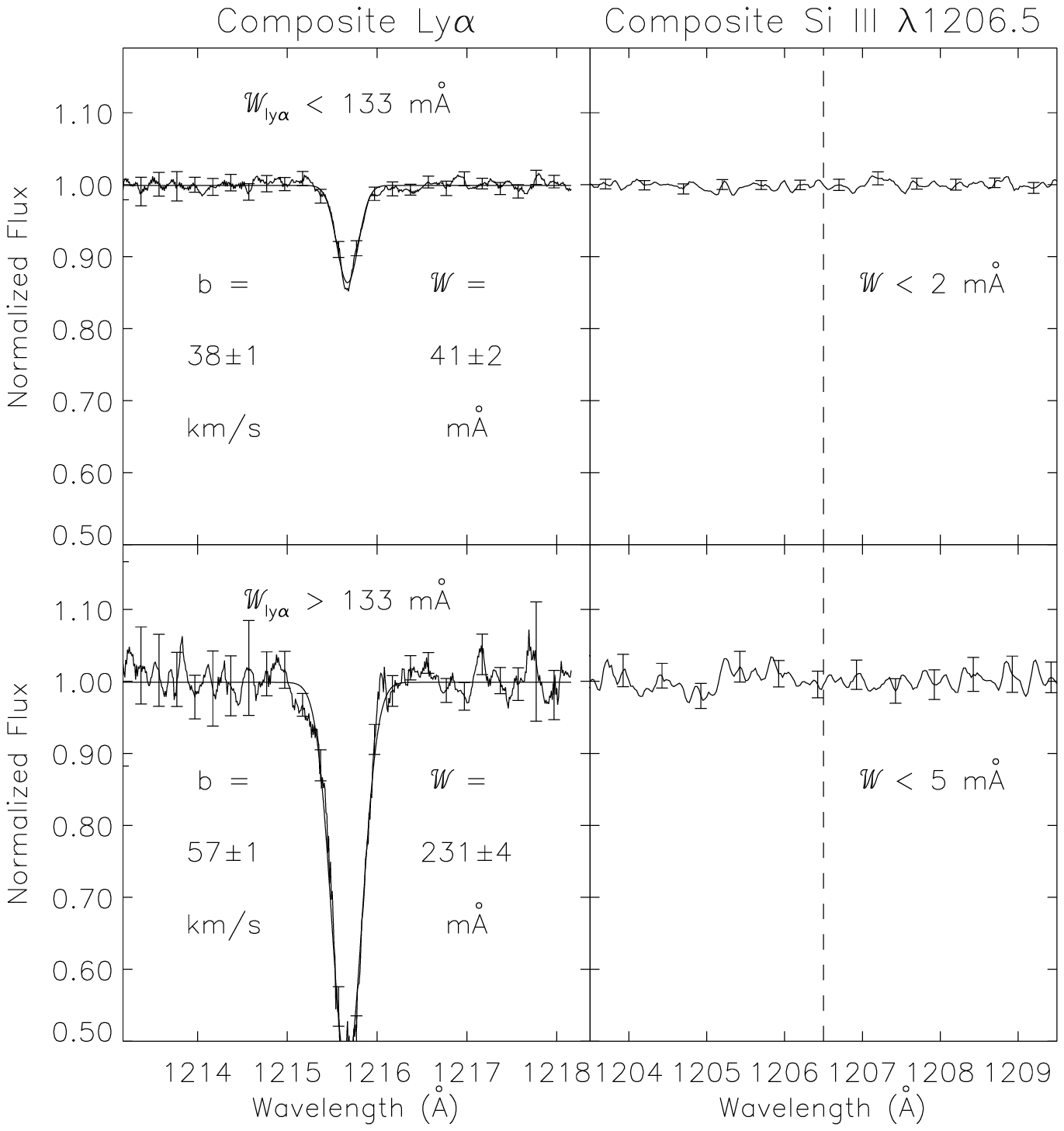}
\caption{\label{composite_fourby} Composite \lya and \Sithree\ spectra for all
absorbers with
\Wlya \lt \Wcutoff\ (upper panels) and \Wlya \gt \Wcutoff\ (bottom panels). }
\end{figure}

%\subsection{The Low-\W Composite Spectrum and Metallicity Limits.}\label{sec:lowW} 
The composite \lya spectrum for our low-\Wno, unsaturated sample (\Wlya\lt\Wcutoff) is shown in the upper
left panel of Figure~\ref{composite_fourby}. 
The measured \W is  41$\pm$2\Mang and the measured \bvalue of the composite
absorption is $\bmsd = 38\pm$1\kmsno. No \Sithree\ absorption is detected in the low-\W composite spectrum down
to the 2\Mang (4\signo) level. The composite \Sithree\ spectrum for the low-\W sample is shown in the upper right panel of
Figure~\ref{composite_fourby}.
 In our sample of unsaturated \lya forest lines, the ratio of the composite equivalent widths is \Wno(\Sithree)/\Wlya\ 
\lt~2\mang/41\Mang \about 0.05. 
For unsaturated lines, the composite absorption should be a good approximation of the average absorber
since \W scales linearly with \Nhno. 

 In the linear regime, the rest-frame equivalent width (\Wno) is related to the column density of absorbing atoms 
along the line of sight, $N_{j}$, by:
\begin{equation} { \W ({\rm cm}) = { {\pi e^2 \over m_e c^2} \lambda^2 N_{j} f_{jk}} = \Exp{8.85}{-13}  \lambda^2 N_{j} f_{jk} }
,\end{equation} where $N_{j}$ is in\percmtwono, $f_{jk}$ is the oscillator 
strength of the transition, and $\lambda$ is the wavelength
of the absorbed photon in cm \citep{Spitzer78}.
With current oscillator strengths \citep[0.4162 for \lyano; 1.68 for \Sithree, ][]{Morton91} and wavelengths, 
the relationships between \W and \nh for these transitions are:
\begin{equation} { \label{EWs}\Wno_{\lya} = 54.43 \left({ N_{\rm HI} \over 10^{13}\percmtwono }\right) \mang,~ {\rm~and~}~\Wno_{\rm SiIII} =
2.164  \left({ N_{\rm SiIII} \over 10^{11}\percmtwono }\right)
\mang \;,}\end{equation} which translates to a ratio of column densities of:
\begin{equation} { \label{Nobs}{N_{\rm SiIII} \over N_{\rm HI} } = 0.2514 ~\left({\Wno_{\rm SiIII} \over \Wno_{\rm HI} }\right)  \lt
0.012
\; ,}\end{equation} given the observed limits.
Determining the actual cloud metallicities ($Z$) requires knowledge of the fractional amounts of \ion{Si}{3} 
($f_{\rm CI}$ = \ion{Si}{3}/Si) and \hone ($f_{\rm HI}$= \ion{H}{1}/H) in the absorbing clouds \citep{Donahue91,pks}.
It is customary to express metallicities ($Z$) relative to solar ($\odot$) values in logarithmic form (denoted by [~]).
The metallicity of the absorbing clouds is \
\begin{equation}{
		{Z_{\rm Si} \over \zsolar} = {{\rm Si \over H} \over {\rm \left(Si \over H\right)_{\odot}} } = {{N_{\rm SiIII} \over N_{\rm HI} } ({f_{\rm HI}
\over f_{\rm SiIII}})
  \over { \rm \left(Si \over H\right)_{\odot}} } {\rm~,~or}}\end{equation}
\begin{equation}{ \label{pre_IC} {\rm \left[ { Si \over H} \right] } = \log\left({Z_{\rm Si} \over \zsolar}\right) =
\log\left({N_{\rm SiIII} \over N_{\rm HI} }\right) +\log\left({f_{\rm HI} \over f_{\rm SiIII}}\right) - \log\left({\rm Si \over H }\right)_{\odot}
{\rm~,~or}}\end{equation}
\begin{equation}{\label{IC}\left[ { \rm Si \over H} \right]  = \log\left({N_{\rm SiIII} \over N_{\rm HI} }\right) +IC~.}\end{equation} 
The ionization correction (IC) in this form accounts for the ratio of fractional abundances corrected for the solar
abundance ratio. For Si, the solar abundance ratio, $\log \left({\rm Si/H}\right)_{\odot} = -4.45$ \citep{Grevesse89}.
The ionization correction (IC) can be made and the metallicity of the clouds determined, only if $f_{\rm SiIII}$ and
$f_{\rm HI}$ are known. Applying the solar abundance of Si and our observational limits equation~(\ref{Nobs}) to equation~(\ref{IC}) provides:
\begin{equation}{ \label{SiH}
	{\rm \left[ { Si \over H} \right] } < \log\left(0.012\right) + IC = -1.91 + 4.45 + \log\left({f_{\rm HI} \over f_{\rm SiIII}}\right) = 2.54 + \log\left({f_{\rm HI} \over
f_{\rm SiIII}}\right) 
~.}\end{equation}
 The determination of the fractional abundances depends on uncertain photoionization modeling of the gas responsible for the absorption. We have
used the photoionization code CLOUDY \citep{Ferland96} to calculate the fractions of \hone and \Sithree\ expected for three
representative models of the gas. The measured \W and \bvalue of the low-\W composite spectrum indicate a composite $\nh = 8\times 10^{12}$\percmtwono.
All models assume that the gas is optically thin, illuminated by a metagalactic radiation
field with a power-law spectrum with spectral index
$\alpha_s = 1.8$, and mean ionizing intensity $\Jo = 10^{-23}$\cgs. We consider three models with total hydrogen densities 
\nofh = $10^{-4}, 10^{-5}$, and $10^{-6}$\percmthree. The resultant ratios of  $\log\left(f_{\rm HI}/f_{\rm SiIII}\right)$ are $-$2.4, $-$1.6, and +1.5,
respectively, implying  [Si/H] ~\lt 0.14, 0.94, and 4.02 solar, or \Zsi/\zsolar \lt~1.4, 8.7, and 10$^{4}$.  
The very high upper limit on metallicity for the
$\nofh= 10^{-6}$\percmthree\ model is due to the transference of the dominant state of Si to \ion{Si}{4}. The
characteristic density for these absorbers remains an unknown quantity,  
and it may be seen from this calculation that a strong constraint on metallicity
is not yet possible regardless of the cloud density.  In  Shull \etl (1998) we were able to place restrictions upon
$n_H$ along the \PKS\ sightline by assuming cloud depths that are constrained by VLA observations of nearby \hone galaxies.
 Typically, observations of two or three separate states of a single element
(e.g., \ion{C}{2}, \ion{C}{3}, and \ion{C}{4}) are required to measure the metallicity accurately. Therefore, the
limited wavelength coverage of our observations prevented a more stringent limit of \lya cloud metallicity.

%\subsection{The High-\W Composite Spectrum and Metallicity Limits.}\label{sec:highW} 
The composite spectrum for our high-\Wno, saturated, \lya sample (\Wlya \gt \Wcutoff) is shown as the lower left
panel of Figure~\ref{composite_fourby}. The measured \W is $231\pm$4\Mang and the measured \bvalue  of the composite absorption, adjusted
for the resolution of the GHRS/G60M, is $\bmsd = 51\pm$1\kmsno. As shown in the lower right panel of Figure~\ref{composite_fourby}, no
\Sithree\ absorption is detected in the \Wlya \gt \Wcutoff\  composite spectrum down to the 5\Mang (4\signo) level. 
Because some absorbers at $\Wno \gt 133\Mang$ may be saturated, the composite value of \Nh is uncertain.
%This level is higher than for the \Wlya \lt
%\Wcutoff\ composite spectrum due to the coaddition of fewer features. 
%In our sample of saturated \lya forest lines, the ratio of the composite equivalent
%widths, \Wno(\Sithree)/\Wlya, is less than~5\mang/231\mang~= 0.022. Above \Wcutoff,  the \lya absorptions 
%are on the logarithmic portion of the curve of growth,
%% where \W scales as $\sqrt{\ln\left(\Nhno\right)}$, 
%so determining limits on metallicity is problematic.  
%In general, the average \W of the saturated \lya absorbers is not a fair estimate of the average \hone column density.
%If the lines all have the same \bvalueno, those lines with \W greater than the composite value have much higher \Nh than those below.
%However, because the lines vary in line widths and  some high \W lines may be blended, it is hard to argue definitively
%that the \Nh is being underestimated. We account for this by evaluating the \hone column density represented by
%the composite \W by using a curve of growth, rather than the unsaturated limit.
%However, by creating the composite spectrum we can safely determine the upper limit of the average \Sithree\ since this
%absorption is unsaturated.  Adopting the observed composite high-\W as the
%fiducial \W (231\mang) and \bvalue (51\kmsno) that correspond to the non-detection of \Sithree,
%the inferred column density is \Nh = \Exp{4.9}{13}\percmtwono. 
If  we adopt  \bb=\myb, then the fiducial \W(231\mang) yields \Nh = \Exp{7.6}{13}\percmtwono, 
whereas we obtain \Nh = \Exp{3.9}{13}\percmtwo for an extension of the linear curve of growth (equation~\ref{EWs}).  
For \Sithree,  as determined from equation~\ref{EWs}, $N_{\rm SiIII} \lt $\Exp{2.3}{11}\percmtwono.
 Applying the same ionization correction as for the low-\W composite spectrum, we
estimate:
\begin{equation}{ \label{SiH_up}
	{ \left[ {\rm Si \over H} \right] }  < \log\left({\Exp{2.3}{11} \over \Exp{7.6}{13}} \right) + IC = -2.52 + 4.45 + \log\left({f_{\rm HI} \over
f_{\rm SiIII}}\right) = 1.93 + \log\left({f_{\rm HI} \over f_{\rm SiIII}}\right). }\end{equation}
For the three CLOUDY models, with assumed hydrogen densities of  \nofh = $10^{-4}, 10^{-5}$, and $10^{-6}$\percmthree,  we find
$\left[Si/H\right] \lt -0.47, 0.33,$ and 3.43 solar, or  $\Zsi \lt 0.33, 2.13$, and \Exp{3}{3} \Zsolar. Again, these
limits are not at all stringent because absorption lines from only a single ionization state of Si fall within our 
observed wavebands.
\section{The Contribution of the Local \lya Forest (\omegacl) to \omegab.}\label{sec:omega}
The contribution of the \lowzya forest to the total number of baryons, as estimated from Big Bang nucleosynthesis,
can be determined from:
\begin{equation}{ 
 		\omegacl = \int_{N_{min}}^{N_{max}} {\phi_0(\Nhno,\impact) \Mcl(\Nhno,\impact,\Jno) \over  \rho_{cr}}~\dNh {\rm\ \  ,}
 }\end{equation}
where $\phi_0(\Nhno,\impact)$ is the comoving space density of the clouds, $\Mcl(\Nhno,\impact,\Jno)$ is the individual cloud mass,
\impact\  is the impact parameter of the line of sight from the cloud center, \Nh is the 
observed  column density, $\rho_{cr} = 3\Hnaught^2 / 8\pi G$ is 
the present-day critical
density necessary to  halt the expansion of the Universe, 
\Jno\ is the specific intensity of the metagalactic ionizing radiation field at  $h\nu_0= 1 $Ryd, in\cgs, and ($N_{min},N_{max}$) is the range of
integration appropriate for the assumed cloud model.  Assuming that all
the \lowzya absorbers are singular isothermal spheres with density distributions of total hydrogen,  $n_H(r) = n_0 (r/r_0)^{-2}$, 
the cloud mass in the absence of metals, but including helium, within radius $r$ = \impact\  is given by:
\begin{equation}{ \label{eq_Mcl}
\Mcl (\Nhno,\impact,\Jno) = \int_0^\impact n_H(r) \mu_b (4\pi r^2) dr = 4\pi n_0 r_0^2 (1.22m_H) \impact\  {\rm \
\ ,} 
}\end{equation}
where $\mu_b=1.22 m_H$ is the average mass per baryon, assuming that the fraction of helium by mass is 
$Y=0.24$ or $\chi=n_{He}/n_H=0.0789$.

Assuming that the gas is  in photoionization equilibrium at $T = 20,000~K$, immersed in an ionizing radiation
field, \Jnuv = \Jo $(\nu/\nu_0)^{-\alpha_s}$, we can equate the observed column density to the expected
\hone column density by integrating through the cloud for a given  impact parameter, \impact, 
\begin{equation}{ 
\Nhno(\impact) =2  \int_{0}^{\infty}~f_{\rm HI}~n_H(r)~dl~~.
}\end{equation}
Here $l$ is the distance along the line of sight measured from the point of closest approach to the
cloud center, \impact, so that $l^2 = r^2 - \impact^2$. The ionization fraction of \hone is given by:
\begin{equation}{ 
\fhI = { n_{\rm HI} \over n_H}  ={ n_e \recom \over \photo } \ ,
}\end{equation} where \recom, the case-A hydrogen recombination rate coefficient,\footnote{Case-A recombination assumes that the energetic
photon emitted in ground-state recombination is not reabsorbed by the \lya cloud.}
 is \Exp{2.51}{-13} cm$^3$ \persecond  at $T = 20,000~$K, and the photoionization rate, \photo, 
in the presence of the described radiation field, \Jnuv, is
\begin{equation}{
\photo = 4 \pi \int_0^{\infty} \left({ \Jnuv \over h \nu }\right) \sigma_{\nu} d\nu = 
 {4 \pi \Jo \sigma_0 \over h \left(\alpha_s + 3\right)} 
= \left(\Exp{2.49}{-14}\persecondno\right)\ \Jm \left({ 4.8 \over \alpha_s + 3 }\right), 
}\end{equation}
where $\alpha_s \sim 1.8$ \citep{Zheng97}, \Jm = \Jo/ ($10^{-23}$\cgs), and 
$\sigma_\nu \approx \sigma_0 \left( \nu/ \nu_0 \right)^{-3}$ is the hydrogen cross section at frequency $\nu$. 
The scaling of \Jo is supported by a recent theoretical estimate of $\Jo = 1.3^{+0.8}_{-0.6} \times 10^{-23}$\cgs\  in the local Universe 
\citep{Shull99b}  and the upper limit of \citet{Donahue95} of \Jo$ <4.2\times 10^{-23}$\cgs\ for $\alpha_s = 1.8$.  
Integrating through the cloud,  using $n_{\rm HI} = n_e n_{\rm H+}
\recom /
\photo$ and $n_ e=\left(1+ 2\chi\right) n_H$, gives the column density,
\begin{equation}{ \label{N_int}
\Nhno\left(\impact\right) = \int_{-\infty}^{\infty} n_{\rm HI}(r)~dl = { 2 n_e \recom \over \photo }
\left(1+ 2\chi \right) \int_0^{\infty}~\left[n_H(r)\right]^2 dl
}\end{equation}
\begin{equation}{ \label{N_int_final}
%\Nhno\left(\impact\right) 
= { \pi n_0^2 r_0^4 \recom \left(1+ 2\chi \right)
\over 2\photo~\impact^3} ~~.}\end{equation}
Spectroscopic  studies of multiple QSO sightlines at \highz \citep{Bechtold94}, and nearest-neighbor studies of \lowzya absorbers 
\citep{Stocke95,Shull96,pks} indicate that \lya
absorbers may have spatial extents of \about100~kpc. However, the sightline pair experiments have found significantly
larger values for individual \lya absorbers \citep{Dinshaw97}.
 We adopt this as a conservative impact parameter  so that  \impact$_{100}$=\impact/(100 kpc),  
recognizing that some of the low-$N_{\rm HI}$ absorbers in voids may be considerable larger.  Solving equation (\ref{N_int_final}) for
$n_0 r_0^2$, and combining this result with equation (\ref{eq_Mcl}), gives the \lya cloud mass as :
\begin{equation}{
\Mcl (\Nhno,\impact,\Jno,\alpha_s)  = 
4\pi \left(1.22 m_H\right) \impact^{5 \over 2} \left[ {2 \photo N_{\rm HI} \over  \pi \recom \left(1+2\chi\right)}\right]^{1\over2} 
}\end{equation}
\begin{equation}{
= \left(\Exp{1.59}{9} M_{\odot}\right)~N^{1\over2}_{14}~\Jm^{1\over2}~\impact^{5\over2}_{100}
\left({ 4.8\over \alpha_s + 3 }\right)^{1\over2} \qquad {\rm ,}}\end{equation}
where  $N_{14}$=\Nhno/(10$^{14}$ \percmtwono). 
At each \Nhno, the space density of absorbers, \phio, is related to \dndz by:
\begin{equation} {
 \phio (\Nhno) =  {\Hnaught \over c \pi \impact^2} \dndzover~.
}\end{equation}
Therefore, assuming that \impact(\Nhno) is a constant\footnote{The \lowz clouds in galaxy voids could have systematically larger impact
parameters for a given column density, making the impact parameter versus column density relationship more complicated.}, we find 
\begin{equation}{  \label{ob}
\omegacl ={  \left(\Exp{1.59}{9} M_{\odot}\right) \Hnaught \Jm^{1\over2}~\impact^{5\over2}_{100}  \left({ 4.8 \over \alpha_s + 3}\right)^{1\over2}\over c \pi
\impact^2 \rho_{cr}} \int_{N_{min}}^{N_{max}} \dndzover N^{1\over2}_{14} \dNh~,
} \end{equation}
where, for an $\Omega_0$ = 1 Universe,  $\rho_{cr} = \left(\Exp{9.21}{-30}\right)~h_{70}^2$ g\percmthree. Performing the integration of
equation~(\ref{ob}) for the \lowzya forest over the high end of our column density range,  \upperrange\ gives:
\begin{equation}{
\omegacl= \left(0.008 \pm 0.001\right) \left[\Jm~\impact_{100} \left({ 4.8 \over \alpha_s + 3 }\right)\right]^{1\over2}~\hsfi~~.
} \end{equation}
This value represents \about20\% of the baryons (\omegab $h_{70}^{2} =0.0388$) as inferred from D/H by \citet{Burles98}.
Decreasing $N_{min}$ to 10$^{12.3}$\percmtwo would increase this estimate by roughly a factor of 2 to $\omegacl= 0.018 \pm 0.003$.
However, the absorbers with \lognh \lt 13.5 are probably uncollapsed matter, for which isothermal spheres of constant impact parameter are
increasingly poor physical approximations. 
In addition, the assumed power-law distribution in column density may not extend to such low values of \Nhno.

There are various corrections to our simple model that could be applied to our estimates. Numerical simulations  \citep{Dave99} suggest that a
flattened geometry is more appropriate for  the higher column density \lya systems \citep{Dinshaw97,Crotts98}. Thus, the absorbers may be denser than assumed,
with smaller ionization corrections. As pointed out in \citet{Rauch95}, \citet{Shull96}, and \citet{Madau96},
 by generalizing from spheres to disk geometries, the value for \omegacl\ is reduced by a factor of
$\langle a \cos{\theta} \rangle^{-{1\over2}} \approx \left(2/a\right)^{1\over2}$, where $a$ (\gt 1) is the disk aspect ratio and
$\theta$ is the disk viewing angle ($\langle \cos{\theta} \rangle=1/2$).  
For example, an aspect ratio of 10:1 would reduce \omegacl\ by a factor of 2.2.
Further complicating the conversion from $N_{\rm HI}$  to \omegacl\  is the possibility that a
substantial portion of the
\lowzya clouds are hot ($10^5 -10^7~K$) and collisionally ionized \citep{Cen99a}. 
Thus, our ionization correction from \hone to total H may be too low for some clouds,
 and others may be undetectable given our sensitivity limits.
Although the results of the \omegacl\  contribution to \omegab\  are model dependent and highly uncertain, 
it is likely that warm, photoionized \lya clouds contain a substantial fraction of the 
baryons, \omegacl = 0.008-0.018, or 20-45\% of \omegab. 
This agrees  with numerical simulations that suggest that a substantial fraction of baryons in the local Universe lie
in these  \lya absorbers.
\clearpage
\section{Conclusions}\label{sec:conc}
\noindent The major conclusions of our physical analysis for the \lowzya forest are:
\begin{itemize}
\item Our observed \bvalue distribution is consistent with that at higher redshift. 
We see little evidence for an increasing median \bvalue with decreasing \z\ 
as reported by \citet{Kim97}. 
We are cautious about inferring that our measured \bvalues  (from line widths) are accurate 
indicators of the true Doppler parameters of the \lowzya forest. 
The unknown contribution of intracloud turbulence or the presence of multiple sub-components 
per absorber leads us to adopt a constant value \bb=\myb\ when
converting equivalent widths (\Wno) into column densities (\Nhno).
\item Our observed \W distribution appears similar to that in higher redshift Keck/HIRES data. 
Applying a careful sensitivity correction, we detect a
significant break in \dndW\ at $\W\leq133$\mang, which probably arises from saturation in the \lya curve of growth. 
Our results for \dndz are consistent with HST/FOS Key Project low-\z\ studies, 
but they extend  to much lower equivalent widths, 
\Wno$\approx$14\Mang for 4$\sigma$ and \Wno$\approx$10\Mang for 3$\sigma$ detections.
\item Using an integrated method (Table~5) for evaluating the functional form of \dtwondzdnhno, we find that, 
for $\lognh \le 14$, \dtwondzdnh $\propto \Nhno^{-\beta}$ with  $\beta = 1.81\pm0.05$. 
There is some evidence for a break at $\lognh = 14$, above which 
$\beta = 1.43\pm0.35$. The reality of this break is suspect due to small number statistics and to the dependence of column density on our
selection of  \bb\ = \myb.
\item We find no \z\ evolution over our small redshift range (\zrange), 
and we observe no unusual deviations in \dndz\ that could arise from large-scale
structure in the local Universe. 
We calculate that the \lowzya forest produces a Lyman-continuum opacity at 1 Ryd 
of \dtaudz$\approx$0.01 for $\lognh \le 13$ and \about0.1 for $\lognh \le 15$.
Most of the intergalactic opacity probably arises from rare, higher-\Nh absorbers.
\item We have compared the \z\ evolution of \dndz (Figure~\ref{comp_dndz}) and \dtwondnhdzno\ (Figure~\ref{dndnh_z}) at low and  high \Nh and \z. 
In both cases, we find significant evolution that can be explained by:
 (1) expansion of the Universe, (2) the rapid decline in \Jnuz\ at \z\lt2, and (3) the merger and dissipation of \lya absorbers. 
Improved statistics over intermediate redshifts ($0.1 \lt \z \lt 1$) and a wide range in \Nh are 
required to extract more meaningful evolutionary results.
However,  these preliminary results agree with hydrodynamical simulations of the \lowzya 
forest by \citet{Dave99}.
\item By comparing the \z\ evolution of \dndz for the low column density (\kimrange) absorbers studied in this paper with 
similar results for higher column density absorbers ($\lognh > 14$) found by the Key Project team, we find no difference in 
the overall \z\ evolution of the low and high \Nh absorbers. This is in disagreement with previous results at both
high and low-\z\ (\citet{Kim97} and \citet{Weymann98}, respectively) who found slower evolution for the low column density absorbers.
However, lacking information at intermediate redshifts, all these results can be in agreement if the dramatic ``break'' in \dndz
occurs later (i.e., at lower redshift) for the lower column density absorbers. 
\item We find a 4\sig  signal in the two-point correlation function (TPCF) of \lowz \lya absorbers 
for velocity separations $\Dv \le 250\kmsno$,
 consistent with results from higher redshift studies. 
Our results at higher velocity separations are limited by relatively poor statistics. In Paper~IV, we will
combine these GHRS observations with our results from 13 STIS sightlines to improve our understanding of the \lowzya TPCF.
%\item We find no evidence for any metals in our composite \lya absorber spectrum, 
%although we are only able to perform a detailed analysis for
%\Sithree. We  divide our \lya lines into samples  with \W above and below 133\mang. 
%For $\Wno \lt 133$\mang,  we find $\Zsi \lt 1.4, 8.7$, and $10^{4}$  \Zsolar\ for  three models assuming absorber density 
%$\nofh = 10^{-6}, 10^{-5},$ and $10^{-4}$ \percmthree,
%respectively.  
%For $\Wno \ge 133$\mang, we find that $\Zsi \lt 0.33, 2.13$, and \Exp{3}{3} \Zsolar\ for the same densities. 
%Along the \PKS\ sightline, there exists a unique cluster of lines for which we are able to place much tighter metallicity limits. 
%As presented in Shull \etl (1998), we obtain metallicity limits of  \Zsi \lt 0.003\Zsolar\ and  Z$_{C}$ \lt 0.005\Zsolar\ 
%for clouds with assumed $\nh = 2 \times 10^{-16}$ \percmtwono.
\item Applying a photoionization correction, we find that the
low-$z$ Ly$\alpha$ forest may contain $\sim20$\% of the total
number of baryons, with closure parameter
$\Omega_{Ly\alpha} = (0.008 \pm 0.001) h_{70}^{-1}$,
for a standard absorber size of 100 kpc and an ionizing
radiation field of intensity
$J_0 = 10^{-23}$\cgs and spectral index $\alpha_s = 1.8$.
\end{itemize}
\acknowledgements
For their assistance in obtaining the HST/GHRS data over several
cycles, we are grateful to the staff at the Space Telescope Science
Institute, particularly Ray Lucas. 
We thank Mark Giroux for helpful discussions and a critical reading
of the manuscript.  Buell Jannuzi and Ray Weymann are thanked for a critical reading of the manuscript and helpful
conversations. This work was supported by HST guest observer 
grant GO-06593.01-95A, the HST COS project (NAS5-98043), and by the Astrophysical Theory Program
(NASA grant NAGW-766 and NSF grant AST96-17073).   
\clearpage
\bibliographystyle{apj}

\end{document}

%% file: table1.tex
\setlength{\tabcolsep}{2mm}
\renewcommand{\arraystretch}{0.8}
\begin{center}
\begin{table}[htbp]
\caption{H~I Column Densities (\Nhno) of Definite \lya  features \label{linelist_nh}}
\tiny
\begin{tabular}{lccccclllc}
\tableline \tableline
Target&Wavelength&Velocity&\bmsd&\bobs\tablenotemark{a}\ &\W&\multicolumn{4}{c}{log[\Nh~(in $cm^{-2}$)]}   \\
        &\AA                  &km~s$^{-1}$ &km~s$^{-1}$&km~s$^{-1}$&m\AA&\bb=20 &\bb=25&\bb=30 & \bb=col 5\tablenotemark{a}\\
\tableline
$*$3C273&1219.786 $\pm$  0.024&  1015 $\pm$     6&  71 $\pm$   5&  69 $\pm$   5& 369 $\pm$  36&15.44&14.76&14.42&13.98\\
$*$3C273&1222.100 $\pm$  0.023&  1586 $\pm$     6&  73 $\pm$   4&  72 $\pm$   4& 373 $\pm$  30&15.48&14.78&14.44&13.98\\
$*$3C273&1224.954 $\pm$  0.029&  2290 $\pm$     7&  56 $\pm$  32&  54 $\pm$  33&  35 $\pm$  30&12.85&12.85&12.84&12.83\\
$*$3C273&1247.593 $\pm$  0.046&  7872 $\pm$    11&  38 $\pm$  15&  34 $\pm$  17&  33 $\pm$  18&12.82&12.82&12.81&12.81\\
$*$3C273&1251.485 $\pm$  0.032&  8832 $\pm$     8&  63 $\pm$   9&  61 $\pm$  10& 114 $\pm$  25&13.48&13.44&13.41&13.36\\
$*$3C273&1255.542 $\pm$  0.069&  9833 $\pm$    17&  66 $\pm$  23&  64 $\pm$  24&  46 $\pm$  22&12.98&12.97&12.96&12.94\\
$*$3C273&1275.243 $\pm$  0.031& 14691 $\pm$     7&  63 $\pm$   8&  61 $\pm$   8& 140 $\pm$  25&13.62&13.56&13.53&13.47\\
$*$3C273&1276.442 $\pm$  0.059& 14987 $\pm$    14&  54 $\pm$  19&  52 $\pm$  20&  46 $\pm$  22&12.98&12.97&12.96&12.95\\
$*$3C273&1277.474 $\pm$  0.136& 15241 $\pm$    33&  89 $\pm$  51&  88 $\pm$  52&  52 $\pm$  40&13.05&13.03&13.02&13.00\\
$*$3C273&1280.267 $\pm$  0.077& 15930 $\pm$    19&  73 $\pm$  27&  71 $\pm$  28&  64 $\pm$  33&13.15&13.13&13.12&13.09\\
$*$3C273&1289.767 $\pm$  0.098& 18273 $\pm$    24&  84 $\pm$  35&  82 $\pm$  36&  47 $\pm$  28&12.99&12.98&12.97&12.95\\
$*$3C273&1292.851 $\pm$  0.051& 19033 $\pm$    12&  48 $\pm$  16&  45 $\pm$  17&  47 $\pm$  22&12.99&12.98&12.97&12.96\\
$*$3C273&1296.591 $\pm$  0.025& 19956 $\pm$     6&  64 $\pm$   4&  62 $\pm$   4& 297 $\pm$  25&14.67&14.28&14.10&13.86\\
AKN120&1232.052 $\pm$  0.034&  4040 $\pm$     8&  36 $\pm$  10&  32 $\pm$  11&  48 $\pm$  18&13.00&12.99&12.98&12.98\\
AKN120&1242.972 $\pm$  0.028&  6733 $\pm$     7&  36 $\pm$   7&  33 $\pm$   8&  53 $\pm$  13&13.05&13.04&13.03&13.02\\
AKN120&1247.570 $\pm$  0.087&  7867 $\pm$    21&  37 $\pm$  32&  34 $\pm$  35&  20 $\pm$  25&12.60&12.59&12.59&12.59\\
AKN120&1247.948 $\pm$  0.023&  7960 $\pm$     5&  32 $\pm$   3&  27 $\pm$   4& 147 $\pm$  22&13.65&13.59&13.56&13.58\\
AKN120&1248.192 $\pm$  0.027&  8020 $\pm$     7&  28 $\pm$   5&  23 $\pm$   6&  65 $\pm$  17&13.16&13.14&13.13&13.14\\
FAIRALL9&1240.988 $\pm$  0.038&  6244 $\pm$     9&  38 $\pm$  12&  35 $\pm$  13&  22 $\pm$   9&12.63&12.63&12.62&12.62\\
FAIRALL9&1244.462 $\pm$  0.034&  7100 $\pm$     8&  42 $\pm$  10&  39 $\pm$  11&  32 $\pm$  10&12.81&12.80&12.80&12.79\\
FAIRALL9&1254.139 $\pm$  0.024&  9487 $\pm$     6&  46 $\pm$   6&  43 $\pm$   6&  84 $\pm$  13&13.29&13.27&13.25&13.23\\
FAIRALL9&1262.864 $\pm$  0.029& 11638 $\pm$     7&  32 $\pm$  12&  28 $\pm$  13&  16 $\pm$   8&12.49&12.48&12.48&12.48\\
FAIRALL9&1263.998 $\pm$  0.041& 11918 $\pm$    10&  45 $\pm$  14&  42 $\pm$  15&  22 $\pm$   9&12.63&12.62&12.62&12.61\\
FAIRALL9&1264.684 $\pm$  0.073& 12087 $\pm$    18&  47 $\pm$  31&  44 $\pm$  33&  30 $\pm$  28&12.78&12.77&12.77&12.76\\
FAIRALL9&1265.104 $\pm$  0.026& 12191 $\pm$     6&  28 $\pm$   6&  23 $\pm$   7&  28 $\pm$   7&12.74&12.73&12.73&12.73\\
FAIRALL9&1265.970 $\pm$  0.117& 12404 $\pm$    29&  36 $\pm$  21&  32 $\pm$  23&  19 $\pm$  23&12.57&12.57&12.56&12.56\\
H1821+643&1245.440 $\pm$  0.023&  7342 $\pm$     5&  49 $\pm$   3&  47 $\pm$   3& 298 $\pm$  20&14.68&14.28&14.11&13.92\\
H1821+643&1246.301 $\pm$  0.036&  7554 $\pm$     9&  44 $\pm$  13&  41 $\pm$  14&  50 $\pm$  24&13.03&13.01&13.00&12.99\\
H1821+643&1247.583 $\pm$  0.029&  7870 $\pm$     7&  25 $\pm$   7&  19 $\pm$   9&  40 $\pm$  17&12.91&12.90&12.89&12.91\\
H1821+643&1247.937 $\pm$  0.033&  7957 $\pm$     8&  37 $\pm$   9&  33 $\pm$  10&  68 $\pm$  38&13.18&13.16&13.15&13.14\\
H1821+643&1265.683 $\pm$  0.025& 12334 $\pm$     6&  32 $\pm$   5&  28 $\pm$   6&  64 $\pm$  15&13.15&13.13&13.12&13.13\\
MARK279&1236.942 $\pm$  0.030&  5246 $\pm$     7&  31 $\pm$   8&  27 $\pm$   9&  30 $\pm$  10&12.78&12.77&12.77&12.77\\
MARK279&1241.509 $\pm$  0.029&  6372 $\pm$     7&  24 $\pm$   3&  18 $\pm$   4&  58 $\pm$   7&13.09&13.08&13.07&13.10\\
MARK279&1241.805 $\pm$  0.023&  6445 $\pm$     6&  26 $\pm$   3&  21 $\pm$   4&  40 $\pm$   7&12.91&12.90&12.89&12.91\\
MARK279&1243.753 $\pm$  0.023&  6925 $\pm$     5&  31 $\pm$   3&  26 $\pm$   3&  65 $\pm$   8&13.16&13.14&13.13&13.14\\
MARK279&1247.216 $\pm$  0.024&  7779 $\pm$     6&  32 $\pm$   4&  28 $\pm$   5&  48 $\pm$   9&13.01&13.00&12.99&12.99\\
MARK279&1247.533 $\pm$  0.042&  7858 $\pm$    10&  38 $\pm$  13&  34 $\pm$  14&  21 $\pm$  10&12.62&12.61&12.61&12.61\\
MARK290&1234.597 $\pm$  0.027&  4667 $\pm$     7&  31 $\pm$   7&  26 $\pm$   8&  60 $\pm$  18&13.12&13.10&13.09&13.10\\
MARK290&1244.408 $\pm$  0.032&  7087 $\pm$     8&  28 $\pm$   9&  23 $\pm$  11&  23 $\pm$  10&12.65&12.64&12.64&12.64\\
MARK290&1245.536 $\pm$  0.025&  7365 $\pm$     6&  19 $\pm$   5&  11 $\pm$   9&  21 $\pm$   7&12.61&12.60&12.60&12.61\\
$*$MARK335&1223.637 $\pm$  0.026&  1965 $\pm$     6&  77 $\pm$   7&  75 $\pm$   7& 229 $\pm$  30&14.12&13.95&13.86&13.71\\
$*$MARK335&1224.974 $\pm$  0.049&  2295 $\pm$    12&  75 $\pm$  17&  73 $\pm$  17&  81 $\pm$  26&13.28&13.25&13.24&13.20\\
$*$MARK335&1232.979 $\pm$  0.057&  4268 $\pm$    14&  53 $\pm$  19&  51 $\pm$  20&  33 $\pm$  16&12.82&12.81&12.81&12.80\\
$*$MARK335&1241.093 $\pm$  0.026&  6269 $\pm$     6&  77 $\pm$   6&  75 $\pm$   6& 130 $\pm$  14&13.56&13.52&13.49&13.42\\
MARK421&1227.977 $\pm$  0.025&  3035 $\pm$     6&  39 $\pm$   5&  35 $\pm$   5&  86 $\pm$  15&13.31&13.28&13.27&13.26\\
$*$MARK501&1234.572 $\pm$  0.039&  4661 $\pm$    10&  62 $\pm$  12&  60 $\pm$  13& 161 $\pm$  43&13.73&13.66&13.62&13.54\\
$*$MARK501&1239.968 $\pm$  0.029&  5992 $\pm$     7&  62 $\pm$  38&  59 $\pm$  39&  55 $\pm$  46&13.07&13.06&13.05&13.03\\
$*$MARK501&1246.177 $\pm$  0.069&  7523 $\pm$    17&  50 $\pm$  25&  48 $\pm$  26&  53 $\pm$  36&13.05&13.04&13.03&13.02\\
$*$MARK501&1251.152 $\pm$  0.029&  8750 $\pm$     7&  78 $\pm$  48&  77 $\pm$  49&  66 $\pm$  57&13.16&13.15&13.13&13.10\\
MARK509&1226.050 $\pm$  0.025&  2560 $\pm$     6&  43 $\pm$   5&  40 $\pm$   5& 209 $\pm$  32&13.99&13.86&13.79&13.72\\
MARK817&1223.507 $\pm$  0.037&  1933 $\pm$     9&  38 $\pm$  12&  34 $\pm$  13&  29 $\pm$  13&12.75&12.75&12.74&12.74\\
MARK817&1224.172 $\pm$  0.023&  2097 $\pm$     5&  44 $\pm$   4&  40 $\pm$   4& 135 $\pm$  15&13.60&13.54&13.51&13.48\\
MARK817&1234.657 $\pm$  0.041&  4682 $\pm$    10&  43 $\pm$  14&  40 $\pm$  15&  23 $\pm$  11&12.65&12.64&12.64&12.63\\
MARK817&1236.303 $\pm$  0.023&  5088 $\pm$     6&  85 $\pm$   4&  84 $\pm$   4& 207 $\pm$  14&13.98&13.85&13.78&13.65\\
MARK817&1236.902 $\pm$  0.027&  5236 $\pm$     7&  29 $\pm$   6&  24 $\pm$   7&  25 $\pm$   7&12.69&12.68&12.68&12.68\\
MARK817&1239.159 $\pm$  0.029&  5793 $\pm$     7&  42 $\pm$  11&  39 $\pm$  12&  34 $\pm$  13&12.84&12.83&12.82&12.82\\
MARK817&1241.034 $\pm$  0.024&  6255 $\pm$     6&  33 $\pm$   5&  29 $\pm$   5&  37 $\pm$   8&12.88&12.87&12.86&12.86\\
MARK817&1245.395 $\pm$  0.051&  7330 $\pm$    13&  53 $\pm$  17&  51 $\pm$  18&  17 $\pm$   7&12.50&12.50&12.50&12.49\\
MARK817&1247.294 $\pm$  0.044&  7799 $\pm$    11&  59 $\pm$  15&  56 $\pm$  16&  28 $\pm$   9&12.74&12.74&12.73&12.72\\
$*$PKS2155-304&1226.345 $\pm$  0.060&  2632 $\pm$    15&  63 $\pm$  31&  61 $\pm$  33&  42 $\pm$  40&12.94&12.93&12.92&12.91\\
$*$PKS2155-304&1226.964 $\pm$  0.065&  2785 $\pm$    16&  66 $\pm$  25&  64 $\pm$  26&  36 $\pm$  22&12.86&12.85&12.84&12.83\\
$*$PKS2155-304&1232.016 $\pm$  0.049&  4031 $\pm$    12&  42 $\pm$  16&  39 $\pm$  17&  21 $\pm$  11&12.62&12.61&12.61&12.61\\
$*$PKS2155-304&1235.748 $\pm$  0.029&  4951 $\pm$     7&  70 $\pm$  14&  68 $\pm$  15&  64 $\pm$  23&13.15&13.14&13.12&13.10\\
$*$PKS2155-304&1235.998 $\pm$  0.029&  5013 $\pm$     7&  61 $\pm$  10&  58 $\pm$  11&  82 $\pm$  22&13.28&13.26&13.24&13.21\\
$*$PKS2155-304&1236.426 $\pm$  0.029&  5119 $\pm$     7&  82 $\pm$   5&  80 $\pm$   5& 218 $\pm$  20&14.05&13.90&13.82&13.68\\
$*$PKS2155-304&1238.451 $\pm$  0.029&  5618 $\pm$     7&  37 $\pm$  12&  33 $\pm$  14&  29 $\pm$  15&12.76&12.75&12.75&12.75\\
$*$PKS2155-304&1238.673 $\pm$  0.031&  5673 $\pm$     8&  34 $\pm$  10&  30 $\pm$  12&  39 $\pm$  16&12.91&12.90&12.89&12.89\\
PKS2155-304&1270.784 $\pm$  0.027& 13591 $\pm$     6&  43 $\pm$   5&  39 $\pm$   6& 101 $\pm$  18&13.41&13.37&13.35&13.33\\
PKS2155-304&1281.375 $\pm$  0.024& 16203 $\pm$     5&  61 $\pm$   3&  58 $\pm$   3& 346 $\pm$  23&15.17&14.59&14.31&13.97\\
PKS2155-304&1281.867 $\pm$  0.061& 16325 $\pm$    15&  52 $\pm$  21&  49 $\pm$  22&  62 $\pm$  34&13.13&13.12&13.11&13.09\\
PKS2155-304&1284.301 $\pm$  0.030& 16925 $\pm$     7&  25 $\pm$  10&  19 $\pm$  13&  43 $\pm$  37&12.95&12.94&12.93&12.95\\
PKS2155-304&1284.497 $\pm$  0.039& 16973 $\pm$     9&  65 $\pm$   6&  63 $\pm$   6& 389 $\pm$  68&15.67&14.91&14.52&14.03\\
PKS2155-304&1285.086 $\pm$  0.038& 17119 $\pm$     9&  89 $\pm$  11&  87 $\pm$  11& 448 $\pm$  79&16.36&15.41&14.86&14.10\\
PKS2155-304&1287.497 $\pm$  0.024& 17713 $\pm$     6&  38 $\pm$   4&  35 $\pm$   5& 139 $\pm$  21&13.62&13.56&13.53&13.51\\
PKS2155-304&1288.958 $\pm$  0.029& 18073 $\pm$     7&  50 $\pm$   7&  47 $\pm$   8&  99 $\pm$  20&13.39&13.36&13.34&13.31\\
Q1230+0115&1221.711 $\pm$  0.026&  1490 $\pm$     6&  26 $\pm$   6&  21 $\pm$   8& 138 $\pm$  42&13.61&13.56&13.52&13.60\\
Q1230+0115&1222.425 $\pm$  0.035&  1666 $\pm$     9&  56 $\pm$   9&  54 $\pm$  10& 385 $\pm$  94&15.62&14.88&14.50&14.07\\
Q1230+0115&1222.747 $\pm$  0.029&  1745 $\pm$     7&  43 $\pm$  11&  40 $\pm$  12& 241 $\pm$  99&14.21&14.00&13.90&13.82\\
Q1230+0115&1223.211 $\pm$  0.051&  1860 $\pm$    13&  50 $\pm$  20&  48 $\pm$  21& 142 $\pm$  81&13.63&13.57&13.54&13.49\\
Q1230+0115&1225.000 $\pm$  0.024&  2301 $\pm$     6&  57 $\pm$   5&  55 $\pm$   6& 439 $\pm$  57&16.26&15.33&14.81&14.16\\
Q1230+0115&1253.145 $\pm$  0.031&  9242 $\pm$     8&  74 $\pm$   8&  72 $\pm$   8& 301 $\pm$  49&14.71&14.30&14.12&13.85\\
\tableline
\end{tabular}
\tablecomments{$*$ Indicates that this absorber was observed with HST/GHRS pre-COSTAR.} 
\tablenotetext{a}{\bvalue after correcting for the GHRS intrumental profile and our pre-fit smoothing (see \S~\ref{sec:Ob}).}
\end{table}
\end{center}
\normalsize

%% file: table2.tex
\setlength{\tabcolsep}{2mm}
\begin{center}
\begin{table}[htbp]
\caption{\hone\ Column Densities (\Nhno) of Possible \lya  features}
\tiny
\label{linelist_poss_nh}
\begin{tabular}{lccccclllc}
\ \\
\tableline \tableline
Target&Wavelength&Velocity&\bmsd&\bobs\tablenotemark{a}\ &\W&\multicolumn{4}{c}{log[\Nh~(in $cm^{-2}$)]} \\
        &\AA                  &km~s$ ^{-1}$ &km~s$ ^{-1}$&km~s$ ^{-1}$&m\AA&\bb=20 &\bb=25&\bb=30 &\bb=col 5\tablenotemark{a}\\
\tableline
$*$3C273&1224.587 $\pm$  0.150&  2199 $\pm$    37&  59 $\pm$  53&  57 $\pm$  55&  29 $\pm$  35&12.77&12.76&12.75&12.74\\
$*$3C273&1234.704 $\pm$  0.029&  4694 $\pm$     7&  71 $\pm$  60&  69 $\pm$  61&  25 $\pm$  29&12.69&12.68&12.68&12.67\\
$*$3C273&1265.701 $\pm$  0.064& 12338 $\pm$    16&  37 $\pm$  24&  33 $\pm$  26&  21 $\pm$  18&12.61&12.61&12.60&12.60\\
$*$3C273&1266.724 $\pm$  0.084& 12590 $\pm$    21&  54 $\pm$  29&  52 $\pm$  30&  24 $\pm$  18&12.68&12.67&12.67&12.66\\
$*$3C273&1268.969 $\pm$  0.076& 13144 $\pm$    19&  45 $\pm$  26&  43 $\pm$  28&  18 $\pm$  15&12.55&12.55&12.55&12.54\\
AKN120&1223.088 $\pm$  0.039&  1829 $\pm$    10&  30 $\pm$  14&  25 $\pm$  17&  64 $\pm$  48&13.15&13.13&13.12&13.13\\
AKN120&1247.267 $\pm$  0.104&  7792 $\pm$    26&  38 $\pm$  29&  34 $\pm$  32&  19 $\pm$  22&12.57&12.57&12.56&12.56\\
ESO141-G55&1249.932 $\pm$  0.036&  8449 $\pm$     9&  23 $\pm$  11&  17 $\pm$  15&  12 $\pm$   8&12.36&12.36&12.36&12.37\\
ESO141-G55&1252.483 $\pm$  0.041&  9078 $\pm$    10&  28 $\pm$  13&  23 $\pm$  16&  12 $\pm$   7&12.35&12.34&12.34&12.34\\
FAIRALL9&1265.407 $\pm$  0.029& 12265 $\pm$     7&  25 $\pm$  13&  19 $\pm$  18&  11 $\pm$   8&12.31&12.31&12.31&12.31\\
H1821+643&1238.014 $\pm$  0.036&  5510 $\pm$     9&  26 $\pm$  11&  21 $\pm$  14&  23 $\pm$  13&12.66&12.65&12.65&12.66\\
H1821+643&1240.569 $\pm$  0.036&  6140 $\pm$     9&  21 $\pm$  10&  14 $\pm$  16&  24 $\pm$  16&12.66&12.66&12.65&12.67\\
H1821+643&1244.966 $\pm$  0.031&  7225 $\pm$     8&  22 $\pm$   8&  15 $\pm$  12&  25 $\pm$  13&12.69&12.68&12.68&12.70\\
H1821+643&1247.362 $\pm$  0.029&  7815 $\pm$     7&  39 $\pm$  31&  36 $\pm$  34&  27 $\pm$  29&12.73&12.73&12.72&12.72\\
H1821+643&1252.477 $\pm$  0.042&  9077 $\pm$    10&  26 $\pm$  12&  21 $\pm$  15&  23 $\pm$  15&12.65&12.65&12.64&12.65\\
H1821+643&1254.874 $\pm$  0.099&  9668 $\pm$    24&  29 $\pm$  26&  24 $\pm$  31&  21 $\pm$  25&12.61&12.61&12.61&12.61\\
MARK279&1237.915 $\pm$  0.029&  5486 $\pm$     7&  34 $\pm$  26&  30 $\pm$  29&  17 $\pm$  18&12.53&12.52&12.52&12.52\\
MARK279&1238.502 $\pm$  0.047&  5631 $\pm$    12&  26 $\pm$  19&  21 $\pm$  23&  18 $\pm$  18&12.54&12.53&12.53&12.54\\
MARK290&1232.797 $\pm$  0.064&  4224 $\pm$    16&  27 $\pm$  23&  21 $\pm$  29&  41 $\pm$  49&12.93&12.92&12.91&12.93\\
MARK290&1235.764 $\pm$  0.044&  4955 $\pm$    11&  30 $\pm$  15&  26 $\pm$  18&  28 $\pm$  19&12.74&12.73&12.73&12.73\\
MARK290&1245.869 $\pm$  0.026&  7447 $\pm$     6&  18 $\pm$   5&   8 $\pm$  12&  18 $\pm$   7&12.54&12.54&12.53&12.55\\
$*$PKS2155-304&1234.767 $\pm$  0.051&  4709 $\pm$    12&  36 $\pm$  21&  32 $\pm$  24&  15 $\pm$  14&12.45&12.45&12.44&12.44\\
$*$PKS2155-304&1246.990 $\pm$  0.029&  7724 $\pm$     7&  35 $\pm$  31&  31 $\pm$  35&  13 $\pm$  16&12.40&12.39&12.39&12.39\\
$*$PKS2155-304&1247.510 $\pm$  0.029&  7852 $\pm$     7&  34 $\pm$  27&  30 $\pm$  31&  13 $\pm$  16&12.41&12.40&12.40&12.40\\
$*$PKS2155-304&1255.084 $\pm$  0.041&  9720 $\pm$    10&  29 $\pm$  13&  24 $\pm$  16&  13 $\pm$   9&12.41&12.40&12.40&12.40\\
$*$PKS2155-304&1256.636 $\pm$  0.042& 10102 $\pm$    10&  30 $\pm$  13&  25 $\pm$  15&  14 $\pm$   8&12.42&12.41&12.41&12.41\\
PKS2155-304&1264.806 $\pm$  0.058& 12117 $\pm$    14&  42 $\pm$  19&  39 $\pm$  20&  31 $\pm$  19&12.78&12.78&12.77&12.77\\
Q1230+0115&1236.045 $\pm$  0.041&  5025 $\pm$    10&  29 $\pm$  12&  24 $\pm$  15&  53 $\pm$  31&13.05&13.04&13.03&13.04\\
Q1230+0115&1242.897 $\pm$  0.044&  6714 $\pm$    11&  25 $\pm$  14&  19 $\pm$  19&  45 $\pm$  35&12.97&12.96&12.95&12.98\\
Q1230+0115&1246.254 $\pm$  0.049&  7542 $\pm$    12&  35 $\pm$  16&  31 $\pm$  18&  44 $\pm$  27&12.96&12.95&12.95&12.94\\
\tableline
\end{tabular}
\tablecomments{ $*$ Indicates that this absorber was observed with HST/GHRS pre-COSTAR.}
\tablenotetext{a}{\bvalue after correcting for the GHRS intrumental profile and our pre-fit smoothing (see \S~\ref{sec:Ob}).}
\end{table}
\end{center}
\normalsize

%% file: table3.tex
\setlength{\tabcolsep}{1.7mm}
\begin{deluxetable}{lccccccc}
\tabletypesize{\footnotesize}
\tablecaption{\label{other_b} Comparison of our \bvalues to other \lya studies}
\tablecolumns{8}
\tablehead{
\colhead{Reference}&
\colhead{\z~range}&
\colhead{\meanz}&
\colhead{$\lambda$ range}&
\colhead{Resolution}&
\colhead{Median \bb}&
\colhead{Mean \bb}&
\colhead{$\sigma_b$\tablenotemark{a}}\\
\colhead{}&
\colhead{}&
\colhead{}&
\colhead{(\noang)}&
\colhead{(\kmsno)}&
\colhead{(\kmsno)}&
\colhead{(\kmsno)}&
\colhead{(\kmsno)}
}
\startdata
\protect{\citet{Lu96}}\tablenotemark{b}      &3.43-3.98&3.70& 5380-6050 &6.6 &27.5(25.9)&34.4(30.6)&29.0(15.3)\\
\protect{\citet{Kim97}}\tablenotemark{c}      &3.20-3.51&3.36& 5105-5484 & 8  &30(27) & NR\tablenotemark{d} & NR \\
\protect{\citet{Hu95}}           &2.54-3.20&2.87& 4300-5100 & 8  &\about 35 &28  & 10 \\
\protect{\citet{Rauch92}}       &2.32-3.40&2.86& 4040-5350 & 23 &33  &36  & 16 \\
\protect{\citet{Kim97}}\tablenotemark{d}      &2.71-3.00&2.86& 4510-4863 & 8  &35.5(30.0) & NR & NR \\
\protect{\citet{Kirkman97}} &2.45-3.05&2.75& 4190-4925 &7.9 &\about 28 &23  &14\\
\protect{\citet{Rauch93}}       &2.10-2.59&2.34& 3760-4360 & \about 8  &26.4 &27.5& NR \\
\protect{\citet{Kim97}}\tablenotemark{d}      &2.17-2.45&2.31& 3850-4195 & 8  &37.7(31.6) & NR & NR \\
\protect{\citet{Carswell84}}     &1.87-2.65&2.26& 3490-4440 & 19 &25.0 & 27.9 & 10.9 \\
\protect{\citet{Khare97}}\tablenotemark{e}         &1.57-2.70&2.13& 3130-4500 & 18 &27.7(30.6)&29.4(32.1)& 7.9 \\
\protect{\citet{Carswell91}}     &1.84-2.15&1.99& 3434-3906 &  9 &33.0 & 34.3 & 14.1 \\
\protect{\citet{Kulkarni96}}\tablenotemark{f} &1.67-2.10&1.88& 3246-3769 & 18 &31.6(29.5)&35.6(31.4)&\about 15\\
\protect{\citet{Savaglio99}}&1.20-2.21&1.70&2670-3900&8.5-50\tablenotemark{g}&28.2&31.1&16.6)\\
\tableline
This Paper (\real)\tablenotemark{h}         &0.002-0.069&0.035&1218-1300&19&34.8(40.7)&38.0(45.4)&15.7(18.6)\\
This Paper (\expanded)\tablenotemark{h}         &0.002-0.069&0.035&1218-1300&19&31.7(35.4)&33.9(40.8)&15.4(18.8)\enddata

\tablenotetext{a}{The standard deviation of the mean, if available.}
\tablenotetext{b}{\protect{\citet{Lu96}} report values for two samples, A and B. The A
sample contains the full set of detected \lya lines, while
the B sample contains only those \lya lines whose values of \Nh~and \bb~ are well
determined. Their results are reported as A(B).}
\tablenotetext{c}{\protect{\citet{Kim97}} report values for two \Nh ranges 13.8 \lt log[\Nh] \lt 16.0 and
13.1 \lt log[\Nh] \lt 14.0. Their results are reported with the second, lower \Nh range in parentheses.
The \protect{\citet{Kim97}} sample was divided into three \z-ranges, with the \meanz = 2.31 and \meanz = 2.86 ranges
reporting the combined results of two sightlines each. The \meanz = 3.36 entry reports the
observations of a single sightline. }
\tablenotetext{d}{NR = Not Reported}
\tablenotetext{e}{\protect{\citet{Khare97}} report values for two samples, A and B. The A
sample contains the set of detected \lya lines that do not show unusual profiles. 
Their B sample includes the full set of detected \lya lines. Their results are reported as A(B).}
\tablenotetext{f}{\protect{\citet{Kulkarni96}} report values for two samples, A and B. The A
sample contains a conservative number of sub-components in blended lines, while
the B sample is more liberal in allowing sub-components. Their results are
reported as A(B).}
\tablenotetext{g} {Savaglio \etl (1999) use a combination of HST/STIS high and medium resolution data, combined
with UCLES/AAT ground-based observations.}
\tablenotetext{h}{We divide both our \real\ and \expanded\ samples into two components. Our A sample does not
include pre-COSTAR  \bvalues since we have found them to be problematic. Our B sample contains all \bvaluesno. Our results
are reported as A(B).}
\end{deluxetable}
\normalsize

%% file: table4.tex
\setlength{\tabcolsep}{2mm}
\begin{center}
\begin{table}[hc]
\caption{\dndzzero~Determinations From Other \lya Studies}
\footnotesize
\label{other_dndz}
\begin{tabular}{llcccc}
 \multicolumn{6}{c}{ } \\
\tableline\tableline
Reference& Instrument/  & \z\ range &  \Wno$_{min}$ & \dndzzero & Our \dndzzero\tablenotemark{a} \\
         & Configuration&          & (\nomang)     &           & at \Wno$_{min}$\\
\tableline
Bahcall \etl 1993a &FOS\tablenotemark{b} 									&0$<\z<$1.3 &320&17.7$\pm$4.3  &18.2$\pm$6.9\\
Bahcall \etl 1996  &FOS\tablenotemark{b}	 								&0$<\z<$1.3 &240&24.3$\pm$6.6  &28.5$\pm$8.6\\
Weymann \etl 1998  &FOS	       &0$<\z<$1.5 &240&32.7$\pm$4.2  &28.5$\pm$8.6\\
Impey \etl 1999    &GHRS/G140L	       &0$<\z<$0.22&240&38.3$\pm$5.3  &28.5$\pm$8.6\\
Tripp  \etl 1998   &GHRS/G140L        &0$<\z<$0.28& 75&  71$\pm$12   &80.0$\pm$14.6\\ 
Tripp  \etl 1998\tablenotemark{c}&GHRS/G140L       &0$<\z<$0.28& 50&102$\pm$16    &134.2$\pm$20.1\\
\tableline
\end{tabular}
\tablenotetext{a}{For comparison to the \dndzzero\ values of the previous studies, this column reports our 
values for  \dndzzero~evaluated at the minimum equivalent width limit (\Wno$_{min}$) of the previous studies.}
\tablenotetext{b}{\protect{\citet{Bahcall93,Bahcall96}} use the G130H, G190H, and G270H HST gratings.}
\tablenotetext{c}{\protect{\citet{Tripp98}} combine their observations with observations of 3C~273
from  \protect{\citet{Morris93}}, who include observations taken with both the GHRS/G140L and GHRS/G160M.}
\end{table}
\end{center}
\normalsize

%% file: table5.tex
\setlength{\tabcolsep}{1mm}
\renewcommand{\arraystretch}{0.8}
\begin{center}
\small
\begin{table}[htbp]
\caption{Differential, \nNh, and integrated, \iNh, determinations of $\beta$ and log[\NofNhno] for the low-\z~\lya absorbers.
\label{int_beta_table}}
\begin{tabular}{lcccccr}
\multicolumn{7}{c}{}\\
%\multicolumn{7}{c}{}\\
\multicolumn{7}{l}{\small 12.3 $\le$ log[\Nh] $\le$ 14.0 }\\
%\multicolumn{7}{c}{}\\
\multicolumn{2}{c}{}&\multicolumn{2}{c}{\small Differential}&\multicolumn{2}{c}{\small Integrated}&\multicolumn{1}{c}{}\\
\tableline
Sample & \bvalue & $\beta$ & log[\NofNh] & $\beta$ & log[\NofNh]  & $N$\\
\tableline
Definite &20&1.83 $\pm$0.16&12.8 $\pm$ 2.1&1.70 $\pm$0.05&11.0 $\pm$ 0.7&          68 \\
Definite &25&1.83 $\pm$0.15&12.8 $\pm$ 1.9&1.72 $\pm$0.06&11.3 $\pm$ 0.7&          71 \\
Definite &30&1.80 $\pm$0.15&12.5 $\pm$ 2.0&1.74 $\pm$0.06&11.5 $\pm$ 0.7&          71 \\
Definite &\bobs&1.84 $\pm$0.14&13.0 $\pm$ 1.9&1.77 $\pm$0.06&11.9 $\pm$ 0.8&          75 \\
Expanded &20&1.81 $\pm$0.13&12.6 $\pm$ 1.7&1.78 $\pm$0.05&12.1 $\pm$ 0.6&          98 \\
Expanded &25&1.81 $\pm$0.12&12.6 $\pm$ 1.6&1.81 $\pm$0.05&12.4 $\pm$ 0.7&         101 \\
Expanded &30&1.80 $\pm$0.12&12.5 $\pm$ 1.6&1.83 $\pm$0.05&12.7 $\pm$ 0.7&         101 \\
Expanded &\bobs&1.83 $\pm$0.11&12.8 $\pm$ 1.5&1.86 $\pm$0.06&13.0 $\pm$ 0.7&         105 \\
 \tableline
\multicolumn{7}{c}{}\\
\multicolumn{7}{l}{\small 14.0 $\le$ log[\Nh] $\le$ 16.0 \tiny}\\
%\multicolumn{7}{c}{}\\
\multicolumn{2}{c}{}&\multicolumn{2}{c}{\small Differential}&\multicolumn{2}{c}{\small Integrated}&\multicolumn{1}{c}{}\\
\tableline
Sample & \bvalue & $\beta$ & log[\NofNh] & $\beta$ & log[\NofNh]  & $N$\\
\tableline
Definite/Expanded &20&1.07 $\pm$0.18& 2.0 $\pm$ 2.7&1.34 $\pm$0.13& 6.2 $\pm$ 1.9&          13 \\
Definite/Expanded &25&1.04 $\pm$0.39& 1.5 $\pm$ 5.7&1.43 $\pm$0.35& 7.4 $\pm$ 5.2&          10 \\
Definite/Expanded &30&1.40 $\pm$0.64& 6.7 $\pm$ 9.5&1.81 $\pm$0.73&12.6 $\pm$10.7&          10 \\
 \tableline
\multicolumn{7}{c}{}\\
\multicolumn{7}{l}{\small 12.3 $\le$ log[\Nh] $\le$ 16.0 \tiny}\\
%\multicolumn{7}{c}{}\\
\multicolumn{2}{c}{}&\multicolumn{2}{c}{\small Differential}&\multicolumn{2}{c}{\small Integrated}&\multicolumn{1}{c}{}\\
\tableline
Sample & \bvalue & $\beta$ & log[\NofNh] & $\beta$ & log[\NofNh]  & $N$\\
\tableline
Definite &20&1.44 $\pm$0.07& 7.6 $\pm$ 1.0&1.53 $\pm$0.04& 8.9 $\pm$ 0.5&          81 \\
Definite &25&1.58 $\pm$0.10& 9.6 $\pm$ 1.4&1.66 $\pm$0.06&10.5 $\pm$ 0.8&          81 \\
Definite &30&1.64 $\pm$0.12&10.3 $\pm$ 1.7&1.71 $\pm$0.08&11.3 $\pm$ 1.0&          81 \\
Definite &\bobs&1.60 $\pm$0.17& 9.8 $\pm$ 2.3&1.77 $\pm$0.11&11.9 $\pm$ 1.4&          81 \\
Expanded &20&1.50 $\pm$0.07& 8.5 $\pm$ 0.9&1.57 $\pm$0.04& 9.5 $\pm$ 0.5&         111 \\
Expanded &25&1.65 $\pm$0.10&10.4 $\pm$ 1.4&1.71 $\pm$0.06&11.3 $\pm$ 0.8&         111 \\
Expanded &30&1.71 $\pm$0.12&11.2 $\pm$ 1.6&1.78 $\pm$0.08&12.1 $\pm$ 1.0&         111 \\
Expanded &\bobs&1.70 $\pm$0.16&11.1 $\pm$ 2.2&1.85 $\pm$0.11&13.0 $\pm$ 1.4&         111 \\
\tableline
\end{tabular}
\end{table}\end{center}\normalsize

%% file: table6.tex
\setlength{\tabcolsep}{2mm}
\begin{center}
\begin{table}[htbp]
\caption{\lya Line Pairs with Velocity Separations of 50--150\kmsno}
\small
\label{linelist_pairs}
\begin{tabular}{cccccccl}
\ \\
\tableline \tableline
$\lam_{cen}$&\Dlam&$\Delta$V&$\Wno_1$ &$\Wno_2$&$\bb_1$ &$\bb_2$
&Sightline\\ (\noang)&(\nomang)&(\nokmsno)&(\nomang)&(\nomang)&(\nokmsno)&(\nokmsno)&\\
\tableline
1247.76& 0.38&          88&          20&         146&          37&          31&AKN120\\
1247.88& 0.62&         145&          20&          64&          37&          28&AKN120\\
1248.07& 0.24&          57&         146&          64&          31&          28&AKN120\\
1264.89& 0.42&          95&          30&          27&          46&          28&FAIRALL9\\
1247.76& 0.35&          82&          39&          67&          25&          36&H1821+643\\
1247.37& 0.32&          74&          48&          21&          32&          37&MARK279\\
1241.66& 0.30&          69&          57&          39&          24&          26&MARK279\\
1236.60& 0.60&         142&         207&          24&          85&          28&MARK817\\
1235.87& 0.25&          59&          64&          81&          69&          60&PKS2155-304\\
1236.21& 0.43&         102&          81&         218&          60&          81&PKS2155-304\\
1238.56& 0.22&          52&          28&          39&          36&          33&PKS2155-304\\
1222.59& 0.32&          78&         384&         241&          56&          42&Q1230+0115\\
1222.98& 0.46&         113&         241&         141&          42&          50&Q1230+0115\\
\tableline
\end{tabular}
\end{table} \end{center} \normalsize

%% file: table7.tex
\setlength{\tabcolsep}{2.5mm}
\begin{table}[htbc]
\begin{center}
\caption{Metallicity upper limits (\foursig) for our composite low-\z~\lya absorber}
\footnotesize
\label{metals}
\begin{tabular}{lc|ccc}
\ \\
\tableline\tableline
&  &\multicolumn{3}{|c}{\W$_{metals}$ (Definite Sample)}\\
Feature&$\lambda$(\Ang)&All&\Wlya$<\Wcutoff$&\Wlya$>\Wcutoff$\\
\tableline
SiIII& 1206.50 & 2 & 2&5\\
MnII & 1201.12 & 2 & 2&7\\
NI   & 1200.71 & 2 & 2&7\\
NI   & 1200.22 & 2 & 2&7\\
NI   & 1199.55 & 2 & 2&8\\
MnII & 1199.39 & 2 & 2&8\\
SiII & 1197.39 & 2 & 3&12\\
MnII & 1197.18 & 2 & 3&12\\
SiII & 1193.29 & 3 & 4&18\\
SiII & 1190.42 & 4 & 5&22\\
\tableline
\end{tabular}
\end{center}
\end{table} \normalsize